\UseRawInputEncoding
\documentclass[aip,twocolumn,10pt]{revtex4}%
\usepackage{amsmath}
\usepackage{amsfonts}
\usepackage{amssymb}
\usepackage{graphics}
\usepackage{graphicx}
\providecommand{\U}[1]{\protect\rule{.1in}{.1in}}

\begin{document}
	
\title{High-fidelity qubit readout using interferometric directional Josephson devices}
\author{Baleegh Abdo}
\author{Oblesh Jinka}
\author{Nicholas T. Bronn}
\author{Salvatore Olivadese}
\author{Markus Brink}
\affiliation{IBM Quantum, IBM Research Center, Yorktown Heights, New York 10598, USA.}
\date{\today}

\begin{abstract}
	Nonreciprocal microwave devices, such as circulators and isolators, are needed in high-fidelity qubit readout schemes to unidirectionally route the readout signals and protect the qubits against noise coming from the output chain. However, cryogenic circulators and isolators are prohibitive in scalable superconducting architectures because they rely on magnetic materials. Here, we perform a fast ($750$ ns) high-fidelity ($95\%$) quantum nondemolition readout of a coherent superconducting qubit ($T_{\rm{1}}=52$ $\mu s$, $T_{\rm{2E}}=35$ $\mu s$) without any nonreciprocal magnetic devices. We employ in our readout chain a microwave-controlled qubit-Readout Multi-Chip Module (qRMCM) that integrates interferometric directional Josephson devices consisting of an isolator and a reconfigurable isolator/amplifier device and an off-chip low-pass filter. Using the qRMCM, we demonstrate isolation up to $45$ dB within $13$ MHz, when both directional devices are operated as isolators, and low-noise amplification in excess of $10$ dB within a dynamical bandwidth of $10$ MHz, when the reconfigurable device is operated as an amplifier. We also demonstrate using the variable isolation of the qRMCM an in-situ enhancement of the qubit coherence times $T_{\rm{\varphi}}$ and  $T_{\rm{2E}}$ by two orders of magnitude (i.e.,  from $T_{\rm{\varphi}}=T_{\rm{2E}}=0.5$ $\mu s$ to $T_{\rm{\varphi}}=90$ $\mu s$ and $T_{\rm{2E}}=50$ $\mu s$). Furthermore, by directly comparing the qRMCM performance to a state-of-art configuration (with $T_{\rm{2E}}\approx 2T_{\rm{1}}$) that employs a pair of wideband magnetic isolators, we find that the excess pure dephasing measured with the qRMCM (for which $T_{\rm{2E}}\approx T_{\rm{1}}$) is likely limited by residual thermal photon population in the readout resonator. Improved versions of the qRMCM could replace magnetic circulators and isolators in large superconducting quantum processors.  
\end{abstract}

\maketitle
\newpage

\section{Introduction}

Nonreciprocity breaks the transmission-coefficient symmetry for light upon exchanging sources and detectors. A common method for  breaking reciprocity is guiding light through magnetic materials exhibiting magneto-optical effects, such as the Faraday effect \cite{FaradyEffect,EMnonreciprocity}. Other nonreciprocity schemes include operating in the nonlinear regime \cite{NonlinearityNR}, employing the quantum Hall effect \cite{circulatorDiVincenzo1,circulatorDiVincenzo2,Hallcirc}, and parametrically modulating a certain physical property of the system \cite{AhranovBohmPhotonic,AhranovBohmMixers,circulatorLehnert,NRAumentado1,NoiselessCirc,BroadbandPhaseMod}. 

Due to their ability to break the transmission symmetry, nonreciprcocal devices play critical roles in a variety of basic science and technology applications, requiring, for example, signal transport control, separation of input from output in reflective or communication setups, and source protection against backscatter or detector backaction. In particular, in the realm of superconducting quantum processors, nonreciprocal microwave devices, such as circulators and isolators, are critical for performing high-fidelity quantum nondemolition (QND) measurements \cite{DevoretScience,QuantumJumps,JPADicarloReset,stabilizetrajectory,SunTrackPhotonJumps,FeedbackJPC}. With relatively low loss, they route readout signals to and from quantum processors in a directional manner and protect qubits against noise coming from the output chain. However, state-of-the-art cryogenic circulators and isolators are prohibitive in scalable architectures because they are bulky and rely on magnetic materials and strong magnetic fields \cite{Pozar,Collin}, which are difficult to integrate on chip and incompatible with superconducting circuits. 

In an attempt to solve this scalability challenge, a variety of viable alternative circulator and isolator schemes have been proposed and realized, which use photonic transitions between coupled resonance modes \cite{ReconfJJCircAmpl,NRAumentado2}, the Hall-effect \cite{Hallcirc}, frequency conversion in nonlinear transmission lines \cite{FreqConvIso}, frequency conversion combined with delay lines \cite{WidelyTunableCirculator,FreqConvAndDelay}, dynamical modulation of transfer switches incorporated with delay lines \cite{CircTransferSwitches}, and reservoir engineered optomechanical interactions \cite{MechOnChipCirc,NonrecipMwOptoMech,NonreciprocalResEng}. However, despite this great progress in the development of alternative directional devices and readout schemes, demonstrating high-fidelity QND readout of a superconducting qubit with relatively high coherence without using any magnetic isolators and circulators in the readout chain remains an outstanding challenge \cite{
	JTWPA,HFwithOnChipPhDetector,effQmeasWithNRAmp,effQmeasWSwitch}. Here, we achieve this milestone by building a quantum Readout Multi-Chip Module (qRMCM), which integrates into a Printed Circuit Board (PCB) interferometric directional Josephson devices, a superconducting directional coupler and a Purcell filter and by incorporating an off-the-shelf low-pass microwave filter at the output of the qRMCM. Also, importantly, our readout scheme operates in continuous-wave mode, which is highly beneficial in scalable architectures because it is compatible with frequency multiplexed readout, unlike, for example, a Superconducting Low-inductance Undulatory Galvanometer (SLUG) microwave amplifier, which exhibits a strong inherent reverse isolation but operates in pulsed mode \cite{SlugReadout}. 

Furthermore, owing to the concatenation of more than one directional device in the qRMCM, we achieve an isolation record in a nonreciprocal superconducting circuit of about $45$ dB at the readout frequency. Moreover, by varying the isolation in-situ, enabled by the qRMCM, we investigate the dependence of qubit dephasing on the isolators performance and noise originating from the $4$ K stage and the High-Electron-Mobility-Transistor (HEMT) amplifier (which is a long-standing, critical microwave component in superconducting qubit readout chains).    

Central to achieving these results are the Multi-Path Interoferometric Josephson ISolator (MPIJIS) \cite{MPIJIS} and a reconfigurable directional device that can be operated as MPIJIS or a Multi-Path Interoferometric Josephson Directional Amplifier (MPIJDA) \cite{JDA,JDAQST}, both of which are key components in the proof-of-principle qRMCM presented here. The MPIJIS is formed by coupling two nominally identical nondegenerate Josephson mixers \cite{JPCreview} via their respective distinct spatial and spectral eigenmodes `a' and `b'. One mode in this scheme, e.g., `a', supports input and output propagating signals within the device bandwidth, while the other, e.g., `b', serves as an internal mode of the system coupled to a dissipation port. By parametrically modulating the inductive coupling between the modes, an artificial gauge-invariant potential for microwave photons is generated, which imprints nonreciprocal phase shifts on the transmitted signals through the mixers, undergoing frequency conversion. By further embedding the mixers in an interferometric setup, a unidirectional transmission of propagating signals is created owing to constructive and destructive wave-intereference taking place between different paths in the device.

However, unlike the MPIJIS device presented in Ref.\,\cite{MPIJIS}, which (1) integrates on-chip superconducting and off-chip normal-metal circuits, and (2) requires, for its operation, two same-frequency phase-locked microwave drives injected through two input lines in the dilution fridge, the present MPIJIS is a superconducting on-chip device that is operated by a single microwave drive fed through one input line. These changes lead to several key advantages, such as (1) eliminating the losses in the normal metal parts, (2) paving the way for size reduction using lumped-element designs, and (3) requiring only one microwave source and input line on a par with Josephson parametric amplifiers. The latter leads to savings in the hardware resources required per qubit, easier tune-up procedures, enhanced stability over time, and simpler control circuitry. These gains become even more pronounced when comparing this single-pump device with nonreciprocal schemes, which require multiple pumps \cite{ReconfJJCircAmpl,NRAumentado1,NRAumentado2}.

Similarly, the MPIJDA device, which is the dual of the MPIJIS, shares the same circuit as the MPIJIS but operates in a different mode. More specifically, the nondegenerate Josephson mixers of the device are operated in low-gain amplification mode instead of frequency conversion without photon gain, which are primarily set by the pump frequency driving the mixers, i.e., the sum of the resonance frequencies of modes 'a' and 'b' in the MPIJDA case and their difference in the MPIJIS case.

\begin{figure*}
	[tb]
	\begin{center}
		\includegraphics[
		width=1.5\columnwidth 
		]%
		{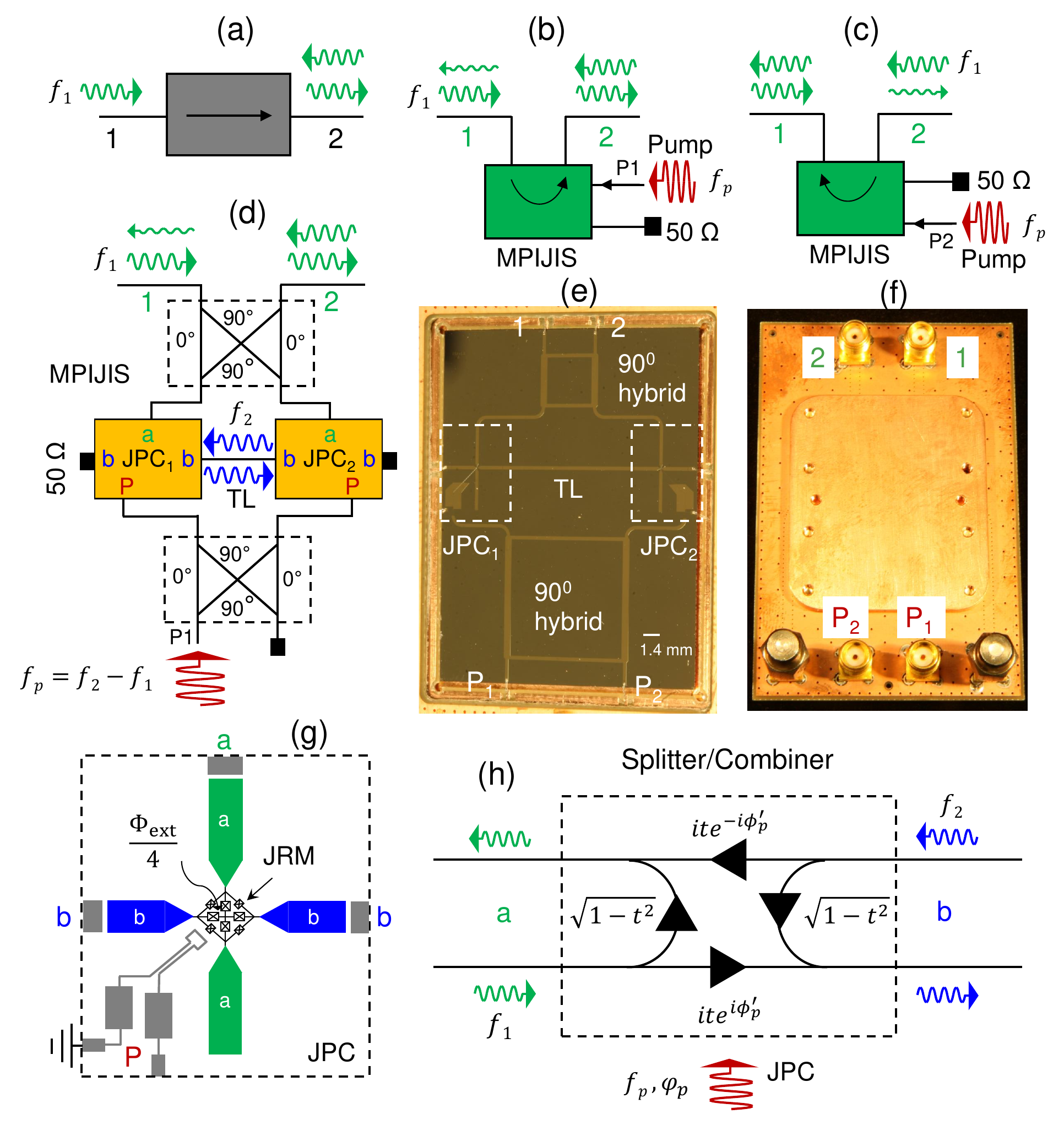}
		\caption{On-chip single-pump Multi-Path Interferometric Josephson ISolator (MPIJIS). (a) Isolator circuit symbol. In ideal isolators, input signals propagating in the direction of the arrow are fully transmitted; signals propagating in the opposite direction are blocked. (b) and (c) circuit symbols for the MPIJIS. The direction of the arrow depends on which pump port of the device is driven. (d) A block diagram of the MPIJIS scheme. It consists of two identical JPCs coupled in an interferometric setup. (e) Photo of the MPIJIS chip. The two JPCs, the signal hybrid, the coupling transmission line (TL), and the pump hybrid are all superconducting and realized on the same chip. (f) Photo of the PCB and bottom cover of the MPIJIS. The length and width of the PCB are $7.62$ cm and $5.08$ cm, respectively. (g) JPC design used in the MPIJIS. (h) Signal flow graph of the JPC operated in frequency conversion mode without photon gain. Signals impinging on ports `a' and `b' undergo frequency conversion and transmission to the other port with amplitude $t$ and get reflected off with amplitude $\sqrt{1-t^2}$. The transmitted signal acquires a nonreciprocal phase shift $\pm\phi^{\prime}_{p}$, where $\phi^{\prime}_{p}$ is the generalized pump phase.    
		}
		\label{Device}
	\end{center}
\end{figure*}

\begin{figure*}
	[tb]
	\begin{center}
		\includegraphics[
		width=1.5\columnwidth 
		]%
		{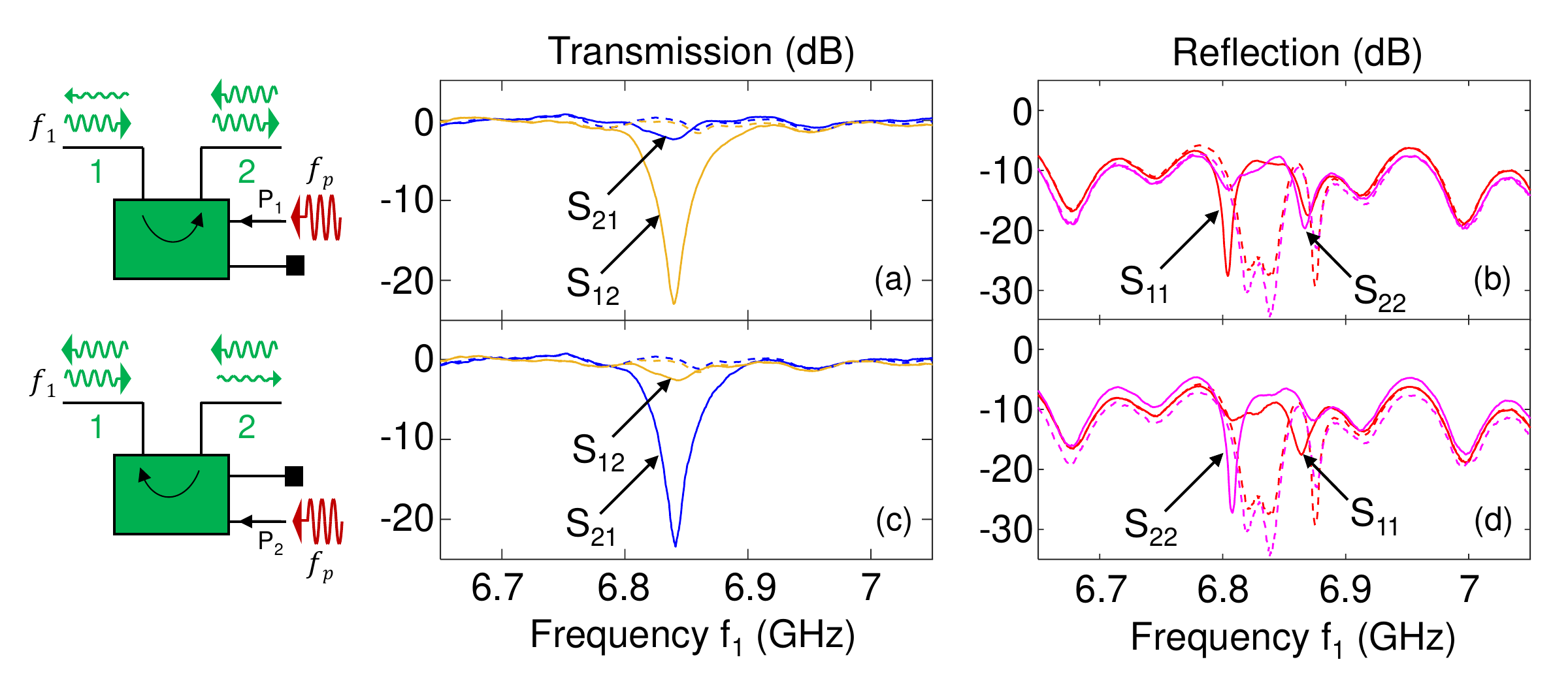}
		\caption{Measured scattering parameters of the MPIJIS. (a) and (b) display transmission parameters, i.e., $|S_{21}|^2$ (blue), $|S_{12}|^2$ (orange) and reflection parameters, i.e., $|S_{11}|^2$ (red), $|S_{22}|^2$ (magenta) measured versus signal frequency for a pump applied to $\rm{P}_1$ at $f_p=2.758$ GHz. The solid and dashed curves correspond to pump on and off, respectively. (c) and (d) are the same as (a) and (b), except the pump is applied to $\rm{P}_2$ instead.     
		}
		\label{wpt}
	\end{center}
\end{figure*}

\begin{figure}
	[tb]
	\begin{center}
		\includegraphics[
		width=\columnwidth 
		]%
		{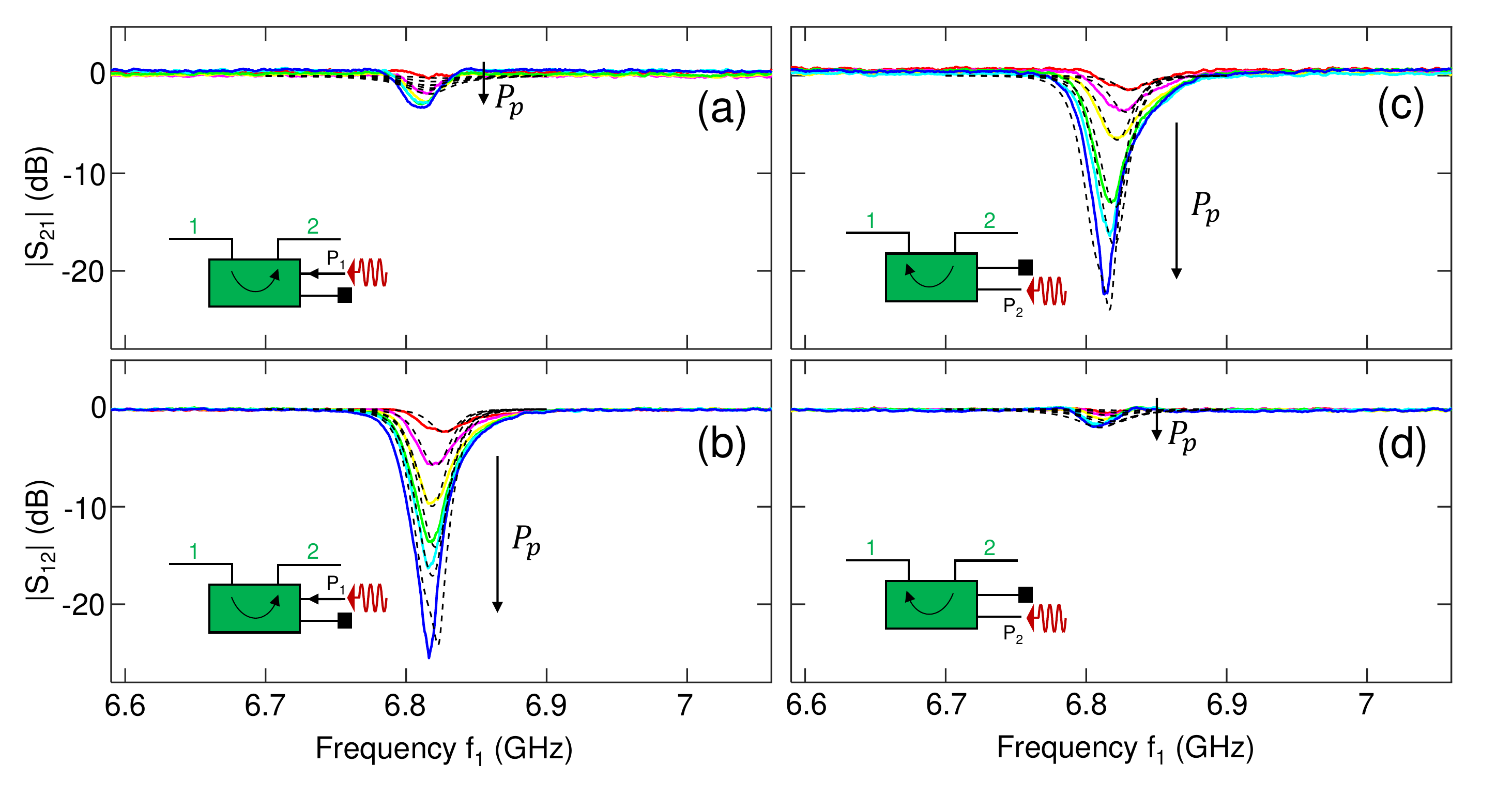}
		\caption{MPIJIS transmission versus pump power. Transmission parameters measured versus signal frequency for varying pump powers applied at $f_p=2.727$ GHz. (a) and (b) exhibit measured transmission parameters $|S_{21}|^2$, $|S_{12}|^2$ versus frequency for varying pump powers applied to $\rm{P}_1$. (c) and (d) are the same as (a) and (b) obtained for varying pump powers applied to $\rm{P}_2$. The black dashed curves represent a calculated response of the MPIJIS. The calculation sequence and the parameters used are included in Appendix B.
		}
		\label{VaryPp}
	\end{center}
\end{figure}

\begin{figure*}
	[tb]
	\begin{center}
		\includegraphics[
		width=1.6\columnwidth 
		]%
		{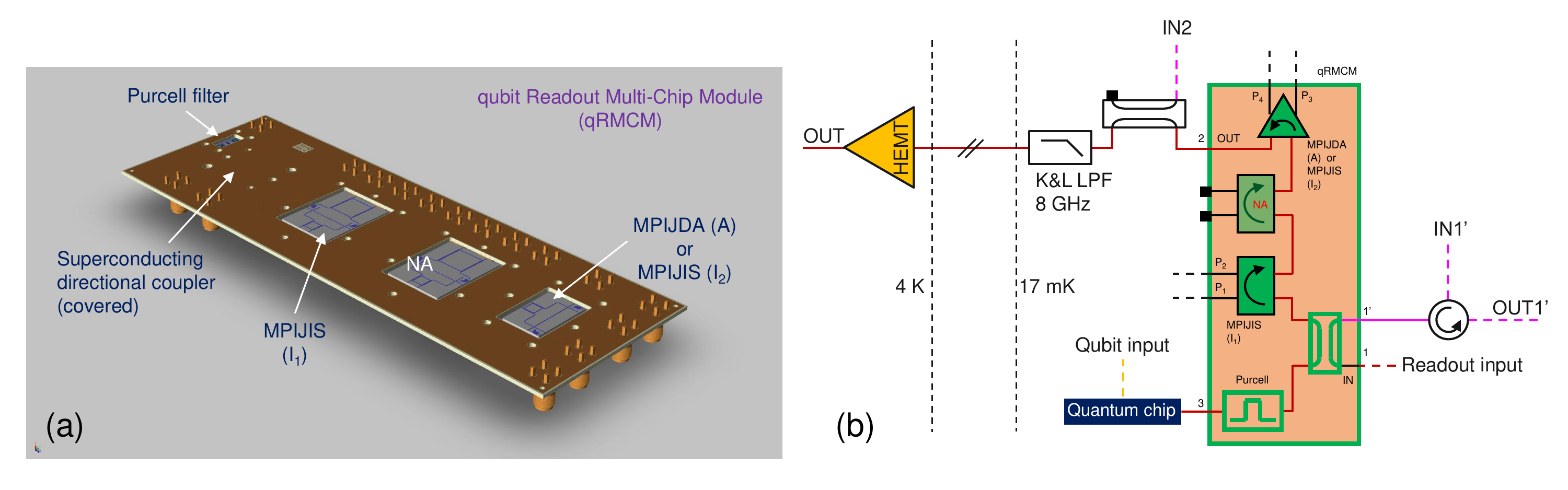}
		\caption{(a) qRMCM image featuring a printed circuit board which integrates several superconducting microwave components: a Purcell filter, a superconducting directional coupler, two MPIJIS devices, and a reconfigurable MPIJDA (A)/MPIJIS ($\rm{I_{2}}$) device. The length and width of the qRMCM are $19.69$ cm and $6.73$ cm, respectively. (b) Simplified schematic of the experimental setup used to obtain the measurement results of Figs.\,\ref{IsoMagData}, \ref{DoubleIsoData}, \ref{DephasingData}. A more detailed schematic of the setup is shown in Fig.\, \ref{Round2FullSetup}. The qRMCM has three main ports that connect to the readout input line, the quantum chip and the readout output line. The pump is fed to \textbf{$\rm{I_{1}}$} via one input line ($\rm{P}_1$ or $\rm{P}_2$) and to the reconfigurable device via two input lines ($\rm{P}_3$ and $\rm{P}_4$). The isolated port of the directional coupler denoted $1'$ is used to measure the transmission through the directional devices of the qRMCM in the forward and backward direction. All other auxiliary ports are terminated by $50$ Ohm loads. The signal path outlined in red represents the qubit readout path. The magenta lines represent auxiliary input and output lines that enable probing the qRMCM transmission in the forward and backward directions.    
		}
		\label{qRMCM}
	\end{center}
\end{figure*}

\begin{figure*}
	[tb]
	\begin{center}
		\includegraphics[
		width=1.8\columnwidth 
		]%
		{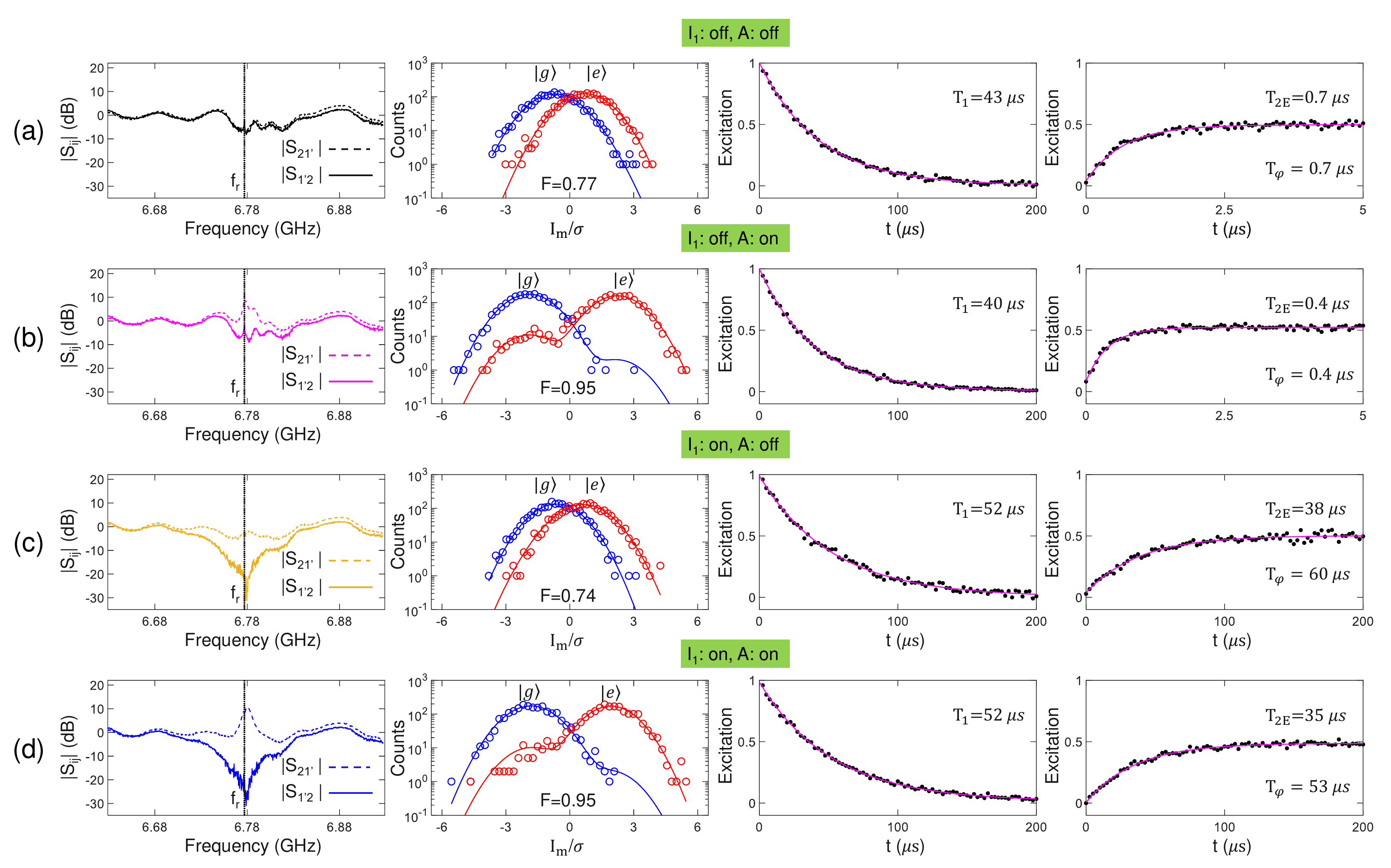}
		\caption{High-fidelity qubit readout without cryogenic magnetic circulators and isolators. A simplified (detailed) version of the experimental setup used for taking this data set is shown in Fig.\, \ref{qRMCM}(b) (Fig.\,\ref{Round2FullSetup}). The columns exhibit, from left to right, the forward ($|S_{21'}|^2$) and backward ($|S_{1'2}|^2$) transmission parameters of the qRMCM, followed by respective measurements of readout fidelity, $T_{\rm{1}}$, and $T_{\rm{2E}}$. The measurements exhibited in (a)-(d) correspond to different $\rm{MPIJIS_1}$ ($\rm{I_1}$) and $\rm{MPIJDA}$ ($\rm{A}$) configurations outlined in the green headings. The vertical dashed black line in the first-column plots indicates the location of the readout frequency.    
		}
		\label{IsoMagData}
	\end{center}
\end{figure*}

\begin{figure*}
	[tb]
	\begin{center}
		\includegraphics[
		width=1.8\columnwidth 
		]%
		{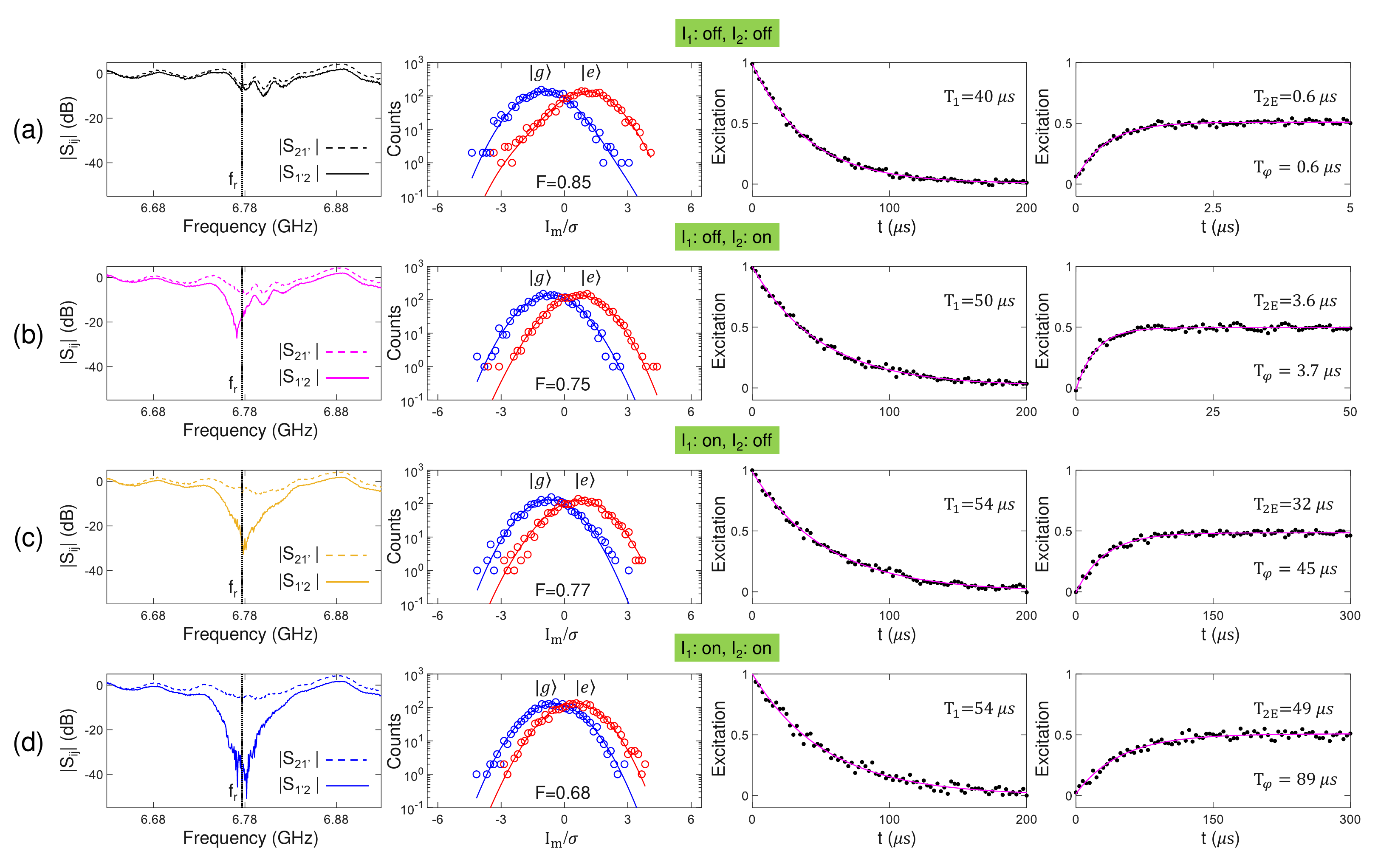}
		\caption{Measured qubit coherence with the reconfigurable Josephson directional device in the qRMCM operated as a second interferometric Josephson isolator. A simplified (detailed) version of the experimental setup used for taking this data set is shown in Fig.\,\ref{qRMCM}(b) (Fig.\,\ref{Round2FullSetup}). The columns exhibit, from left to right, the forward ($|S_{21'}|^2$) and backward ($|S_{1'2}|^2$) transmission parameters of the qRMCM, followed by respective measurements of readout fidelity, $T_{\rm{1}}$, and $T_{\rm{2E}}$. The measurements exhibited in (a)-(d) correspond to different $\rm{MPIJIS_1}$ ($\rm{I_1}$) and $\rm{MPIJIS_2}$ ($\rm{I_2}$) configurations outlined in the green headings. The vertical dashed black line in the first-column plots indicates the location of the readout frequency.    
		}
		\label{DoubleIsoData}
	\end{center}
\end{figure*}

\begin{figure*}
	[tb]
	\begin{center}
		\includegraphics[
		width=1.8\columnwidth 
		]%
		{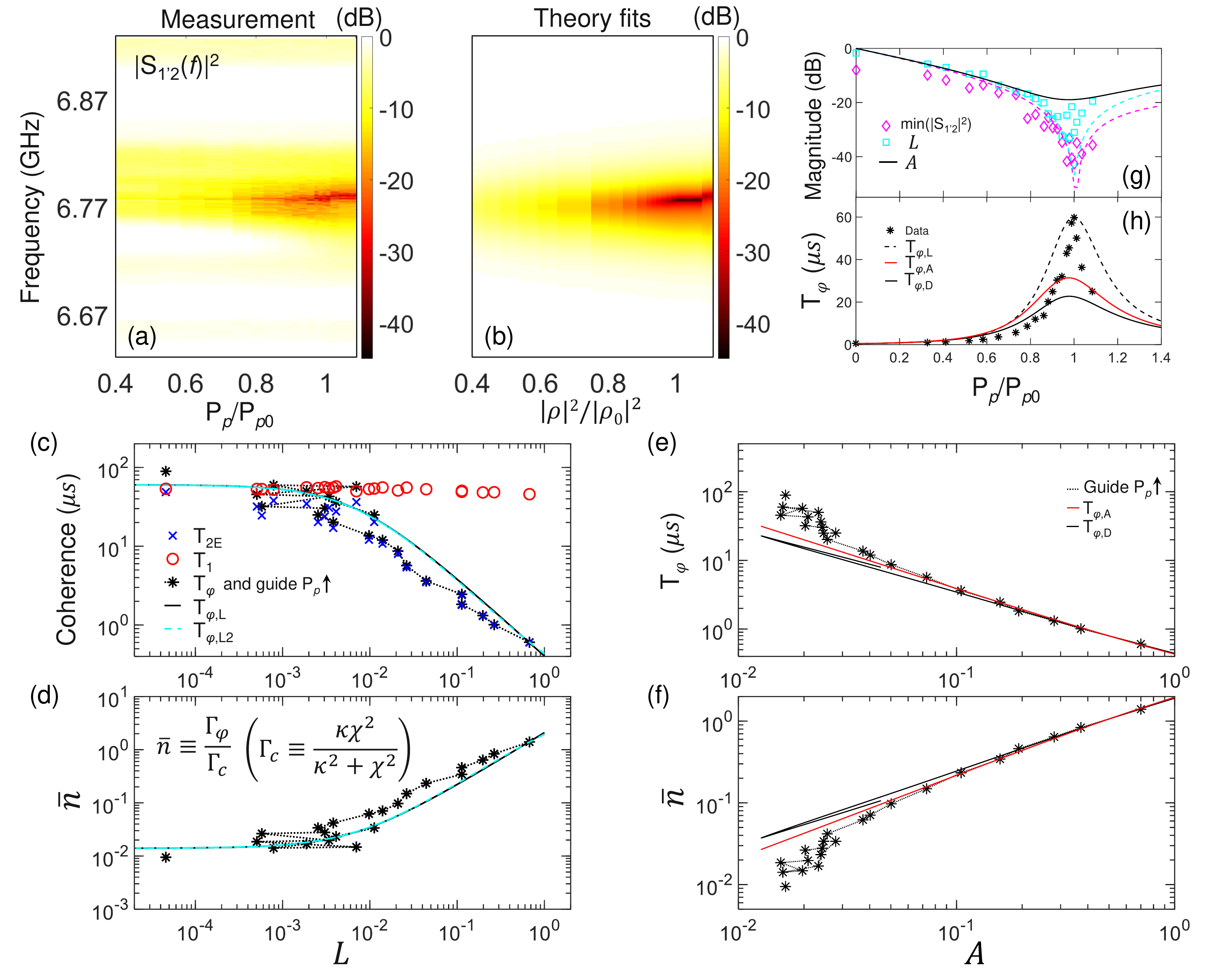}
		\caption{Measured (a) and calculated (b) isolation curves $|S_{1'2}|^2$ of MPIJIS ($\rm{I_1}$) versus frequency and normalized pump power. The calculated response uses the effective two-port model of the MPIJIS presented in Appendix B, with the parameters $\gamma_a/2\pi=100$ MHz, $\gamma_b/2\pi=140$ MHz, $\omega_a/2\pi=6.765$ GHz, $\omega_p/2\pi=2.712$ GHz, and $|\alpha|=0.4$. (c) Qubit coherence times measured versus the isolation magnitude at the readout frequency $L$. (d) Extracted $\bar{n}$ versus $L$. The solid black and dashed cyan fits in (c) and (d) are calculated using Eq.\, (\ref{DephasingRate}) and Eq.\,(\ref{DephasingRate2}), respectively.  Measured $T_{\rm{\varphi}}$ (e) and extracted $\bar{n}$ (f) (black stars) versus the combined filtering parameter $A$ of the MPIJIS and readout resonator. The dotted black line in (c)-(f) is a guide to the eye for increasing $P_p$, and the data points unconnected by the dotted line belongs to configuration (d) of the MPIJIS-MPIJIS experiment (Fig.\,\ref{DoubleIsoData}(d)). The solid red and black curves in (e) and (f) represent calculated fits (see text for details). The black curve accounts for the saturation effect of the MPIJIS due to pump depletion. (g) Measured isolation magnitude at the minima points (diamond) and at the readout frequency (squares) versus normalized pump power. The corresponding theory fits are plotted as dashed magenta and cyan curves. The solid black curve exhibits the calculated parameter $A$. (h) Measured $T_{\rm{\varphi}}$ (black stars) versus normalized pump power. The dashed black, solid red, and black curves represent theory fits based on three models of the MPIJIS impact on $T_{\rm{\varphi}}$ (see text for details).      
		}
		\label{DephasingData}
	\end{center}
\end{figure*}

\begin{figure*}
	[tb]
	\begin{center}
		\includegraphics[
		width=1.6\columnwidth 
		]%
		{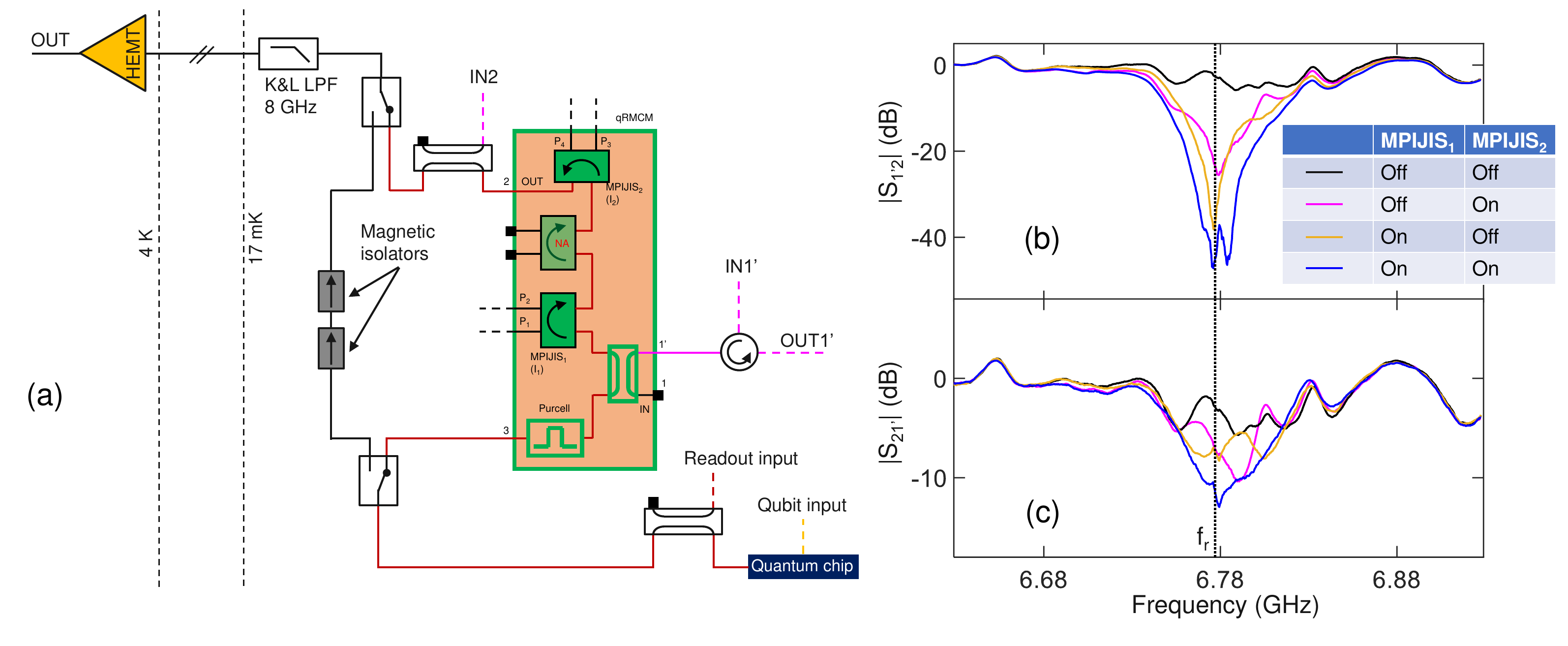}
		\caption{(a) Simplified schematic of the experimental setup used for directly comparing, in the same cooldown, the performance of the qRMCM and two commercial magnetic isolators connected in series as they can be switched in and out of the qubit output chain. This setup is used to obtain the measurement result of Fig.\,\ref{magIsoJISComp}. A more detailed schematic is shown in Fig.\,\ref{IsoMagComFullSetup}. The signal path outlined in red represents the qubit readout path with the qRMCM switched in. The magenta lines represent auxiliary input and output lines that enable probing the qRMCM transmission in the forward and backward directions. (b) and (c) exhibit the measured isolation $|S_{1'2}|^2$ and transmission $|S_{21'}|^2$ of the qRMCM versus frequency for the different operation configurations of the qRMCM outlined in the table. The qubit coherence measurements corresponding to the three bottom configurations (in which at least one of the MPIJIS is on) are shown in Fig.\,\ref{magIsoJISComp}. The vertical dashed black line in plots (b) and (c) indicates the location of the readout frequency.
		}
		\label{magIsolJISsetupSimple}
	\end{center}
\end{figure*} 

\begin{figure*}
	[tb]
	\begin{center}
		\includegraphics[
		width=1.8\columnwidth 
		]%
		{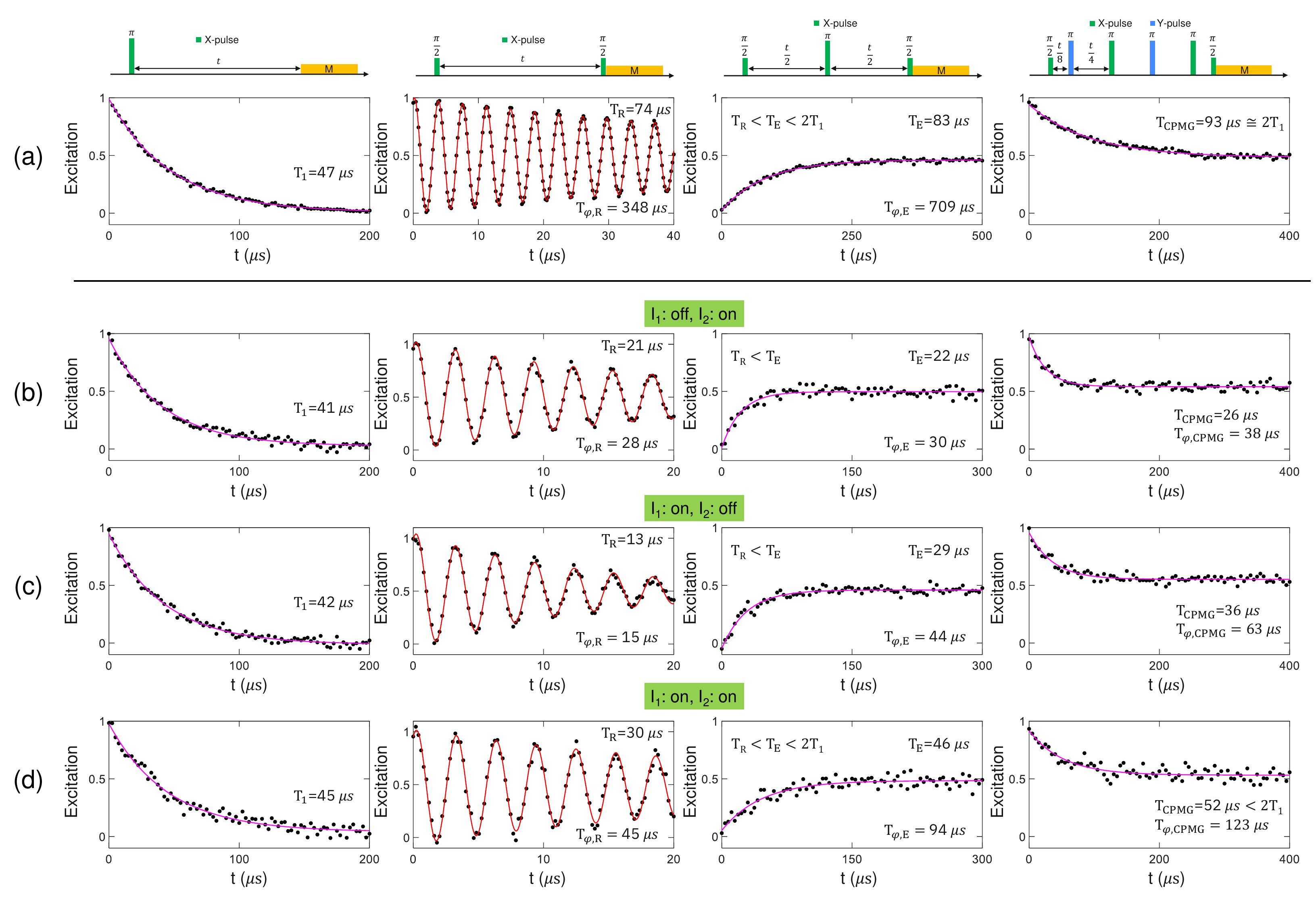}
		\caption{(a) Qubit coherence measurements taken with the output chain incorporating two commercial wideband magnetic isolators. (b)-(d) Similar coherence measurements taken in the same cooldown with the output chain incorporating the qRMCM. A simplified (detailed) version of the experimental setup used for taking this data set is shown in Fig.\,\ref{magIsolJISsetupSimple}(a) (Fig.\,\ref{IsoMagComFullSetup}). The columns from left to right exhibit $T_{\rm{1}}$, $T_{\rm{R}}$, $T_{\rm{E}}$, and $T_{\rm{CPMG}}$ measurement results, respectively. The pulse sequence employed in the various cases is shown at the top. The measurements exhibited in (b)-(d) correspond to different $\rm{MPIJIS_1}$ ($\rm{I_1}$) and $\rm{MPIJIS_2}$ ($\rm{I_2}$) configurations outlined in the green headings. The corresponding qRMCM transmission parameters measured for these configurations are shown in Fig.\,\ref{magIsolJISsetupSimple} (b), (c).       
		}
		\label{magIsoJISComp}
	\end{center}
\end{figure*}

\section{The MPIJIS device}

Isolators, whose circuit symbol is shown in Fig.\,\ref{Device}(a), are two-port microwave devices, which transmit microwave signals propagating in the direction of the arrow, i.e., from port 1 to 2, at frequency $f_1$ (within the device bandwidth), while blocking signals propagating in the opposite direction. Here, we realize an on-chip superconducting MPIJIS that requires a single pump for its operation, as illustrated in Fig.\,\ref{Device}(b),(c). The basic building block of the device, whose photos are shown in Fig.\,\ref{Device}(e),(f), is the Josephson parametric converter (JPC) \cite{JPCreview,hybridLessJPC,microstripJPC,JPCnature}, which functions as a lossless nondegenerate three-wave mixing device. The JPC, as illustrated in Fig.\,\ref{Device}(g), is comprised of two half-wavelength, microstrip resonators denoted `a' and `b', which intersect in the middle at an inductively-shunted Josephson ring modulator (JRM) \cite{Roch}, serving as a dispersive nonlinear medium. The resonators are characterized by flux-tunable resonance frequencies $\omega_{a,b}(\Phi_{\rm{ext}})/2\pi$, where $\Phi_{\rm{ext}}$ is the external magnetic flux threading the JRM loop, and bandwidths $\gamma_{a,b}/2\pi$ set by the capacitive coupling to the external feedlines. As illustrated in Fig.\,\ref{Device}(g), resonator `a' is open ended and has one feedline, resonator `b' has one feedline on each side, and the pump drive is fed to the device via a separate on-chip feedline (marked P). The three-wave mixing operation of the JPC is captured by the leading nonlinear term in the system Hamiltonian given by $\mathcal{H}_{\rm{3wave}}={\hbar}g_{3}(\textit{a}+\textit{a}^{\dagger})(\textit{b}+\textit{b}^{\dagger})(\textit{c}+\textit{c}^{\dagger})$ \cite{JPCreview}, where $g_{3}$ is a flux-dependent coupling strength, \textit{a} and \textit{b} are the annihilation operators for the differential modes \textit{a} and \textit{b}, while \textit{c} is the annihilation operator for mode \textit{c} common to both resonators. When applying a strong, coherent, off-resonant, common drive at  $\omega_{c}\equiv\omega_{p}=\omega_{b}\pm\omega_{a}$ the JPC functions as a nondegenerate quantum-limited amplifier \cite{JPCreview,JPCnature} or a lossless frequency converter between modes \textit{a} and \textit{b} \cite{JPCreview,Conv,QuantumNode}, respectively. Under the latter classical drive corresponding to the frequency difference, we obtain in the rotating wave approximation $\mathcal{H}_{\rm{3wave}}={\hbar}|g_{ab}|(e^{i\phi^{\prime}_{p}}\textit{a}\textit{b}^{\dagger}+e^{-i\phi^{\prime}_{p}}\textit{a}^{\dagger}\textit{b})$, where $g_{ab}$ is a pump-amplitude-dependent coupling strength ($\propto g_3$), and  $\phi^{\prime}_{p}$ is a generalized pump phase, which is related to the applied pump phase $\phi_{p}$ by $\phi^{\prime}_{p}=\phi_{p}+n_{g}\pi$, where

\begin{equation}
n_{g}=\begin{cases} 
0 & \varphi_{\rm{ext}}\leq 0 \\
1 & \varphi_{\rm{ext}}>0
\end{cases}.
\label{ng}
\end{equation}

The added phase to $\phi_{p}$, i.e., $0$ ($n_{g}=0$) or $\pi$ ($n_{g}=1$), depends on whether $g_{3}$ is positive or negative, respectively, which is determined by the sign of $\varphi_{\rm{ext}}=2\pi\Phi_{\rm{ext}}/\Phi_0$, where $\Phi_0=h/2e$ is the flux quantum or alternatively the sign of the dc-current circulating in the JRM loop. Equation\,(\ref{ng}) assumes that the JPCs are flux-biased at the primary flux lobe centered around zero flux. 

On resonance, the transmission amplitude associated with the frequency conversion process is given by $t=2\rho/(1+\rho^2)$, where $\rho=2|g_{ab}|/\sqrt{\gamma_{a}\gamma_{b}}$ is a dimensionless pump amplitude, which varies between $0$ (total reflection) and $1$ (full conversion). Notably, increasing $\rho$ beyond 1 is possible in conversion mode but it results in lower $t$ than its maximum at $\rho=1$.   

As seen in the block diagram of Fig.\,\ref{Device}(d), the MPIJIS is formed by coupling two nominally identical JPCs in an interferometric setup, where mode `a' of the JPCs is coupled via a $90^{\circ}$ hybrid, and mode `b' is coupled to external $50$ Ohm cold terminations and an intermediate transmission line (TL). The device also includes a $90^{\circ}$ hybrid for the pump that is connected to the pump feedlines of the JPCs. Since this hybrid splits the pump evenly between the two stages and imposes, by design, the required phase difference for nonreciprocity, i.e., $\pm\pi/2$, the device operates using a single drive. 

Assuming symmetric coupling between modes `b' of the JPCs and the cold terminations, the transmission parameters of the MPIJIS on resonance are given by (see Appendix A)

\begin{equation} 
S_{2\leftrightarrows1}=i\frac{\sqrt{1-t^{2}%
	}\mp\sqrt{2}t^{2}\sin(\varphi)}{1+t^{2}},
\label{S12specialcase}
\end{equation}

\noindent and the reflection parameters read 

\begin{equation} 
S_{11}=S_{22}=-i\frac{\sqrt{2}t^{2}\cos(\varphi)}{1+t^{2}},
\label{S11specialcase}
\end{equation}

\noindent where $\varphi \equiv \varphi_{p}+p\pi$, $\varphi_{p}\equiv\phi_{p1}-\phi_{p2}$ is the phase difference between the same-frequency pumps feeding the two stages and $p\equiv (n_{g1}+n_{g2})$ $\rm{mod}$ $2$ is the parity of the applied fluxes threading the two JRMs. 

From these definitions and Eq.\,(\ref{ng}), it follows that (1) $\varphi_{p}=\mp\pi/2$ when ports $\rm{P}_1$ and $\rm{P}_2$ are driven, respectively, (2) $p=0$ when the signs of $\varphi_{\rm{ext1}}$ and $\varphi_{\rm{ext2}}$ are the same, i.e., even parity, and (3) $p=1$ when their signs are opposite, i.e., odd parity (for further details see Appendices A-B, F-G).  

By operating the two JPCs in frequency conversion mode at the 50:50 beam splitting point, i.e., $t=1/\sqrt{2}$, in which half of the signal input on each port is reflected and half is transmitted with frequency conversion to the other port (as illustrated in the signal flow graph in Fig.\,\ref{Device}(h)), and by setting, for example, the generalized phase difference to $\varphi=\varphi_{p}=-\rm{\pi}/2$, attained in the case of pumping through $\rm{P}_1$ while $p=0$, we generate a nonreciprocal response, in which input signals on resonance propagating in the direction of the arrow shown in Fig.\,\ref{Device}(b), are transmitted with near unity transmission, i.e., $|S_{21}|=2\sqrt{2}/3\backsimeq0.94$ (equivalent to about $0.5$ dB loss in signal power), whereas signals propagating in the opposite direction are blocked $S_{12}=0$ (i.e., routed to the cold terminations), and reflections vanish $S_{11}=S_{22}=0$. 
 
\section{MPIJIS measurements}

\subsection{Scattering parameters}
Figure\,\ref{wpt} exhibits a measurement of the scattering parameters of the MPIJIS operated in two modes of operation at fixed applied fluxes, for which $p=0$. In the first, the MPIJIS is driven through $\rm{P}_1$, setting its directionality from port 1 to 2, while in the second, it is driven through $\rm{P}_2$, which reverses its directionality as illustrated in the block diagrams in the left column. The input power applied in the scattering parameter measurements of this work, unless stated otherwise, is about $-140$ dBm, which is much lower than the saturation power of the MPIJIS. In Fig.\,\ref{wpt}(a), we show the transmission parameters $|S_{21}|^2$ (blue) and $|S_{12}|^2$ (orange) corresponding to the first case. The dashed and solid lines correspond to the MPIJIS being off (no pump) and on (with pump), respectively. Similarly, Fig.\,\ref{wpt}(c), exhibits the reversed transmission response corresponding to the second case. In both cases, the MPIJIS response yields on resonance, at $f_{a}=\omega_{a}/2\pi=6.84$ GHz, about $2$ dB attenuation in the forward direction and $23$ dB in the backward (isolated) direction with a dynamical bandwidth of $8$ MHz. The corresponding $f_{b}=\omega_{b}/2\pi=9.598$ GHz is given by $f_{a}+f_{p}$, where $f_{p}=2.758$ GHz is the applied pump frequency in this measurement. Figures \ref{wpt}(b),(d) depict the reflection parameters $S_{11}$ (red) and $S_{22}$ (magenta) measured for the respective directionality cases. The magnitude of the measured transmission and reflection parameters of the device are calibrated using the experimental procedure outlined in Appendix H. As seen in Figs.\,\ref{wpt}(b),(d) the reflections off the two ports are small when the pump is on and off, depicted as solid and dashed lines, respectively. In addition to demonstrating that the on-chip single-pump MPIJIS works as intended, the results of Fig.\,\ref{wpt} directly confirm the theory prediction that the phase gradient condition for nonreciprocity is $\pm\pi/2$, imposed by the pump hybrid. 

Since the JPCs in the MPIJIS are operated in frequency conversion mode without photon gain, they are not required to add noise to the processed signal \cite{Caves,QuantumNoiseIntro}. However, any added noise by the MPIJIS is primarily set by its insertion loss in the forward direction, e.g., $|S_{ij}|^2$ (assuming $j \rightarrow i$), and is given by $n_{\rm{add}}=(1-|S_{ij}|^2)/2|S_{ij}|^2$, where $n_{\rm{add}}$ is the noise-equivalent-input-photons at the signal frequency \cite{MPIJIS}. For the measurement of Fig.\,\ref{wpt}, which exhibits about $2$ dB attenuation in the forward direction, we get $n_{\rm{add}}=0.29$. In the ideal case scenario, i.e., with $0.5$ dB of attenuation, this figure is expected to be much lower $n_{\rm{add}}=0.06$.  

\subsection{Pump power dependence}

In Fig.\,\ref{VaryPp}, we measure, for a fixed pump frequency and fluxes (for which $p=0$), $|S_{21}|^2$ and $|S_{12}|^2$ of the MPIJIS, as we vary the pump power. Figures \,\ref{VaryPp}(a),(b) show measurements of $|S_{21}|^2$ and $|S_{12}|^2$ when $\rm{P}_1$ is driven (i.e., directionality $1\rightarrow2$), whereas Figs.\,\ref{VaryPp}(c),(d) show measurements of the same transmission parameters when $\rm{P}_2$ is driven instead (i.e., directionality $2\rightarrow1$). The colored solid curves are measured data, whereas the black dashed curves represent a calculated response of the device using the theory model presented in Appendix B. As seen in these measurements, the device response varies with the pump power in a monotonic and stable manner similar to the observed response of JPCs \cite{JPCnature} and JPAs \cite{JPAsquidarray}. Other important characterization and measurement results of the MPIJIS can be found in Appendix I.     

\subsection{Flux parity dependence}
In Appendix G we present experimental data that demonstrates the dependence of the MPIJIS transmission/isolation direction on the parity of the magnetic fluxes threading its two JRMs. In particular, we show that the JIS directionality can be
reversed for the same pump by changing the orientation
parity of the applied fluxes. This effect could allow for example
the detection of the orientation parity of pairs of weak
magnetic sources using microwave transmission measurements (see Appendix K). 

\section{qubit Readout Multi-Chip Module (qRMCM)}
     
After successfully demonstrating the operation of the MPIJIS as a standalone device, we now integrate it with other microwave components to form a qRMCM devoid of magnetic materials and strong magnetic fields. An image of the qRMCM and block diagram of its components and setup are shown in Fig.\,\ref{qRMCM}(a), and Fig.\,\ref{qRMCM}(b), respectively. It integrates: a Purcell filter \cite{MPIJIS,Bronn2015b}, a superconducting wideband directional coupler \cite{MPIJIS}, two MPIJIS devices (sidenote: MPIJIS 2 got damaged during wirebonding and, therefore, is not operational in this work), and a reconfigurable directional Josephson device that is identical to the MPIJIS circuit but does not include an on-chip pump hybrid, thus enabling us to operate it as MPIJDA or MPIJIS, depending on the applied pump frequency. The qRMCM has a few pump ports employed for powering the Josephson devices. While only one pump is used for operating the MPIJIS, the reconfigurable device requires for its operation two same-frequency pumps, whose phase difference between $\rm{P}_3$ and $\rm{P}_4$ is $\pm\pi/2$. The qRMCM has three main ports connecting to the readout input line, the quantum chip (containing a qubit coupled to a readout resonator), and the readout output line. The output line employed here includes a commercial low-pass filter with a cut-off frequency of $8$ GHz at the base stage, a HEMT amplifier at the $4$ K stage and a superconducting NbTi coax cable connecting the two stages. The qRMCM setup also includes several auxiliary lines (such as IN2, IN1', OUT') and auxiliary components (such as a commercial directional coupler on the output line and a cryogenic circulator on the IN1' line) as seen in Fig.\,\ref{qRMCM}(b) (and the detailed setup diagram in Fig.\,\ref{Round2FullSetup}) that allow us to probe the transmission through the qRMCM in both directions at various working points. We also use a separate input line for injecting the qubit pulses as seen in  Fig.\,\ref{qRMCM}(b) that directly connects to the qubit chip. All other auxiliary ports of the directional Josephson devices (for example those that are coupled to their internal modes) are terminated with cryogenic $50$ Ohm loads.   

Using the qRMCM, we conduct four main qubit experiments, whose results are exhibited in Figs.\,\ref{IsoMagData}, \ref{DoubleIsoData}, \ref{DephasingData} (taken in the same cooldown) and Figs.\,\ref{magIsolJISsetupSimple}(b)-(c), \ref{magIsoJISComp} (taken in a separate cooldown, whose setup diagram is shown in Fig.\,\ref{magIsolJISsetupSimple}(a)). In attempt to optimize the distribution of the attenuation and filtering on the various lines, we carried out two additional cooldowns of the qRMCM. The results of these brief cooldowns are not shown here but are fully consistent with the reported data.      

In the qubit measurements of the first three experiments (i.e., Figs.\,\ref{IsoMagData}, \ref{DoubleIsoData}, \ref{DephasingData}), the readout pulse duration and integration time are $0.75$ $\mu$s. In the fourth experiment,  comparing the qRMCM with magnetic isolators  (i.e., Fig.\,\ref{magIsoJISComp}), these times are set to $1.5$ $\mu$s. 

\subsection{MPIJIS-MPIJDA experiment}

In this experiment, we perform a fast QND readout measurement with high fidelity using the qRMCM, while maintaining high coherence times for the qubit. 

We operate the first MPIJIS ($\rm{I_1}$) of the qRMCM as an isolator and the reconfigurable directional device as a near quantum-limited amplifier (A) (i.e., MPIJDA). In Fig.\,\ref{IsoMagData}, we present the main results, measured for the four possible configurations of the qRMCM specified in the headings of rows (a)-(d). In the first column, we present the forward and backward transmission parameters of the qRMCM measured between ports 1' and 2 as defined in Fig.\,\ref{qRMCM}(b). In the second, we present the measured readout fidelity histograms and the corresponding assignment fidelity. In the third and fourth, we exhibit the measured qubit $T_1$ and $T_{\rm{2E}}$, respectively. The dephasing time in the various cases is calculated using the relation   $T_{\rm{\varphi}}^{-1}=T_{\rm{2E}}^{-1}-(2T_{\rm{1}})^{-1}$. 

When both $\rm{I_1}$ and A are off, $|S_{21'}|^2$ and $|S_{1'2}|^2$ overlap and exhibit an insertion loss of about $2$ dB at the readout frequency. We measure a readout fidelity of F=0.77, $T_1=43$ $\mu$s and $T_{\rm{2E}}=T_{\rm{\varphi}}=0.7$ $\mu$s. Turning on A in configuration (b) while keeping $\rm{I_1}$ off, results in about $12$ dB of gain in the forward direction and $-1$ dB in the backward direction, an increase in the fidelity to $0.95$, and a slight drop in $T_1=40$ $\mu$s and $T_{\rm{2E}}=T_{\rm{\varphi}}=0.4$ $\mu$s due to excess backaction noise of the MPIJDA. Turning on $\rm{I_1}$ instead as shown in configuration (c), results in large isolation of more than $30$ dB in the backward direction and insertion loss of about $4$ dB in the forward direction, a similar fidelity F=0.74 as in the off-off case, but a significant enhancement of the coherence times $T_1=52$ $\mu$s, $T_{\rm{2E}}=38$ $\mu$s, $T_{\rm{\varphi}}=60$ $\mu$s due to the increased qubit protection against output noise. Lastly, turning both $\rm{I_1}$ and A on, results in a forward gain of about $10$ dB, reverse isolation of about $22$ dB, a high readout fidelity similar to case (b) of F=0.95, $T_1=52$ $\mu$s (similar to case (c)), and only a slight drop of $T_{\rm{2E}}=35$ $\mu$s, $T_{\rm{\varphi}}=53$ $\mu$s compared to case (c) for which only $\rm{I_1}$ is on.    

\subsection{MPIJIS-MPIJIS experiment}

Next, we operate the reconfigurable directional Josephson device in the qRMCM as a second MPIJIS ($\rm{I_2}$) in series with $\rm{I_1}$ and measure the impact of added isolation on the qubit coherence. The main results of this experiment are exhibited in Fig.\,\ref{DoubleIsoData}. Similar to the MPIJIS-MPIJDA experiment, we measure for the four configurations of $\rm{I_1}$ and $\rm{I_2}$ (being on or off), outlined in the headings of rows (a)-(d), the forward $|S_{21'}|^2$ and backward $|S_{1'2}|^2$ transmission through the qRMCM plotted in the first column from the left, the readout fidelity F plotted in the second column, and the qubit coherence times $T_{\rm{1}}$ and $T_{\rm{2E}}$ plotted in the third and fourth columns, respectively. 

With both MPIJIS devices off in configuration (a), we measure an insertion loss (through $\rm{I_1}$ and $\rm{I_2}$) of about $3$ dB at $f_r$, F=0.85, $T_1=40$ $\mu$s, and $T_{\rm{2E}}=T_{\rm{\varphi}}=0.6$ $\mu$s. Turning on $\rm{I_2}$ alone, results in $18$ dB of isolation at $f_r$ and $8$ dB of insertion loss in the forward direction, a drop in F to $0.7$5, and an increase in the qubit coherence times $T_1=50$ $\mu$s, $T_{\rm{2E}}=3.6$ $\mu$s (6-fold enhancement compared to case (a)), and $T_{\rm{\varphi}}=3.7$ $\mu$s. Turning on $\rm{I_1}$ instead, in configuration (c), yields $31$ dB of isolation and $5$ dB of insertion loss, F=0.77, a full recovery of $T_1=54$ $\mu$s (which falls within the qubit typical range), and considerable enhancement in $T_{\rm{2E}}=32$ $\mu$s (53-fold higher than case (a)) and $T_{\rm{\varphi}}=45$ $\mu$s. Lastly, turning on both $\rm{I_1}$ and $\rm{I_2}$, gives an isolation of $44$ dB at $f_r$ and insertion loss similar to case (b), F=0.68, and achieves $T_1=54$ $\mu$s (similar to (c)) and a maximum in $T_{\rm{2E}}=49$ $\mu$s (82-fold higher than case (a)) and $T_{\rm{\varphi}}=89$ $\mu$s.        

\subsection{Variable isolation experiment}
Here, we investigate the dependence of the qubit coherence times on the isolator response by varying the isolation in-situ via the applied pump power.  

In Fig.\,\ref{DephasingData}(a), we plot the measured isolation curves $L_0(f)=|S_{1'2}(f)|^2$ of the single-pump MPIJIS as a function of the normalized pump power, where the isolation at the readout frequency $L=L_0(f_r)$ is minimal at $P_{p0}$. In Fig.\,\ref{DephasingData}(b), we plot the corresponding theory fits calculated using the effective two-port model of the MPIJIS derived in Appendix B.  

In Fig.\,\ref{DephasingData}(c), we plot the corresponding measured coherence times of the qubit (i.e., $T_1$, $T_{\rm{2E}}$, $T_{\rm{\varphi}}$) as a function of $L$. The black dotted line is a guide to the eye for increasing $P_p$. We also added on the same plot the highest coherence points obtained in the MPIJIS-MPIJIS experiment in Fig.\,\ref{DoubleIsoData}(d) corresponding to $L<10^{-4}$ (not connected by the dotted line). 

Since thermal photon population in the readout resonator is the likely dominant dephasing mechanism in our dispersively coupled qubit-resonator system exposed to thermal noise coming from the $4$ K stage, we use the dephasing rate equation derived in Refs.\cite{ClerkShotNoise,ChadShotNoise}, given by

\begin{equation}
\Gamma_{\rm{\varphi}}=\dfrac{\kappa}{2}\rm{Re}\left(\sqrt{\left( 1+i\dfrac{\chi}{\kappa}\right)^2+4i\dfrac{\chi}{\kappa}\bar{n}} -1\right).\label{DephasingRate}\\
\end{equation}

In Eq.\,(\ref{DephasingRate}), $\kappa$ is total photon decay rate of the fundamental mode of the readout resonator with angular frequency $\omega_r=2\pi f_r$ (here $\kappa$ is dominated by the coupling rate to the external feedline $\kappa\cong\kappa_{\rm{e}}$), $\chi$ is the qubit-state-dependent frequency shift of the readout resonator, and $\bar{n}=\Sigma_{i}F_i\kappa_i\bar{n}_i/\kappa$ is the average thermal photon number in the resonator \cite{AdamShotNoise}, where  $\bar{n}_i=1/\left(e^{\left(\hbar\omega_r/k_BT_i\right)}-1\right)$ is the Bose-Einstein population of a $50$ Ohm load noise source $i$ at effective temperature $T_i$, $\kappa_i$ is the readout resonator relaxation rate through source $i$, and $F_i$ is the linear power attenuation between source $i$ and the port through which it couples to the readout resonator (e.g., qubit or readout port).  

To fit the measured dephasing time, given by $T_{\rm{\varphi}}\equiv\Gamma_{\rm{\varphi}}^{-1}$, we model the MPIJIS as a cold filter that attenuates thermal noise coming from the output chain. Considering the case of a constant filter with attenuation $L$, where $L$ is the power isolation at $f_r$, we express $\bar{n}$ as $\bar{n}=L\bar{n}_o+\bar{n}_r$, where $\bar{n}_o$ is the thermal photon number acted upon by the MPIJIS devices, while $\bar{n}_r$ is a residual thermal photon population due to out-of-band noise or sources that lie outside the isolation path (e.g., couple to the MPIJIS pump ports or qubit port). Substituting $\bar{n}$ in the inverse of Eq.\,(\ref{DephasingRate}), we get the black curve fit $T_{\rm{\varphi,L}}$ exhibited in Fig.\,\ref{DephasingData}(c) as a function of $L$. An almost identical fit $T_{\rm{\varphi,L2}}$ corresponding to the dashed cyan curve in Fig.\,\ref{DephasingData}(c) is obtained by substituting $\bar{n}$ in the approximation of Eq.\,(\ref{DephasingRate}) (in the limit $\bar{n}\ll1$) \cite{CavityAtten}, given by

\begin{equation}
\Gamma_{\rm{\varphi}}=\Gamma_c\bar{n},\label{DephasingRate2}\\
\end{equation}     

\noindent where $\Gamma_c=\kappa\chi^2/(\kappa^2+\chi^2)$.
The values of $\bar{n}_o=2.08$ and $\bar{n}_r=0.014$ in these fits are set to give  the dephasing times measured for $P_p=0$ and $P_p=P_{p0}$. Similarly, in Fig.\,\ref{DephasingData}(d), we plot $\bar{n}\equiv\Gamma_{\rm{\varphi}}/\Gamma_c$ measured versus $L$ (black stars) and the corresponding theory fits using Eq.\,(\ref{DephasingRate}) (black) and Eq.\,(\ref{DephasingRate2}) (cyan). In both Figs.\,\ref{DephasingData}(c) and (d), we see that the theory fits, in this case, overestimates the dephaing time in the intermediate pump power range between $0$ and $P_{p0}$ or alternatively underestimates $\bar{n}$ in the resonator. Using $\bar{n}_o$ and $\bar{n}_r$ and the Bose-Einstein population expression $\bar{n}_i$, we evaluate the effective temperature of the noise source seen by the qubit in the $L=1$ and $L\ll1$ cases, which is about $1$ K and $70$ mK, respectively.

In Figs.\,\ref{DephasingData}(e) and (f), we further analyze the dephasing time measurements. We assume that the noise is solely coming through the MPIJIS path and consider the combined filtering effect, denoted $A$, of the frequency-dependent MPIJIS isolation $L_0(\omega)=|S_{12}|^2$ (using Eq.\,(\ref{Eff_S12})) and the Lorentzian response of the resonator given by $R(\omega)=\left( \kappa/2\pi\right)/\left[\left(\omega-\omega_r \right) ^2 +\left(\kappa/2\right) ^2 \right]$. Hence, we express $\bar{n}=A\bar{n}_{\rm{out}}$, where $A\equiv\int\limits_{\omega_{c1}}^{\omega_{c2}}L_0(\omega)R(\omega)\,\rm{d}\omega$ and $\bar{n}_{\rm{out}}$ represents an effective photon number of the output-line noise in the relevant frequency range $[\omega_{c1},\omega_{c2}]$. 

Using this model with $\bar{n}_{\rm{out}}=2.1$, we plot, in Figs.\,\ref{DephasingData}(e) and (f), the measured dephasing time and $\bar{n}$ (black stars) versus the parameter $A$, which we numerically calculate using the measured MPIJIS curves exhibited in Fig.\,\ref{DephasingData}(a). We also plot with solid red lines the corresponding theory fits for the dephasing time $T_{\varphi,\rm{A}}$ and $\bar{n}$ using Eq.\,(\ref{DephasingRate}) and the parameter $A$ evaluated using the calculated response of the MPIJIS, featured in Fig.\,\ref{DephasingData}(b). Similar to Fig.\,\ref{DephasingData}(c), the dotted black line in Figs.\,\ref{DephasingData}(d)-(f) is a guide to the eye for increasing $P_p$, and the data point in each figure that is unconnected by the dashed line belonging to configuration (d) of the MPIJIS-MPIJIS experiment (Fig.\,\ref{DoubleIsoData}(d)). As seen in Figs.\,\ref{DephasingData}(e),(f), the theory fits drawn as red lines yield a good agreement with the measured data in the low-mid pump power range $P_p<P_{p0}$. Deviations up to a factor of $2$ between the fits and the data are observed around and above $P_{p0}$. Notably, we observe an enhancement of $T_{\varphi}$ (or decrease in $\bar{n}$) even for similar values of the parameter $A$ near $10^{-2}$, where it plateaus. However, as revealed by the dashed black line, higher $T_{\varphi}$, corresponding to similar $A$ values, seem to correlate with higher applied pump powers. To account for this pump power dependency, we update the parameter $A$ (as done in Appendix D and E) to include the effect of input-power saturation of the MPIJIS (i.e., dynamic range) due to pump depletion. Substituting the result of Eq.\,(\ref{LRcorrwIntSimple}) into the inverse of Eq.\,(\ref{DephasingRate}) with $\bar{n}_{\rm{out}}=1.7$, yields the updated theory fit $T_{\rm{\varphi,\rm{D}}}$ drawn as a solid black curve in Figs.\,\ref{DephasingData}(e). Importantly, although the fits $T_{\rm{\varphi,\rm{A}}}$, $T_{\rm{\varphi,\rm{D}}}$ drawn in Figs.\,\ref{DephasingData}(e) yield similar quantitative agreement with the data, $T_{\varphi,\rm{D}}$ qualitatively reproduces the semi-U-turn feature of the measured dephasing time with respect to $A$. Similar characteristics are seen in the corresponding fits of $\bar{n}$ in Fig.\,\ref{DephasingData}(f). It is important to emphasize here the implications of the data drawn in Figs.\,\ref{DephasingData}(e) and (f). They show (see Eq.\,(\ref{LRcorrwInt})) that for similar $A$ values of the MPIJIS that generally correspond to similar isolation response versus frequency, the qubit dephasing time increases when the ratio of the incoming noise photon per unit time to the input pump photon per unit time decreases. Furthermore, the fact that in the double isolation experiment (represented by the top and bottom data points in Figs.\,\ref{DephasingData}(e) and (f), respectively), the measured dephasing time is higher than the maximum achieved with one isolation stage, despite having similar $A$ parameter value, might be due to the presence of two pump inputs feeding the two isolation stages, which further reduce the noise to pump photon ratio.    

To complete the picture, we plot in Fig.\,\ref{DephasingData}(g), the measured isolation at the minima points (diamonds) and at the readout frequency  (squares), and the calculated parameter $A$ (solid black curve) as a function of the normalized pump power $P_p/P_{p0}$. The corresponding theory fits for the isolation at the global minima points and the readout frequency (i.e., $L$) are plotted using dashed magenta and cyan curves, respectively. We also plot in Fig.\,\ref{DephasingData}(h) along the same x-axis, the measured dephasing time (black stars) and the theory fits $T_{\rm{\varphi,\rm{L}}}$ (dashed black curve), $T_{\rm{\varphi,\rm{A}}}$ (solid red curve), and $T_{\rm{\varphi,\rm{D}}}$ (solid black curve) discussed earlier.

\subsection{qRMCM vs. two magnetic isolators experiment}

In this experiment, we modify the experimental setup inside the fridge to enable a direct comparison between the protection provided by the qRMCM and two commercial magnetic isolators connected in series as illustrated in Fig.\,\ref{magIsolJISsetupSimple}(a), showing the main components (a detailed diagram of the setup is displayed in Fig.\,\ref{IsoMagComFullSetup}). In this setup, we connect the readout input line to a commercial directional coupler to enable the passage of the reflected readout signals off the quantum chip through the qRMCM or magnetic isolators depending on the state of a first cryogenic switch, which connects to either the Purcell filter port of the qRMCM or to the input of the magnetic isolators. We also add a second cryogenic switch connected to the rest of the output line, which depending on its state, connects to either the output of the qRMCM (i.e., the auxiliary directional coupler connected to IN2) or the output of the magnetic isolators. In both cases, the readout and qubit input lines are the same as well as the output line that includes the low-pass filter and the HEMT. 

The two commercial magnetic isolators employed in this comparison are $4-12$ GHz isolators. Separate characterization of this kind of broadband isolator at $10$ mK in a different fridge (data not shown), shows that it gives more than $20$ dB of isolation in the range $1-12$ GHz and even stronger isolation, exceeding $60$ dB, at the readout frequency of this experiment. 

In Fig.\,\ref{magIsolJISsetupSimple} (b) and (c), we plot the transmission in the backward $|S_{1'2}|^2$ and forward $|S_{21'}|^2$ direction measured through the two MPIJIS devices, i.e., $\rm{I_{1}}$ and $\rm{I_{2}}$, corresponding to the four different configurations listed in the inset table. Similar to the double-isolation experiment in Fig.\,\ref{DoubleIsoData}, we obtain at this new working point an isolation of about $45$ dB at $f_r$ but a higher insertion loss in the forward direction $10$ dB instead of $8$ dB, which underscores the relatively large tuneup parameter space for Josephson parametric devices that possess several degrees of freedom, such as fluxes and pump drives (frequency and power). In the following section, we discuss how this tuneup parameter space can be reduced. 

In Fig.\,\ref{magIsoJISComp}, we plot the qubit coherence times measured with the magnetic-isolator setup (a), and the qRMCM configurations (b)-(d) specified in the headings. From left to right, we plot $T_1$ (relaxation), $T_{\rm{R}}$ (Ramsey), $T_{\rm{E}}$ (Echo), and $T_{\rm{CPMG}}$ (CPMG-like decoherence measurement). An illustration of the pulse sequence applied in the different measurements is shown at the top. In (a), we measure for the magnetic-isolator setup, $T_1=47$ $\mu$s, $T_{\rm{R}}=74$ $\mu$s, $T_{\rm{E}}=83$ $\mu$s, and $T_{\rm{CPMG}}=93$ $\mu$s. The depahsing time $T_{\rm{\varphi,i}}$, associated with the various decoherecne measurements $i=\rm{R,E,CPMG}$, is calculated using the generalized relation $T_{\rm{\varphi,i}}^{-1}=T_{\rm{i}}^{-1}-(2T_{\rm{1}})^{-1}$. From the results in (a), we find that in the 2-magnetic isolator case, the qubit decoherence is mainly limited by low-frequency noise since $T_{\rm{R}}<T_{\rm{E}}<T_{\rm{CPMG}}$, which can be filtered out by adding one $\pi$ pulse in the echo measurement and four in the CPMG-like measurement applied here. We also achieve in this input-output line configuration, the maximum attainable decoherence time $\cong 2T_1$ (i.e., limited by $T_1$). Thus, forming an ideal benchmark configuration for evaluating the qRMCM performance.           

Turning now to the coherence results obtained with the qRMCM shown in Fig.\,\ref{magIsoJISComp} (b)-(d). In the baseline case, where both MPIJIS devices $\rm{I_{1,2}}$ are off (data not shown), we obtain $T_1=28$ $\mu$s, $T_{\rm{R}}=0.7$ $\mu$s, and $T_{\rm{E}}=0.6$ $\mu$s. In (b), where only $\rm{I_{2}}$ is on, we obtain $T_1=41$ $\mu$s, $T_{\rm{R}}=21$ $\mu$s, $T_{\rm{E}}=22$ $\mu$s, and $T_{\rm{CPMG}}=26$ $\mu$s. Similarly, in (c), where only $\rm{I_{1}}$ is on, we obtain $T_1=42$ $\mu$s, $T_{\rm{R}}=13$ $\mu$s, $T_{\rm{E}}=29$ $\mu$s and $T_{\rm{CPMG}}=36$ $\mu$s. Lastly, in (d), where both $\rm{I_{1,2}}$ are on, we measure the highest coherence times, compared to the off case and to (b) and (c), i.e., $T_1=45$ $\mu$s, $T_{\rm{R}}=30$ $\mu$s, $T_{\rm{E}}=46$ $\mu$s and $T_{\rm{CPMG}}=52$ $\mu$s. While this case gives $T_1=45$ $\mu$s that is effectively equal to $47$ $\mu$s obtained with the 2-magnetic isolator setup and yields a considerable enhancement in the decoherence times with the addition of $\pi$ pulses as indicated by $T_{\rm{R}}<T_{\rm{E}}<T_{\rm{CPMG}}$, it only achieves $T_{\rm{E}}$ and $T_{\rm{CPMG}}$ that are slightly higher than $T_1$ and much shorter than $2T_1$. This result suggests that in the qRMCM case, the decoherence time is likely limited by high-frequency noise, such as residual thermal photon noise in the readout resonator that cannot be filtered out by applying a small number of $\pi$ pulses \cite{FluxQubitRevisited}.             

\section{Discussion}

The qubit-qRMCM experiment differs from the qubit-MPIJIS experiment reported in Ref.\,\cite{MPIJIS} in several important aspects, it (1) realizes an on-chip MPIJIS device operated with a single pump, (2) introduces a working qRMCM operated in continuous mode, which integrates a Purcell filter, superconducting directional coupler, MPIJIS devices, and a reconfigurable MPIJDA/MPIJIS device, (3) demonstrates a high-fidelity, high-coherence, QND qubit measurement without any magnetic isolators and circulators in the output chain, (4) achieves an isolation of more than $40$ dB at the readout frequency using two MPIJIS devices in series, and (5) investigates the dependence of the qubit coherence on the MPIJIS response, varied in-situ using the pump tone.  

In addition to enabling high-fidelity QND readout, the 
qRMCM scheme presented here has two main advantages: (1) it is fully compatible with frequency multiplexed readout, which is useful in scalable architectures (provided that the bandwidth and saturation power of the MPIJIS/MPIJDA can be significantly enhanced as we discuss below), (2) its isolation and amplification stages are inherently compatible due to their shared circuitry, fabrication process, and mode of operation. This compatibility could be particularly useful in large systems, which benefit from standardization, reliability, and matching.     

The measured response of the MPIJIS shown in Figs.\,\ref{wpt}, \ref{VaryPp} differ from the ideal theoretical case scenario in two notable aspects. The first is the MPIJIS attenuation on resonance in the forward direction of about $2$ dB, which is higher than the $0.5$ dB, predicted for a device, whose JPCs are operated around the 50:50 beamsplitting point, and whose modes `b' are equally coupled to each other and to the cold terminations. The likely cause for this effect, as revealed from the calculated response in Fig.\,\ref{VaryPp}, is that modes `b' of the JPCs are coupled more strongly to the cold terminations than to each other (see Appendix B). Consequently, this suggests it is possible to reduce the forward attenuation of the device by adjusting the unintentional asymmetry in the couplings in future designs. The second is the slight frequency detuning between the dips in the reflection (Fig.\,\ref{wpt}(b),(d)) and transmission (Fig.\,\ref{wpt}(a),(c)) parameters when the MPIJIS is on. This effect could be due to a phase imbalance in the signal hybrid, which results in a slightly different frequency condition for constructive/destructive intereferences in the two cases.  

As to the elevated insertion loss ($2-3$ dB) of the MPIJIS and MPIJDA devices in the off state, seen for example in Figs.\,\ref{IsoMagData}(a), \ref{DoubleIsoData}(a), \ref{magIsolJISsetupSimple}(b),(c), it originates from mismatches between the resonances of their JPC building blocks at the given flux biasing points and the phase and amplitude imbalance of their signal hybrids. Both of which can be minimized to about $0$ dB (as seen for example in Fig.\,\ref{wpt}(a) and (b) for the standalone device biased at a higher frequency) by increasing the uniformity of their JPCs, operating them near their maximum frequencies versus flux, and better match the center frequency of their signal hybrids to the intended operation frequency.

Following this work, there are several avenues to explore going forward, for example (1) realizing a single-pump, on-chip MPIJDA, which employs an adapted hybrid for the pump in an analogous manner to the single-pump MPIJIS, (2) pining down the source of the residual high-frequency noise limiting $T_{\rm{2E}}$ in the qRMCM case in comparison to the conventional magnetic isolator setup (see Fig.\,\ref{magIsoJISComp}). In particular, investigate whether it originates from out-of-band noise coming from the output chain (i.e., lies outside the narrow bandwidth of the MPIJIS devices) or from incoming noise that lies outside the isolation path of the MPIJIS devices (e.g., thermal noise entering through the pump-line circuitry), (3) reducing the size of the qRMCM components using lumped-element implementations of the JPCs \cite{LumpedJPC} and hybrids \cite{Lumpedhybrids}, and (4) enhancing the bandwidth and saturation power of the MPIJIS and MPIJDA to support multiplexed qubit readout in scalable architectures. This could potentially be achieved by first enhancing the bandwidth and saturation power of JPCs, which constitute the bottleneck. One possible way to enhance the JPC bandwidth is by implementing an impedance matching network between its Josephson junctions and external feedlines, as was successfully demonstrated in the case of single-port Josephson parametric amplifiers \cite{JPAimpedanceEng,StrongEnvCoupling}. Enhancing the saturation power of JPCs on the other hand, requires getting a better handle on the Kerr and higher order nonlinearities exhibited by the device \cite{JPChigherNon,OptJPC,UnderstandingSatPow}, which are dependent on the inductance ratios of the Josephson junctions to the linear inductances inside and outside the ring \cite{OptJPC}. For example, the analysis of Ref. \cite{OptJPC} shows that there exists an experimentally feasible inductance-ratio parameter space in which JPCs have saturation powers as high as $-104$ dBm.

Furthermore, enhancing the bandwidth of these directional Josephson devices and reducing their footprint are expected to yield another benefit. That is reducing their tuneup parameter space, by further relaxing the matching requirement of their JPCs and enabling them to be flux biased using one coil or flux line versus two. This will further simplify and decrease the variability of their tuneup procedure, and reduce the number of flux lines and sources needed per device.    

Finally, it is important to emphasize that the qRMCM concept introduced here constitutes an interface block between the qubit system and the input and output lines that is modular and versatile, and, therefore, it is not limited to specific superconducting components or readout schemes. For example, the qRMCM could potentially serve in the future readout schemes, which rely on single microwave photon detectors \cite{PhotonCounterJJ,QubitReadoutPhotonCounter,PhotonCounterProp} instead of parametric amplifiers. 

\section{Conclusion}
We realize a qubit readout multi-chip module (qRMCM) devoid of magnetic materials and strong magnetic fields. The qRMCM includes a Purcell filter, a superconducting directional coupler, two MPIJIS devices, and one reconfigurable directional Josephson device integrated into one PCB. We use the qRMCM alongside an off-the-shelf lowpass filter connected at its output to read out a superconducting qubit without any cryogenic magentic isolators or circulators in the output chain. With the reconfigurable device operated as an MPIJDA and one of the MPIJIS devices turned on, we demonstrate a fast ($T_{\rm{r}}=0.75$ $\mu$s), high-fidelity ($F>0.95$), QND measurement of a coherent qubit with $T_1=52$ $\mu$s and $T_{\rm{2E}}=35$ $\mu$s. We further enhance the qubit coherence up to $T_{\rm{2E}}=49$ $\mu$s by operating the reconfigurable device as a second MPIJIS, with a total qRMCM isolation of $45$ dB at the readout frequency and a bandwidth of about $13$ MHz. Moreover, by varying the isolation of the qRMCM with the pump power, we demonstrate an in-situ enhancement of $T_{\rm{2E}}$ and $T_{\rm{\varphi}}$ by a factor of $80$ and $150$, respectively (up to a maximum of $T_{\rm{2E}}=49$ $\mu$s and $T_{\rm{\varphi}}=89$ $\mu$s). A direct comparison with an output chain that includes two commercial wideband magnetic isolators in the output chain and achieves $T_{\rm{2E}}\approx2T_1$, shows that the qubit dephasing time measured with the qRMCM (for which $T_{\rm{2E}}\approx T_1=45$ $\mu$s) is likely limited by residual thermal photon population in the readout resonator that is not acted upon by the MPIJIS devices.   

One key component enabling these results is the single-pump, on-chip MPIJIS realized in this work, which is comprised of two nominally identical nondegenerate, three-wave Josephson mixers that are coupled in an interferometric setup and operated in frequency conversion mode. The microwave drive, giving rise to nonreciprocity, is fed through an on-chip quadrature hybrid, which equally splits the drive between the mixers and imposes the required phase difference, i.e., $\pm\pi/2$, between the split drives. The MPIJIS yields on resonance attenuation of about $2$ dB in the forward direction and $23$ dB in the backward direction with a dynamical bandwidth of $8$ MHz. The device has a tunable bandwidth of about $300$ MHz with isolation larger than $18$ dB and input saturation power of about $-120$ dBm at $21$ dB of isolation (see Appendix I). 

An improved and smaller version of this qRMCM, which integrates large-bandwidth and high-saturation MPIJIS and MPIJDA devices could enable frequency multiplexed readout of multiple qubits in scalable quantum processor architectures.    

\section*{ACKNOWLEDGMENTS} 
B.A. expresses gratitude to Jerry M. Chow and Pat Gumann for enabling this work. Fruitful discussions with David Lokken-Toyli, Luke Govia, Ted Thorbeck, and Oliver Dial are highly appreciated. The authors are also grateful to William Shanks, Vincent Arena, Thomas McConkey, and Serafino Carri for important technical support. The authors thank the group of David Pappas at NIST for fabricating the directional coupler chip and its package. Work pertaining to the development of the Purcell filter was supported by IARPA under contract W911NF-10-1-0324 and to the development of the pogo-pin packaging by IARPA under contract W911NF-16-1-0114-FE. Contribution of the U.S. Government, not subject to copyright.

\appendix

\section{MPIJIS scattering parameters} 
\begin{figure*}
	[tb]
	\begin{center}
		\includegraphics[
		width=1.5\columnwidth 
		]%
		{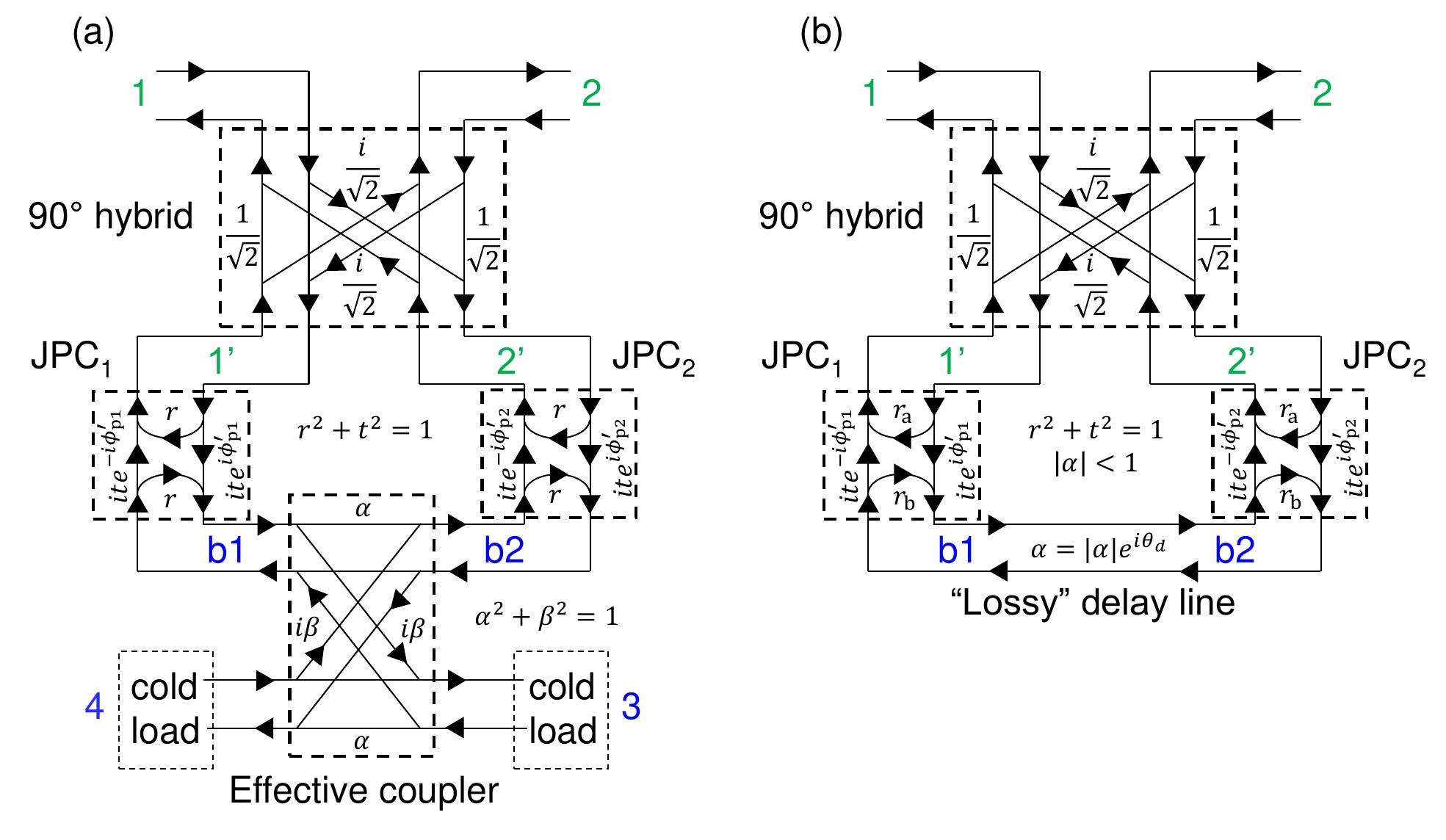}
		\caption{Signal flow graphs for the MPIJIS. The graphs in (a) and (b) exhibit two JPCs operated in frequency conversion mode on resonance, i.e., $f_{\rm{1}}=f_{a}$, $f_{\rm{2}}=f_{b}$. Ports `a' of the JPCs, denoted as $1^{\prime}$, $2^{\prime}$, are coupled via a $90^{\circ}$ hybrid, while their internal `b' ports, denoted as b1 and b2, are coupled via a fictitious coupler in (a) or equivalently a lossy delay line in (b), which model the attenuation present in the internal `b' channel of the MPIJIS due to dissipation in the $50$ $\Omega$ cold terminations. In (a) the coupler coefficients $\alpha$ and $\beta$ are taken to be real, satisfying the condition ${\alpha}^2+{\beta}^2=1$. The transmitted signals between ports `a' and `b' of the JPC undergo frequency conversion and acquire a nonreciprocal phase shift, which depends on the phase of the drive.      
		}
		\label{SignalFlow}
	\end{center}
\end{figure*}

To derive analytical expressions for the scattering parameters of the MPIJIS, we solve the effective signal-flow graph of the device exhibited in Fig.\,\ref{SignalFlow}(a), which includes the coupled JPCs operated in frequency-conversion mode. On-resonance signals at $f_{\rm{1}}=f_{a}$ or $f_{\rm{2}}=f_{b}$ input on port `a' (e.g., $1^{\prime}$ or $2^{\prime}$) or `b' (e.g., b1 or b2) are reflected off with a reflection-parameter $r$ and transmitted with frequency-conversion to the other port with a transmission-parameter $t$, where $r$ and $t$ are determined by the pump drive amplitude and satisfy the energy conservation condition $r^2+t^2=1$. More specifically, in the stiff pump approximation, $r$ and $t$ can be written as \cite{JPCreview,Conv}

\begin{align}
\begin{array}
[c]{cc}%
r=\dfrac{1-\rho^2}{1+\rho^2}, \\
t=\dfrac{2\rho}{1+\rho^2},  
\end{array}
\label{r_s_res}%
\end{align}

\noindent where $0\leq\rho$ is a dimensionless pump amplitude. For $\rho=0$, the JPC acts as a perfect mirror, whereas for $\rho=1$, the JPC operates in full frequency conversion mode between ports `a' and `b'.

\begin{figure}
	[tb]
	\begin{center}
		\includegraphics[
		width=\columnwidth 
		]%
		{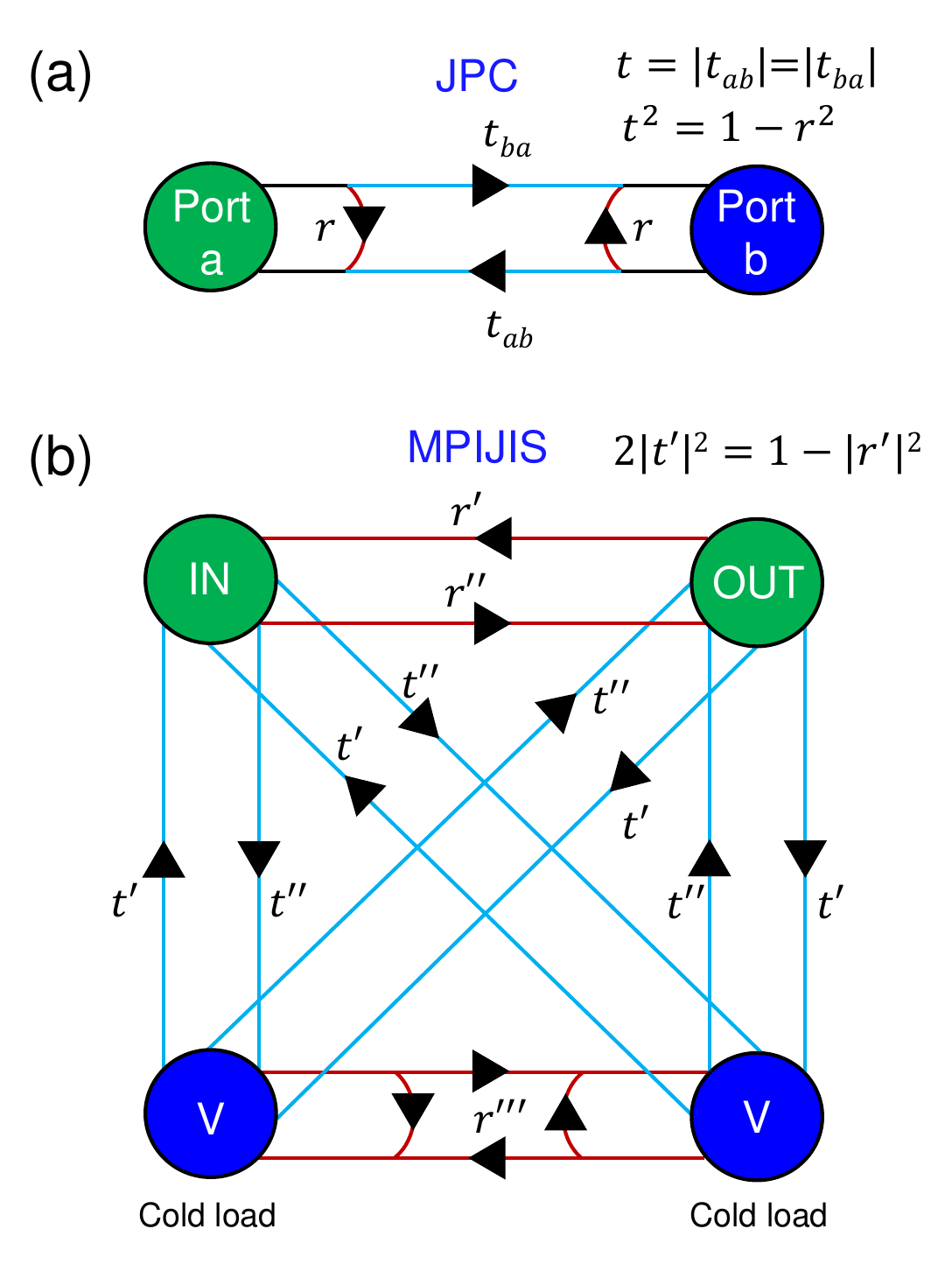}
		\caption{Simplified signal flow graph for a JPC operated in frequency conversion mode (a) and for the MPIJIS (b). See Appendix C for details.      
		}
		\label{JPCMPIJIScomp}
	\end{center}
\end{figure}

In the derivation, we assume that the two JPCs are balanced, i.e., their reflection and transmission parameters are equal. Figure\,\ref{SignalFlow}(a) also includes flow-graphs of two couplers coupling the `a' and `b' ports of the JPCs; one represents the $90^{\circ}$ hybrid, which couples between the `a' ports of the JPCs, while the other is fictitious, coup\-ling the `b' ports. The latter models the amplitude attenuation $\alpha$ present on the `b' port due to signal absorption in the $50$ $\Omega$ cold loads. Due to the structural symmetry of the device, we consider a coupler with real coefficients $\alpha$ and $\beta$, which satisfy the condition ${\alpha}^2+{\beta}^2=1$. For an ideal symmetric coupler (i.e., $90^{\circ}$ hybrid), $\alpha=\beta=1/\sqrt{2}$ \cite{Pozar}.

A detailed derivation of the scattering parameters of MPIJIS appears in the supplementary information of Ref. \cite{MPIJIS}. For completeness, we will list in this section the main results of Ref. \cite{MPIJIS} and update the equations to account for the effect of the applied fluxes in the JRMs on the isolation direction of the MPIJIS. More specifically, we generalize the nonreciprocal phases $\phi_{p1,p2}$ acquired by the frequency-converted transmitted signals between ports `a' and `b', which in Ref. \cite{MPIJIS} represent only the phases of the pump drives feeding $\rm{JPC_1}$ and $\rm{JPC_2}$ at frequency $f_{p}$, to $\phi^{\prime}_{p1,p2}=\phi_{p1,p2}+in_{g}\pi$, which account for the sign of the real coupling constant $g_{ab}$ (i.e., $\pm 1$ obtained for $n_{g}=0,1$, respectively) set by the sign of the applied flux in the JRM or alternatively the direction of the circulating current \cite{FlaviusThesis}. Consequently, the multiplication of the transmission parameters through the two JPCs results in multiplication of the signs of the coupling constants or addition of the corresponding phases, i.e., $e^{ip\pi}$ where $p=(n_{g1}+n_{g2})$ mod $2$. After introducing this update and expressing the MPIJIS scattering parameters in terms of the parameter $t$, we obtain on resonance  

\begin{equation}
S_{21}=\frac{i}{1+\frac{\alpha^{2}}{\beta^{2}}t^{2}}\left[  \sqrt{1-t^{2}%
}-\frac{\alpha}{\beta^{2}}t^{2}\sin\varphi\right]  ,\label{S21}\\
\end{equation}

\begin{equation}
S_{12}=\frac{i}{1+\frac{\alpha^{2}}{\beta^{2}}t^{2}}\left[  \sqrt{1-t^{2}%
}+\frac{\alpha}{\beta^{2}}t^{2}\sin\varphi\right]  ,\label{S12}\\
\end{equation}

\begin{equation}
S_{11}=S_{22}=-\frac{i\alpha}{\beta^{2}}\frac{t^{2}}{1+\frac{\alpha^{2}%
	}{\beta^{2}}t^{2}}\cos\varphi,\label{S11}\\
\end{equation}

\begin{equation}
S_{33}=S_{44}=-\frac{\sqrt{1-t^{2}}}{1+\frac{\alpha^{2}}{\beta^{2}}t^{2}%
},\label{S33}\\
\end{equation}

\begin{equation}
S_{34}=S_{43}=\frac{\alpha}{\beta^{2}}\frac{t^{2}}{1+\frac{\alpha^{2}%
	}{\beta^{2}}t^{2}},\label{S34}\\
\end{equation}

\begin{equation}
S_{13}=-\frac{te^{-i\varphi_{\rm{s}}/2+i\rm{\pi}/4}%
}{\sqrt{2}\beta\left(1+\frac{\alpha^{2}}{\beta^{2}}t^{2}\right)  }\left[
\sqrt{1-t^{2}}\alpha e^{i\frac{\varphi}{2}+i\frac{\rm{\pi}}{4}}+e^{-i\frac
	{\varphi}{2}-i\frac{\rm{\pi}}{4}}\right]  ,\label{S13}\\
\end{equation}

\begin{equation}
S_{14}=-\frac{te^{-i\varphi_{\rm{s}}/2+i\rm{\pi}/4}%
}{\sqrt{2}\beta\left(1+\frac{\alpha^{2}}{\beta^{2}}t^{2}\right)  }\left[e^{i\frac{\varphi}{2}+i\frac{\rm{\pi}}{4}}+\sqrt{1-t^{2}}\alpha e^{-i\frac
	{\varphi}{2}-i\frac{\rm{\pi}}{4}}\right]  ,\label{S14}\\
\end{equation}

\begin{equation}
S_{23}=-\frac{te^{-i\varphi_{\rm{s}}/2+i\rm{\pi}/4}%
}{\sqrt{2}\beta\left(1+\frac{\alpha^{2}}{\beta^{2}}t^{2}\right)  }\left[
\sqrt{1-t^{2}}\alpha e^{i\frac{\varphi}{2}-i\frac{\rm{\pi}}{4}}+e^{-i\frac
	{\varphi}{2}+i\frac{\rm{\pi}}{4}}\right]  ,\label{S23}\\
\end{equation}

\begin{equation}
S_{24}=-\frac{te^{-i\varphi_{\rm{s}}/2+i\rm{\pi}/4}%
}{\sqrt{2}\beta\left(1+\frac{\alpha^{2}}{\beta^{2}}t^{2}\right)  }\left[
e^{i\frac{\varphi}{2}-i\frac{\rm{\pi}}{4}}+\sqrt{1-t^{2}}\alpha e^{-i\frac
	{\varphi}{2}+i\frac{\rm{\pi}}{4}}\right]  ,\label{S24}\\
\end{equation}

\begin{equation}
S_{31}=-\frac{te^{i\varphi_{\rm{s}}/2+i\rm{\pi}/4}%
}{\sqrt{2}\beta\left(1+\frac{\alpha^{2}}{\beta^{2}}t^{2}\right)  }\left[
\sqrt{1-t^{2}}\alpha e^{-i\frac{\varphi}{2}+i\frac{\rm{\pi}}{4}}%
+e^{i\frac{\varphi}{2}-i\frac{\rm{\pi}}{4}}\right]  ,\label{S31}\\
\end{equation}

\begin{equation}
S_{32}=-\frac{te^{i\varphi_{\rm{s}}/2+i\rm{\pi}/4}%
}{\sqrt{2}\beta\left(1+\frac{\alpha^{2}}{\beta^{2}}t^{2}\right)  }\left[
\sqrt{1-t^{2}}\alpha e^{-i\frac{\varphi}{2}-i\frac{\rm{\pi}}{4}}%
+e^{i\frac{\varphi}{2}+i\frac{\rm{\pi}}{4}}\right]  ,\label{S32}\\
\end{equation}

\begin{equation}
S_{41}=-\frac{te^{i\varphi_{\rm{s}}/2+i\rm{\pi}/4}%
}{\sqrt{2}\beta\left(1+\frac{\alpha^{2}}{\beta^{2}}t^{2}\right)  }\left[
e^{-i\frac{\varphi}{2}+i\frac{\rm{\pi}}{4}}+\sqrt{1-t^{2}}\alpha
e^{i\frac{\varphi}{2}-i\frac{\rm{\pi}}{4}}\right]  ,\label{S41}\\
\end{equation}

\begin{equation}
S_{42}=-\frac{te^{i\varphi_{\rm{s}}/2+i\rm{\pi}/4}%
}{\sqrt{2}\beta\left(1+\frac{\alpha^{2}}{\beta^{2}}t^{2}\right)  }\left[
e^{-i\frac{\varphi}{2}-i\frac{\rm{\pi}}{4}}+\sqrt{1-t^{2}}\alpha
e^{i\frac{\varphi}{2}+i\frac{\rm{\pi}}{4}}\right]. \label{S42}
\end{equation}

While Eqs.\,(\ref{S21})-(\ref{S42}) have the same form as those derived in Ref. \cite{MPIJIS}, the phases $\varphi$ and $\varphi_s$ are different. In this case, they are given by $\varphi\equiv\varphi_{p}+p\pi$ and $\varphi_{\rm{s}}\equiv\phi_{p1}+\phi_{p2}+p\pi$, where $\varphi_{p}=\phi_{p1}-\phi_{p2}$. 

Before we outline the effect of the parity parameter $p$ on the MPIJIS response, let us consider two important cases:

1) No applied pump, i.e., $t=0$. In this case, the scattering matrix reduces into

\begin{equation}
\left[  S\right]  =\left(
\begin{array}
[c]{cccc}%
0 & i & 0 & 0\\
i & 0 & 0 & 0\\
0 & 0 & -1 & 0\\
0 & 0 & 0 & -1
\end{array}
\right) \cdot\label{S_mat_t_zero}
\end{equation}

This result shows that when the MPIJIS is off, it is transparent for propagating signals and effectively behaves as a lossless transmission line with an added reciprocal phase shift of $\rm{\pi}/2$ for transmitted signals within the bandwidth of the $90^{\circ}$ hybrid.

2) The JPCs are biased at the 50:50 beam splitter working point, i.e., $r=t=1/\sqrt{2}$, the `b' mode coupler is symmetrical $\alpha=\beta=1/\sqrt{2}$, the phase difference is $\varphi=-\rm{\pi}/2$, and the sum is  $\varphi_{\rm{s}}=\rm{\pi}/2$. In this case, the scattering matrix becomes

\begin{equation}
\left[  S\right]  =\left(
\begin{array}
[c]{cccc}%
0 & 0 & -\frac{1}{\sqrt{2}} & -\frac{1}{\sqrt{2}}\\
\frac{i2\sqrt{2}}{3} & 0 & -\frac{i}{3\sqrt{2}} & \frac{i}{3\sqrt{2}}\\
-\frac{1}{3\sqrt{2}} & -\frac{i}{\sqrt{2}} & -\frac{\sqrt{2}}{3} & \frac{\sqrt{2}}{3} \\
\frac{1}{3\sqrt{2}} & -\frac{i}{\sqrt{2}} & \frac{\sqrt{2}}{3} & -\frac{\sqrt{2}}{3}
\end{array}
\right), \label{S_mat_t_gh}
\end{equation}

\noindent which shows that, under the above conditions, the MPIJIS functions as an isolator with almost unity transmission in the forward direction, i.e., $|S_{21}|=2\sqrt{2}/3\cong0.943$ (which corresponds to an insertion loss of about $0.5$ dB in the signal power), total isolation in the opposite direction $|S_{12}|=0$, and vanishing reflections $|S_{11}|=|S_{22}|=0$. Furthermore, it shows that the cold loads on ports $3$ and $4$ play a similar role to internal ports of standard magnetic isolators. They dissipate the energy of back-propagating signals $|S_{32}|=|S_{42}|=1/\sqrt{2}$ and emit noise (e.g., vacuum noise) towards the input $|S_{13}|=|S_{14}|=1/\sqrt{2}$. 

Although Eqs.\,(\ref{S21})-(\ref{S42}) are derived for on-resonance signals, it is straightforward to generalize them for signals that lie within the JPC dynamical bandwidth. This is done by substituting \cite{JPCreview}

\begin{align}
t[\omega_{\rm{1}}]=\dfrac{2\rho}{\chi_{a}^{-1}\chi_{b}^{-1}+\rho^2}, \label{t_param_vs_freq} 
\end{align}

\noindent where $\chi's$ are the bare response functions of modes \textit{a} and \textit{b} (whose inverses depend linearly on $f_{\rm{1}}$ and $f_{\rm{2}}$): 

\begin{align}
\chi_{a}^{-1}[\omega_{\rm{1}}]=1-2i\dfrac{\omega_{\rm{1}}-\omega_{a}}{\gamma_{a}}, \nonumber \\ 
\chi_{b}^{-1}[\omega_{\rm{1}}]=1-2i\dfrac{\omega_{\rm{1}}-\omega_{a}}{\gamma_{b}}, 
\label{Chi_params}%
\end{align}

\noindent where $\omega_{\rm{1}}=2\pi f_{1}$, $\omega_{a}=2\pi f_{a}$ and the applied pump angular frequency is given by $\omega_{p}=\omega_{b}-\omega_{a}=\omega_{\rm{2}}-\omega_{\rm{1}}$, where $\omega_{b}=2\pi f_{b}$ and $\omega_{\rm{2}}=2\pi f_{2}$.

This generalization holds under the assumption that the bandwidth of the $90^{\circ}$ hybrid is much larger than the $3-12$ MHz dynamical bandwidths of the JPCs, which is generally the case because transmission-line-based hybrids typically exhibit bandwidths of a few hundreds of megahertz \cite{CPWhybrids}. 

\section{Effective two-port model}
Here, we derive an effective two-port model of MPIJIS, which calculates the scattering parameters of ports 1 and 2 only. In this model, we replace the effective coupler shown in Fig.\,\ref{SignalFlow}(a) by a lossy delay line, coupling mode `b' of the two JPCs as depicted in Fig.\,\ref{SignalFlow}(b). 
The transmission coefficient of this delay line is frequency dependent and can be expressed as $\alpha[\omega_{\rm{1}}]=|\alpha|e^{i\theta_d}$ where the phase delay reads $\theta_d[\omega_{\rm{1}}]=\omega_2 \sqrt{\epsilon_{\rm{eff}}}l_d/c=(\omega_1+\omega_p)\sqrt{\epsilon_{\rm{eff}}}l_d/c$, $l_d=11.283$ mm is the length of the microstrip transmission line coupling the two JPCs, $c$ is the speed of light, $\epsilon_{\rm{eff}}=7.418$ is the effective dielectric constant of the microstrip transmission line. The amplitude $0 \leq |\alpha| \leq 1$ represents the amplitude attenuation of the internal mode `b' due to coupling to the cold terminations.

To calculate the device scattering parameters at ports $1$ and $2$, we start off by writing the scattering parameters of the inner device defined by ports $1^{\prime}$ and $2^{\prime}$, which excludes the $90^{\circ}$ hybrid. Using the signal flow graph shown in Fig.\,\ref{SignalFlow}(b), we obtain 

\begin{align}
s_{1^{\prime}2^{\prime}}[\omega_{\rm{1}}] & =-\dfrac{\alpha t^2 e^{-i \varphi}}{1-r^{2}_{b} \alpha^2}, \\
s_{2^{\prime}1^{\prime}}[\omega_{\rm{1}}] & =-\dfrac{\alpha t^2 e^{i \varphi}}{1-r^{2}_{b} \alpha^2}, \\
s_{1^{\prime}1^{\prime}}[\omega_{\rm{1}}] & =s_{2^{\prime}2^{\prime}}[\omega_{\rm{1}}]=r_{a}-\dfrac{{r_{b} \alpha}^2 t^2 }{1-r^{2}_{b} \alpha^2}, 
\end{align}

\noindent where the JPC transmission coefficient is given by Eq.\,(\ref{t_param_vs_freq}), $r_{a}$ and $r_{b}$ are the JPC reflection parameters at ports `a' and `b' given by

\begin{align}
r_a[\omega_{\rm{1}}]=\dfrac{\chi_{a}^{-1*}\chi_{b}^{-1}-\rho^2}{\chi_{a}^{-1}\chi_{b}^{-1}+\rho^2}, \label{r_a_param_vs_freq} \\  
r_b[\omega_{\rm{1}}]=\dfrac{\chi_{a}^{-1}\chi_{b}^{-1*}-\rho^2}{\chi_{a}^{-1}\chi_{b}^{-1}+\rho^2}. 
\label{r_b_param_vs_freq} 
\end{align}

\noindent The inverse bare response functions $\chi_{a,b}^{-1}$ of modes `a' and `b' are given by Eq.\,(\ref{Chi_params}).

After including the wave interference effect introduced by the $90^{\circ}$ hybrid, we finally arrive at the following expressions for the scattering matrix of the device 

\begin{align}
S_{21}[\omega_{\rm{1}}]=\dfrac{1}{2} \left(is_{1^{\prime}1^{\prime}}
+is_{2^{\prime}2^{\prime}} +s_{2^{\prime}1^{\prime}}-s_{1^{\prime}2^{\prime}} \right), \label{Eff_S21} \\ 
S_{12}[\omega_{\rm{1}}]=\dfrac{1}{2} \left ( is_{1^{\prime}1^{\prime}}+is_{2^{\prime}2^{\prime}} +s_{1^{\prime}2^{\prime}}-s_{2^{\prime}1^{\prime}}\right ), \label{Eff_S12} \\ 
S_{11}[\omega_{\rm{1}}]=\dfrac{1}{2} \left ( s_{1^{\prime}1^{\prime}}-s_{2^{\prime}2^{\prime}} +is_{2^{\prime}1^{\prime}}+is_{1^{\prime}2^{\prime}}\right ), \label{Eff_S11} \\ 
S_{22}[\omega_{\rm{1}}]=\dfrac{1}{2} \left ( s_{2^{\prime}2^{\prime}}-s_{1^{\prime}1^{\prime}}+is_{2^{\prime}1^{\prime}}+is_{1^{\prime}2^{\prime}}\right ). \label{Eff_S22}
\end{align}

To calculate the MPIJIS transmission response versus frequency that matches the measured curves in Fig.\,\ref{VaryPp}, we evaluate $|S_{21}|^2$, $|S_{12}|^2$ based on Eqs.\,(\ref{Eff_S21}), (\ref{Eff_S12}), while substituting $\omega_{p}/2\pi=2.727$ GHz (applied in the experiment) and $\varphi=-\pi/2$ ($\varphi=\pi/2$) when $\rm{P}_1$ ($\rm{P}_2$) is driven. We also substitute $\gamma_a/2\pi=40$ MHz (measured), $\gamma_b/2\pi=100$ MHz (set to match the bandwidths of the measured curves), and $|\alpha|=0.51$. The latter parameter, i.e., $|\alpha|$, is evaluated by varying $\rho$ and $|\alpha|$ to simultaneously best match one pair of $|S_{21}|^2$, $|S_{12}|^2$ values measured on-resonance for the same pump power and pump port. This is because, on resonance, the device response in both directions is solely dependent on these two parameters. Next, for each pair of curves $|S_{21}|^2$, $|S_{12}|^2$ measured for the same pump power and pump port, we substitute $\omega_{a}/2\pi$ corresponding to the frequency of the respective isolation dip. This is done because the theoretical model here does not account for shifts in the JPC resonance frequency due to the Kerr effect. Finally, for each pair of curves ($|S_{21}|^2$, $|S_{12}|^2$ corresponding to the same pump power and pump port), we substitute $\rho$ that gives the respective isolation dip magnitude on resonance.  

One likely physical reason for the deviation of the parameter value $|\alpha|=0.51$ in our device from $1/\sqrt{2}$ expected for an ideal symmetric coupler (in which $\alpha$ the coupling coefficient between the `b' modes of the two JPCs is equal to $\beta$ the coupling coefficient of each JPC to the cold termination, see Fig.\,\ref{SignalFlow}(a)) is the presence of two coupling capacitors in series between the two JPCs, which couple the JPCs to the intermediate delay line (see the device configuration shown in Figs.\,\ref{Device} (d),(g)). Having these two coupling capacitors in-series in the path between the two JPCs, instead of one, effectively reduces the coupling for wave amplitudes by  $\sqrt{2}$, thus giving $|\alpha|=0.5$, which matches well the estimated value of $0.51$ in our device.   

\section{Common attributes of the MPIJIS and JPC}

To better understand the scattering parameters of the MPIJIS, we outline in Fig.\,\ref{JPCMPIJIScomp}(a) and (b), the on-resonance signal flow graph for a JPC operated in frequency conversion and a MPIJIS operated in the forward direction, respectively. If we suppress the phases of the various scattering parameters between the ports and focus on their magnitude, then the JPC can be characterized by two parameters: a reflection parameter $r$ and a transmission parameter $t$ given by Eq.\,(\ref{r_s_res}) and satisfy the condition $t^2=1-r^2$. Similarly, in the MPIJIS case, we can reduce its various scattering parameters between the input (1), output (2), and vacuum (V) ports (3,4) given by Eqs.\,(\ref{S21})-(\ref{S42}), into five main parameters, denoted as $r', r'', r''', t', t''$ in Fig.\,\ref{JPCMPIJIScomp}(b). Assuming the two JPCs in the MPIJIS are uniform and balanced, we express these parameters as,   

\begin{equation}
r'=\dfrac{r-\alpha}{1-\alpha r}  ,\label{rp}\\
\end{equation}

\begin{equation}
r''=\dfrac{r+\alpha}{1+\alpha r}  ,\label{rpp}\\
\end{equation}

\begin{equation}
r'''=\dfrac{r}{1+\dfrac{\alpha^2}{1-\alpha^2} t^2}  ,\label{rppp}\\
\end{equation}

\begin{equation}
t'=\dfrac{1}{\sqrt{2}}\dfrac{\sqrt{1-\alpha^2}}{1-\alpha r}t  ,\label{tp}\\
\end{equation}

\begin{equation}
t''=\dfrac{1}{\sqrt{2}}\dfrac{\sqrt{1-\alpha^2}}{1+\alpha r}t  ,\label{tpp}\\
\end{equation}

\noindent where $r$, $t$ are the scattering parameters of one JPC stage and $\alpha$ is the amplitude attenuation of the internal mode.

In the special case where $r=t=\alpha=1/\sqrt{2}$, we retrieve the result of Eq.\,(\ref{S_mat_t_gh}). It is also straightforward to see that in the limit $r\rightarrow \alpha$, we get $r'\rightarrow 0$, $r''\gg r'$ and $t'\gg t''$.

Interestingly, the first three parameters denoted $r'$, $r''$, and $r'''$ exhibit reflection-like dependence on $\rho$ despite representing transmission parameters between ports, in particular $r'$ and $r''$, which are the MPIJIS transmission in the backward and forward direction, respectively. This can be seen for example by substituting Eq.\,(\ref{r_s_res}) into Eqs.\,(\ref{rp}),(\ref{rpp}), which gives:

\begin{equation}
r'=\dfrac{1-\alpha'\rho^2}{1+\alpha'\rho^2}  ,\label{rp_rho}\\
\end{equation}

\begin{equation}
r''=\dfrac{1-\alpha''\rho^2}{1+\alpha''\rho^2}  ,\label{rpp_rho}\\
\end{equation}   

\noindent where $\alpha'=(1+\alpha)/(1-\alpha)$, and $\alpha''=1/\alpha'$. Casting $r'$ and $r''$ in this form implies that they generally behave as $r$ of a JPC (see Eq.\,(\ref{r_s_res})) with rescaled pump amplitudes. And since $\alpha'\geq 1$ ($\alpha''\leq 1$), we find that in the limit $\rho^2 \rightarrow \alpha''$, we get $r'\approx 0$ and $r''\approx (1-\alpha''^2)/(1+\alpha''^2)$. 

Extending this observation to the scattering parameters of the MPIJIS and JPC, we find that they generally exhibit a reflection-like dependence on $\rho$ (indicated by red lines in Fig.\,\ref{JPCMPIJIScomp}) if they link between the same mode \textit{a} or \textit{b} regardless of their physical ports, and transmission-like dependence if they link between different modes (indicated by cyan lines in Fig.\,\ref{JPCMPIJIScomp}).   

Furthermore, by inspecting the signal flow graph of Fig.\,\ref{JPCMPIJIScomp}(b), we derive a useful energy conservation condition for the MPIJIS 

\begin{equation}
T\equiv 2|t'|^2=1-|r'|^2=\frac{t^2}{1-\alpha^2} ,\label{tprp}\\
\end{equation}

\noindent which states that the energy of the signal entering through the output port of the MPIJIS is either dissipated in the loads or passed to the input. To a large extent, this is analogous to the condition of a JPC operated in frequency conversion, namely $T\equiv t^2=1-r^2$. 

Equation (\ref{tprp}) is important for two reasons: (1) it shows that for a signal entering the output, the MPIJIS with $t\rightarrow 1-\alpha^2$, mimics a JPC with  $t\rightarrow 1$ (operated in full conversion mode). But unlike the JPC, in which a small portion of the signal is reflected back to the same port, in the MPIJIS case, it is transmitted to a different port, i.e., the input. (2) It allows us to infer the amount of energy of the signal coming from the output that gets dissipated in the loads, i.e., $2|t'|^2$, by measuring the isolation of the MPIJIS, i.e., $|r'|^2$.  

\section{MPIJIS saturation due to pump depletion}

To evaluate the saturation of the MPIJIS due to pump depletion effects, we rely on the equivalence, established in Appendix C, between processing of incoming signals through the output port of a MPIJIS and operating a JPC in frequency conversion mode. We also apply, in our calculation, the same method and definitions of  Ref. \cite{JPCreview} used for evaluating the saturation of a JPC due to pump depletion effects in the amplification mode.

We start off with the quantum Langevin equation for the pump mode \textit{c} of a JPC operated in conversion mode, which reads

\begin{equation}
\dfrac{\rm{d}}{\rm{dt}}c=-i\omega_cc-ig_3a^{\dagger}b-\dfrac{\gamma_c}{2}c+\sqrt{\gamma_c}\tilde{c}^{\rm{in}}(t) ,\label{LangevinPump}\\
\end{equation}

\noindent where the input field is given by

\begin{equation}
\tilde{c}^{\rm{in}}(t)=\dfrac{1}{\sqrt{2\pi}} \int\limits_{0}^{\infty} c^{\rm{in}}[\omega]e^{-i\omega t} \,\rm{d}\omega ,\label{c_in_t}\\
\end{equation}

\noindent and satisfies the commutation relation $[c^{\rm{in}}[\omega],c^{\rm{in}}[\omega']]=\rm{sgn}((\omega-\omega')/2)\delta(\omega+\omega')$.

Taking the average value for the field \textit{c} gives:
\begin{equation}
\dfrac{\rm{d}}{\rm{dt}}\left\langle c \right\rangle =-i\omega_c \left\langle c \right\rangle -ig_3 \left\langle a^{\dagger}b \right\rangle-\dfrac{\gamma_c}{2} \left\langle c\right\rangle+\sqrt{\gamma_c}\left\langle \tilde{c}^{\rm{in}}(t) \right\rangle.\label{LangevinAvg}\\
\end{equation}

In steady state and using the rotating-wave-approximation (RWA) we obtain:

\begin{equation}
ig_3 \left\langle a^{\dagger}b \right\rangle+\dfrac{\gamma_c}{2} \left\langle c\right\rangle=\sqrt{\gamma_c}\left\langle \tilde{c}^{\rm{in}}(t) \right\rangle.\label{LangevinSteadyState}\\
\end{equation} 

In the limit of no input ($\left\langle  a^{\dagger}b \right\rangle=0$) we get:

\begin{equation}
\left\langle c\right\rangle=\dfrac{2}{\sqrt{\gamma_c}}\left\langle \tilde{c}^{\rm{in}}(t) \right\rangle.\label{cinNoIn}\\
\end{equation} 

In this case the average number of photons in the resonator \textit{c} is

\begin{equation}
\lim_{\left\langle  a^{\dagger}b \right\rangle\rightarrow 0}\bar{n}_c=\left| \left\langle c\right\rangle\right|^2=\dfrac{4}{\gamma_c}\left| \left\langle \tilde{c}^{\rm{in}}(t) \right\rangle\right|^2=\dfrac{4}{\gamma_c}\bar{n}^{\rm{in}}_c,\label{avgPumpPhNoIn}\\
\end{equation} 

\noindent where $\bar{n}^{\rm{in}}_c$ is the average input pump photons per unit time. 

In the presence of input $\left\langle  a^{\dagger}b \right\rangle\neq0$, the pump drive experiences an additional decay channel associated with the photon conversion process taking place in the device. Thus, we define an effective decay rate of pump photons $\gamma_{\rm{eff}}$ given by

\begin{equation}
ig_3\left\langle  a(t)^{\dagger}b(t) \right\rangle=\dfrac{\gamma_{\rm{eff}}}{2} \left\langle c(t) \right\rangle.\label{GammaEffDef}\\
\end{equation} 

To calculate $\gamma_{\rm{eff}}$, we first evaluate $\left\langle  a(t)^{\dagger}b(t) \right\rangle$ in the frame rotating with the pump phase:

\begin{equation}
\left\langle  a(t)^{\dagger}b(t) \right\rangle=\dfrac{1}{2\pi} \int\limits_{-\infty}^{\infty}\int\limits_{-\infty}^{\infty} \left\langle a\left[ \omega\right] b\left[ \omega'\right]\right\rangle e^{-i(\omega+\omega')t}\,\rm{d}\omega\,\rm{d}\omega'. \label{AvgAdb}\\
\end{equation}

Using the JPC scattering parameters in conversion mode \cite{JPCreview}, and the input-output relations given by:

 \begin{equation}
\sqrt{\gamma_a}a\left[ \omega\right]=a^{\rm{in}}\left[ \omega\right] +a^{\rm{out}}\left[ \omega \right] ,\label{InOutA}\\
 \end{equation}

\begin{equation}
\sqrt{\gamma_b}b\left[ \omega' \right]=b^{\rm{in}}\left[ \omega'\right] +b^{\rm{out}}\left[ \omega' \right] ,\label{InOutB}\\
\end{equation} 

\noindent we obtain 

\begin{equation}
\left\langle a\left[ \omega\right] b \left[ \omega'\right] \right\rangle=\dfrac{-iT\left( \Delta\omega\right)}{\sqrt{\gamma_a\gamma_b}\rho} \left[ \left\langle a^{\rm{in}}\left[ \omega \right] a^{\rm{in}}\left[ \omega'\right] \right\rangle + \left\langle b^{\rm{in}}\left[ \omega \right] b^{\rm{in}}\left[ \omega'\right] \right\rangle\right]  ,\label{AvgabFreq}\\
\end{equation} 

\noindent where $T(\Delta\omega)=|t|^2=4\rho^2/|\chi_{a}^{-1}\chi_{b}^{-1}+\rho^2|^2$.

Substituting the anticommutator for the noise field given by

\begin{align}
\left\langle \left\lbrace a^{\rm{in}}\left[ \omega\right],a^{\rm{in}}\left[ \omega'\right] \right\rbrace \right\rangle_T  & 
=\left\langle \left\lbrace b^{\rm{in}}\left[ \omega\right],b^{\rm{in}}\left[ \omega'\right] \right\rbrace \right\rangle_T\nonumber \\
&=2\mathcal{N}_T \left( \dfrac{\omega-\omega'}{2} \right)\delta\left( \omega+\omega'\right) \nonumber  ,\label{Anticommutator}\\
\end{align} 
 
\noindent into Eqs.\,(\ref{AvgabFreq}) and (\ref{AvgAdb}), yields

\begin{equation}
\left\langle  a(t)^{\dagger}b(t) \right\rangle=\dfrac{-2i}{\sqrt{\gamma_a\gamma_b}\rho}\dfrac{1}{2\pi} \int\limits_{0}^{\infty} \left[ \mathcal{N}^{\rm{in}}_a(\omega)+\mathcal{N}^{\rm{in}}_b(\omega)\right] T(\Delta\omega)\,\rm{d}\omega, \label{AvgAdbwNoise}\\
\end{equation}
   
\noindent where $\mathcal{N}_T$ is the photon spectral density given by

\begin{align}
\mathcal{N}_T[\omega] & =\dfrac{\rm{sgn}(\omega)}{2}\rm{coth}\left( \dfrac{\hbar\omega}{2k_BT}\right)  \nonumber\\
&=\rm{sgn}(\omega)\left[\textit{N}_\textit{T}\left[ |\omega|\right]+\dfrac{1}{2} \right], \label{PhotonSpectralNoise}\\ \nonumber
\end{align}

\noindent and $N_T\left[ \omega \right]=1/(e^{\hbar\omega/k_BT}-1)$ is the Bose-Einstein distribution.  
  
Neglecting vacuum noise which is much smaller than the thermal noise coming from the $4$ K stage and couples to mode \textit{a}, we obtain

\begin{equation}
\left\langle  a(t)^{\dagger}b(t) \right\rangle =\dfrac{-2i}{\sqrt{\gamma_a\gamma_b}\rho}\dfrac{1}{2\pi} \int\limits_{0}^{\infty}  N^{\rm{in}}_a(\omega) \left[ 1-L(\omega)\right] \,\rm{d}\omega, \label{AvgANoise}\\
\end{equation}    

\noindent where we applied the relation (\ref{tprp}) that links the transmission of a JPC $T(\Delta\omega)$ to the reverse transmission of a MPIJIS device $L(\omega)$.

Substituting Eq.\,(\ref{AvgANoise}) in Eq.\,(\ref{GammaEffDef}) along with Eq.\,(\ref{avgPumpPhNoIn}) and the relation $\rho=2g_3\sqrt{\bar{n}_c}/\sqrt{\gamma_a \gamma_b}$, yields the simple relation

\begin{equation}
\gamma_{\rm{eff}}=\gamma_c\dfrac{\bar{n}^{\rm{in}}_n}{4\bar{n}^{\rm{in}}_c}, \label{gammeEffsimple}\\
\end{equation} 

\noindent where $\bar{n}^{\rm{in}}_n \equiv \dfrac{1}{\pi}\int\limits_{0}^{\infty}  N^{\rm{in}}_a(\omega) \left[ 1-L(\omega)\right]\,\rm{d}\omega$ represents the average input noise photons per unit time that are transferred to the MPIJIS loads.

Next, we note from Eq.\,(\ref{avgPumpPhNoIn}) that for a given setpoint $L_0$, the average pump photons in the device with no applied noise is 

\begin{equation}
\bar{n}_c\left(L_0, \bar{n}^{\rm{in}}_n=0\right) =\dfrac{4}{\gamma_c}\bar{n}^{\rm{in}}_c. \label{ncNoNoise}\\
\end{equation} 

Whereas, in the presence of noise or input signal, it becomes

\begin{equation}
\bar{n}_c\left(L_0, \bar{n}^{\rm{in}}_n\right) =\dfrac{4\gamma_c}{\left( \gamma_c+\gamma_{\rm{eff}}(L_0)\right) ^2}\bar{n}^{\rm{in}}_c. \label{ncWithNoise}\\
\end{equation} 
 
Thus, for a fixed input pump power (i.e., fixed $\bar{n}^{\rm{in}}_c$), and using Eqs.\,(\ref{gammeEffsimple}), (\ref{ncNoNoise}), (\ref{ncWithNoise}), we get 
		
\begin{align}
\dfrac{\bar{n}_c\left(L_0, \bar{n}^{\rm{in}}_n\right)}{\bar{n}_c\left(L_0, \bar{n}^{\rm{in}}_n=0\right)} & =
\dfrac{1}{\left(1+\dfrac{\bar{n}^{\rm{in}}_n(L_0)}{4\bar{n}^{\rm{in}}_c}\right) ^2} \nonumber \\
&\cong \left(1- \dfrac{\bar{n}^{\rm{in}}_n(L_0)}{2\bar{n}^{\rm{in}}_c}\right). \label{ncRatio}\\ \nonumber
\end{align} 		

On the other hand, on resonance we have

\begin{equation}
\dfrac{\bar{n}_c\left(L_0, \bar{n}^{\rm{in}}_n\right)}{\bar{n}_c\left(L_0, \bar{n}^{\rm{in}}_n=0\right)}=\dfrac{\dfrac{1-\sqrt{L}}{1+\sqrt{L}}}{\dfrac{1-\sqrt{L_0}}{1+\sqrt{L_0}}}, \label{ncRatioRes}\\ 
\end{equation} 		

\noindent where we used the relation $\rho^2=\dfrac{1}{\alpha'}\dfrac{1-\sqrt{L}}{1+\sqrt{L}}$ derived from Eq.\,(\ref{rp_rho}) (in which $r'\equiv\sqrt{L}$) and the dependence of $\rho$ on $\bar{n}_c$.

Combining the results of Eq.\,(\ref{ncRatio}) and Eq.\,(\ref{ncRatioRes}) gives for $L\ll1$

\begin{equation}
2\sqrt{L_0}\left(\sqrt{\dfrac{L}{L_0}}-1 \right) =\dfrac{\bar{n}^{\rm{in}}_n(L_0)}{2\bar{n}^{\rm{in}}_c}. \label{SatCreteria1}\\ 
\end{equation} 	

Hence, if we limit the deviation of $L$ compared to $L_0$ to $L/L_0<1+\epsilon$, where $\epsilon\ll1$, then the device saturation due to pump depletion can be considered small if the ratio of the input noise to input pump photons satisfy the inequality
 
\begin{equation}
\dfrac{\bar{n}^{\rm{in}}_n(L_0)}{2\bar{n}^{\rm{in}}_c}<\epsilon\sqrt{L_0}. \label{SatCreteria2}\\ 
\end{equation} 
 
\section{The effect of pump depletion on the dephasing rate}
 
To model the observed decrease in the qubit dephasing rate with the MPIJIS applied pump power, we update the filtering parameter $A$ of the resonator-MPIJIS system to include the effect of power saturation of the MPIJIS due to pump depletion, i.e., $A\rightarrow A(L)$. 

Starting from the on-resonance relation (\ref{SatCreteria1}), we obtain the following approximation

\begin{equation}
L=L_0+\sqrt{L_0}\dfrac{\bar{n}^{\rm{in}}_n(L_0)}{2\bar{n}^{\rm{in}}_c}. \label{Lcorr}\\ 
\end{equation}

Extending it to frequencies within the device bandwidth, gives

\begin{equation}
L(\omega)=L_0(\omega)+\sqrt{L_0(\omega)}\dfrac{\bar{n}^{\rm{in}}_n(L_0)}{2\bar{n}^{\rm{in}}_c}. \label{Lcorrw}\\ 
\end{equation}

Multiplying Eq.\,(\ref{Lcorrw}) by the filter response of the readout-resonator in the frequency domain, i.e., $R(\omega)$, yields 

\begin{equation}
L(\omega)R(\omega)=L_0(\omega)R(\omega)+\sqrt{L_0(\omega)}R(\omega)\dfrac{\bar{n}^{\rm{in}}_n(L_0)}{2\bar{n}^{\rm{in}}_c}. \label{LRcorrw}\\ 
\end{equation}

Integrating the combined response over the frequency range spanned by the cut-off angular frequencies $\omega_{c1}$ and $\omega_{c1}$, gives 

\begin{equation}
A(L)=A+\dfrac{\bar{n}^{\rm{in}}_n(L_0)}{2\bar{n}^{\rm{in}}_c}\int\limits_{\omega_{c1}}^{\omega_{c2}}\sqrt{L_0(\omega)}R(\omega)\,\rm{d}\omega, \label{LRcorrwInt}\\ 
\end{equation}
 
\noindent where $A(L)\equiv\int\limits_{\omega_{c1}}^{\omega_{c2}}L(\omega)R(\omega)\,\rm{d}\omega$, $A\equiv\int\limits_{\omega_{c1}}^{\omega_{c2}}L_0(\omega)R(\omega)\,\rm{d}\omega$, and $\bar{n}^{\rm{in}}_n(L_0) \equiv \dfrac{1}{\pi}\int\limits_{\omega_{c1}}^{\omega_{c2}}  N^{\rm{in}}_a(\omega) \left[ 1-L_0(\omega)\right]\,\rm{d}\omega$. 

Furthermore, since $N^{\rm{in}}_a(\omega)$ is a slowly varying function in the relevant frequency range  $[\omega_{c1},\omega_{c2}]$, we introduce 
an effective noise photon flux (per unit time) $\gamma_n\bar{n}_n$ such that $\bar{n}^{\rm{in}}_n(L_0)=\gamma_n\bar{n}_nT_t$, where $T_t=\int\limits_{\omega_{c1}}^{\omega_{c2}} \left[ 1-L_0(\omega)\right]\,\rm{d}\omega$.

This allows us to cast Eq.\,(\ref{LRcorrwInt}) in a simpler form 

\begin{equation}
A(L)=A+C\dfrac{T_t}{\rho^2}\int\limits_{\omega_{c1}}^{\omega_{c2}}\sqrt{L_0(\omega)}R(\omega)\,\rm{d}\omega, \label{LRcorrwIntSimple}\\ 
\end{equation}

\noindent where $C=(2\gamma_n/\pi\gamma_c)\rho^2_n$, and $\rho^2_n\equiv4g^2_3\bar{n}_n/\gamma_a\gamma_b$. 

When calculating the fit $T_{\rm{\varphi,D}}$ drawn in Figs.\,\ref{DephasingData} (e),(h), we take $C=0.02$, which is within the range of our system parameters. 

\section{JPC Hamiltonian}

\begin{figure*}
	[tb]
	\begin{center}
		\includegraphics[
		width=1.4\columnwidth 
		]%
		{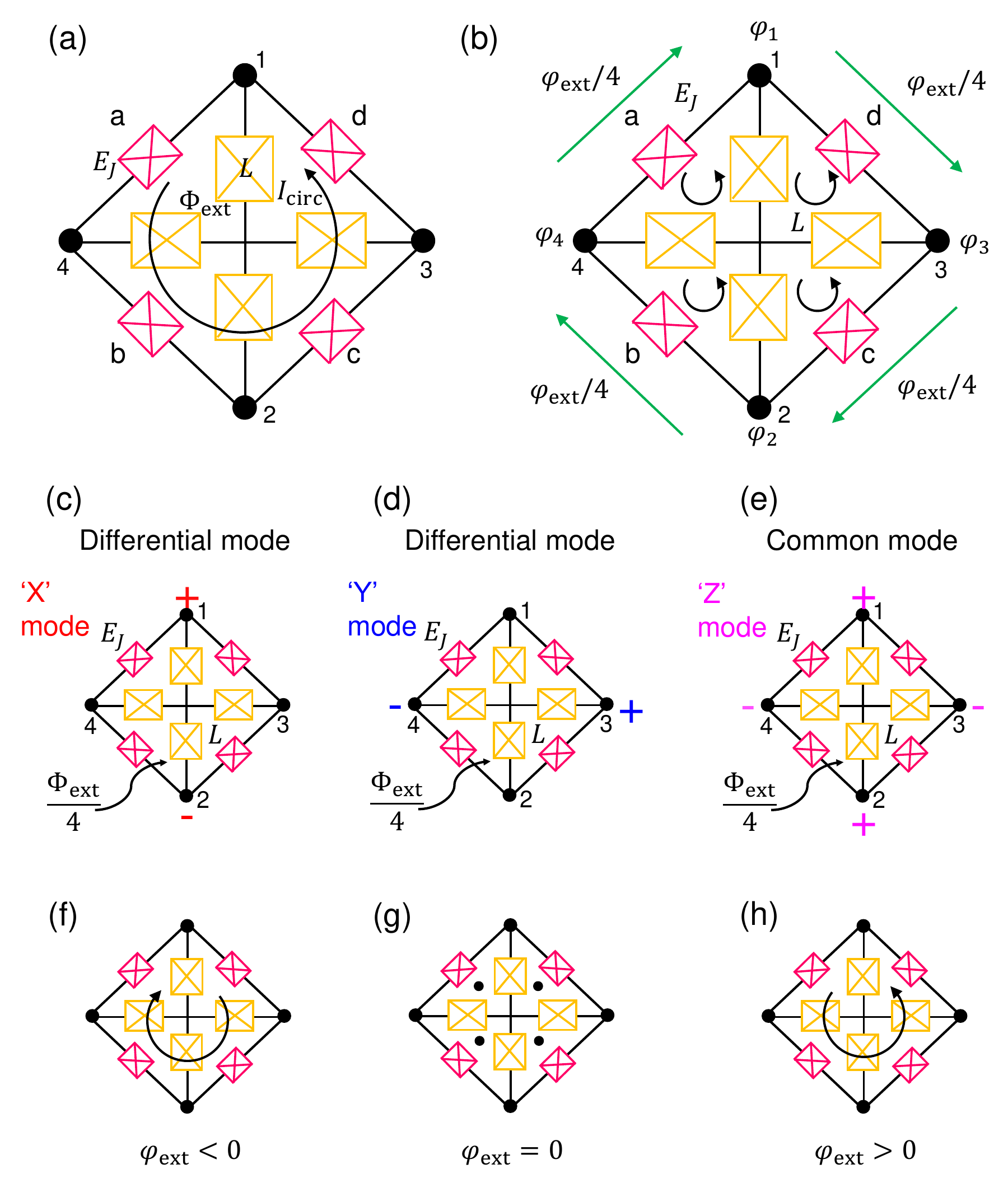}
		\caption{JRM structure, eigenmodes, and circulating currents. (a) The JRM consists of four Josephson junctions a, b, c, d with energy $E_{J}$, which are arranged in a Wheatstone bridge configuration between nodes 1, 2, 3, 4. The four large inner junctions, each with inductance $L$, function as linear shunt inductors for the JRM junctions. An externally applied magnetic flux threading the JRM loop gives rise to a circulating dc-current $I_{\rm{circ}}$. (b) The diagram shows the reduced node fluxes $\varphi_1$, $\varphi_2$, $\varphi_3$, $\varphi_4$ used in the calculation of the JRM eigenmodes. The four inner loops are assumed to be equal in size, thus each loop receives a quarter of $\Phi_{\rm{ext}}$. When the JRM is biased at a primary flux lobe, no dc current flows in the shunt inductances inside the JRM (i.e., the circulating currents in the inner branches cancel each other). (c), (d), (e) exhibit the polarity of the rf node-voltages corresponding to the eigenmodes of the JRM. (c) and (d) show the polarity patterns for the differential excitation modes `X', `Y', respectively. (e) shows the pattern for the common excitation mode `Z'. (f), (g), (h) illustrate the correspondence between the circulating current direction and the sign of the reduced external magnetic flux $\varphi_{\rm{ext}}$. The black discs in (g) indicate no circulating currents.
		}
		\label{JRM}
	\end{center}
\end{figure*}

\begin{figure*}
	[tb]
	\begin{center}
		\includegraphics[
		width=1.5\columnwidth 
		]%
		{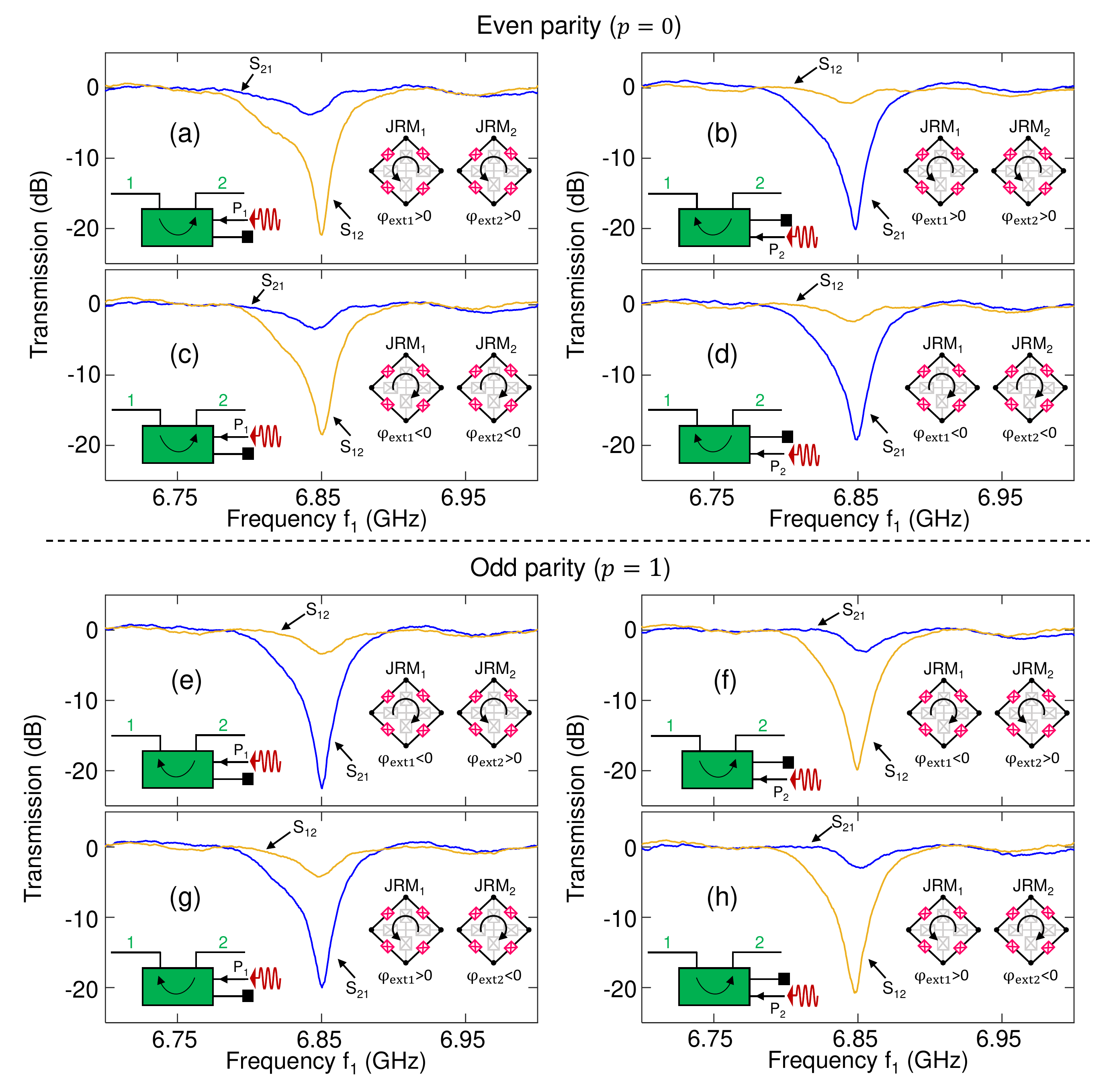}
		\caption{Dependence of the transmission/isolation direction on the parity of the fluxes threading the two JRMs. Transmission parameters of MPIJIS measured versus signal frequency. The measurement results shown in the left (right) column are taken while applying the pump drive to $\rm{P}_1$ ($\rm{P}_2$) at $f_p=2.783$ GHz. In the upper half of the figure, representing the even parity cases ($p=0$), the magnetic fluxes threading the two JRMs are either $\varphi_{\rm{ext1}}>0$, $\varphi_{\rm{ext2}}>0$ in (a) and (b) or $\varphi_{\rm{ext1}}<0$, $\varphi_{\rm{ext2}}<0$ in (c) and (d). In the bottom half of the figure, representing the odd parity cases ($p=1$), the magnetic fluxes threading the two JRMs are either $\varphi_{\rm{ext1}}<0$, $\varphi_{\rm{ext2}}>0$ in (e) and (f) or $\varphi_{\rm{ext1}}>0$, $\varphi_{\rm{ext2}}<0$ in (g) and (h). The magnitude of the flux used is $|\varphi_{\rm{ext1}}|=|\varphi_{\rm{ext2}}|=2\pi\cdot1.12$ (i.e., $\Phi_{\rm{ext}}=1.12\Phi_0$), which corresponds to a magnetic field magnitude of about $|B_{\rm{ext}}|=|\Phi_{\rm{ext}}|/A_{\rm{JRM}}\cong4\cdot10^{-6}$ T at the JRM plane, where $A_{\rm{JRM}}=(24 \mu m)^2$ is the area of the JRM loop. 
		}
		\label{ParityMag}
	\end{center}
\end{figure*}

The bare Hamiltonian of a nondegenerate three-wave mixing device, comprised of three parallel LC resonators coupled via a dispersive nonlinear medium, can be written as \cite{JPCreview,FlaviusThesis}

\begin{align}
H_{0} & =\dfrac{\Phi^{2}_{\rm{X}}}{2L_{a}}+\dfrac{\Phi^{2}_{\rm{Y}}}{2L_{b}}+\dfrac{\Phi^{2}_{\rm{Z}}}{2L_{c}}+\dfrac{Q^{2}_{\rm{X}}}{2C_{a}}+\dfrac{Q^{2}_{\rm{Y}}}{2C_{b}}+\dfrac{Q^{2}_{\rm{Z}}}{2C_{c}} \nonumber\\
&+K\Phi_{\rm{X}}\Phi_{\rm{Y}}\Phi_{\rm{Z}}, 
\label{H0} 
\end{align}

\noindent where $\Phi_{\rm{X,Y,Z}}$ are the generalized flux variables of the three resonators and $Q_{\rm{X,Y,Z}}$ are the corresponding charge variables. The first six terms of $H_0$ represent the energy of three independent harmonic oscillators a, b, and c, having angular frequencies $\omega_{a,b,c}=1/\sqrt{L_{a,b,c}C_{a,b,c}}$ and characteristic impedances $Z_{a,b,c}=\sqrt{L_{a,b,c}/C_{a,b,c}}$. Notably, the last term represents a trilinear mixing interaction between modes X, Y, and Z with a coefficient $K$.  

Assuming that the angular frequencies of the three modes satisfy $\omega_{c}<\omega_{a}<\omega_{b}$, $\omega_{c}=\omega_{b}-\omega_{a}$, the resonators are well in the underdamped regime $\gamma_{a,b,c}\ll\omega_{a,b,c}$, where $\gamma_{a}$, $\gamma_{b}$, $\gamma_{\rm{c}}$ are the corresponding photon escape rates, and $\gamma_{a}+\gamma_{b}\ll\omega_{b}-\omega_{a}$, which supposes that the envelopes of the drive signals exciting these modes are slow compared to the respective drive frequencies. 

Equivalently, the Hamiltonian $H_0$ can be expressed in terms of bosonic operators    

\begin{align}
H_{0} & =\hbar\omega_{a}a^{\dagger}a+\hbar\omega_{b}b^{\dagger}b+\hbar\omega_{c}c^{\dagger}c \nonumber\\
&+\hbar g_3 \left (a^{\dagger}+a \right)\left (b^{\dagger}+b \right)\left (c^{\dagger}+c \right), \label{H0_abc} 
\end{align}

\noindent where $a^{\dagger}$, $a$, $b$, $b^{\dagger}$, $c^{\dagger}$, $c$ are the raising and annihilation operators associated with the three modes, which commute with each other and satisfy the standard commutation relations $[a,a^{\dagger}]=1$, $[b,b^{\dagger}]=1$, $[c,c^{\dagger}]=1$. The coupling constant between the modes $g_3$ is assumed to be much smaller than the angular frequencies $\omega_{a,b,c}$ and decay rates $\gamma_{a,b,c}$. 

By further working in the framework of the rotating wave approximation and assuming a classical coherent drive (i.e., pump) at $\omega_{p}=\omega_{c}=\omega_{b}-\omega_{a}$ and $g_3\geq 0$, we obtain

\begin{equation}
H_{\rm{3wave}}={\hbar}|g_{ab}|(e^{i\phi_{p}}\textit{a}\textit{b}^{\dagger}+e^{-i\phi_{p}}\textit{a}^{\dagger}\textit{b}), \label{H_int} 
\end{equation}

\noindent where $g_{ab}=g_3\sqrt{\bar{n}_{c}}e^{-i\phi_{p}}$. In the derivation of Eq.\,(\ref{H_int}), we replaced the annihilation operator $c$ by its average value in the coherent state produced by the pump, where $\bar{n}_{c}$ is the average pump photon number and $\phi_{p}$ is the pump phase.  

To find $g_3$, we express the mode amplitudes, i.e., $\Phi_{\rm{X,Y,Z}}$, in terms of the corresponding bosonic operators: $\Phi_{\rm{X}}=\Phi^{0}_{\rm{X}}(a+a^{\dagger})$, $\Phi_{\rm{Y}}=\Phi^{0}_{\rm{Y}}(b+b^{\dagger})$, $\Phi_{\rm{Z}}=\Phi^{0}_{\rm{Z}}(c+c^{\dagger})$, where $\Phi^{0}_{\rm{X,Y,Z}}=\sqrt{\hbar Z_{a,b,c}/2}$ are the zero-point-fluctuations of the flux. Using these relations and Eqs.\,(\ref{H0}), (\ref{H0_abc}), we obtain the following link between $g_3$ and $K$ \cite{JPCreview}

\begin{equation}
{\hbar}g_3=K\Phi^{0}_{\rm{X}}\Phi^{0}_{\rm{Y}}\Phi^{0}_{\rm{Z}}. \label{g_3_K} 
\end{equation}

In the case of the JPC, the mixing term $K\Phi_{\rm{X}}\Phi_{\rm{Y}}\Phi_{\rm{Z}}$ originates from the JRM, which functions as a dispersive nonlinear mixing element. As seen in Fig.\,\ref{JRM}(a), the JRM consists of four nominally identical Josephson junctions with energy $E_{J}=\varphi_{0}I_0$, i.e., a, b, c, d, arranged in a Wheatstone bridge configuration, where $\varphi_0=\Phi_0/2\pi$ is the reduced flux quantum ($\Phi_0=h/2e$), and $I_0$ is the critical current. The JJs of the JRM are shunted by linear inductors in the form of large inner Josephson junctions, each with inductance $L$. Externally applied magnetic flux threading the JRM $\Phi_{\rm{ext}}$, induces a dc current $I_{\rm{circ}}$ circulating in the outer loop. 

For a symmetrical JRM, in which the areas of the four inner loops are equal, as shown in Fig.\,\ref{JRM}(b), the external flux threading them is $\Phi_{\rm{ext}}/4$ and for $0\leq|\Phi_{\rm{ext}}|\leq 1.4\Phi_0$ (i.e., located on the primary flux lobe), no dc currents flow in the inner branches. It is straightforward to show that the JRM supports 4 spatial eigenmodes, three which resonate at microwave frequencies X, Y, Z (shown in Fig.\,\ref{JRM}(c),(d),(e)) and a fourth (not shown) which is at dc \cite{FlaviusThesis}. As seen  Fig.\,\ref{JRM}(c),(d),(e), two of the eigenemodes `X' and `Y' are differential, whereas Z is a common excitation. As inferred from Fig.\,\ref{JRM}(b)-(e), the reduced branch fluxes representing these eigenmodes ($\varphi^{\prime}_{\rm{X,Y,Z}}\equiv\Phi^{\prime}_{\rm{X,Y,Z}}/\varphi_0$) can be expressed as orthogonal linear combinations of the reduced node fluxes of the JRM (i.e., $\varphi_i=\Phi_i/\varphi_0$, $i=1,2,3,4$): $\varphi^{\prime}_{\rm{X}}=\varphi_1-\varphi_2$, $\varphi^{\prime}_{\rm{Y}}=\varphi_3-\varphi_4$, $\varphi^{\prime}_{\rm{Z}}=\left (\varphi_1+\varphi_2-\varphi_3-\varphi_4 \right )/2$. Since the branch fluxes across the JRM $\Phi^{\prime}_{\rm{X,Y,Z}}$ are only a fraction of the total generalized fluxes $\Phi_{\rm{X,Y,Z}}$ of the oscillators in Eq.\,(\ref{H0}), they satisfy $\Phi^{\prime}_{\rm{X,Y,Z}}=p_{a,b,c}\Phi_{\rm{X,Y,Z}}$, where $p_{a,b,c}\approx L_{\rm{J0}}/L_{a,b,c}$ are participation ratios representing the fraction of the spatial modes contained in the JRM and $L_{J0}=\varphi_0/I_0$ is the linear inductance of the outer JJs. 

For small $\varphi^{\prime}_{\rm{X,Y,Z}}\ll 1$, the Hamiltonian of the shunted JRM reads \cite{FlaviusThesis,Roch}

\begin{align}
H_{\rm{JRM}} & =-E_{\rm{J}} \rm{sin} \left (\dfrac{\varphi_{\rm{ext}}}{4} \right )\varphi^{\prime}_{\rm{X}} \varphi^{\prime}_{\rm{Y}} \varphi^{\prime}_{\rm{Z}} \nonumber \\
& +\left (\dfrac{E_{\rm{L}}}{2}+E_{\rm{J}} \rm{cos} \left (\dfrac{\varphi_{\rm{ext}}}{4} \right ) \right ) \left (\dfrac{{\varphi^{\prime}_{\rm{X}}}^2}{2}+\dfrac{{\varphi^{\prime}_{\rm{Y}}}^2}{2} \right ) \nonumber \\
& +2 \left (\dfrac{E_{\rm{L}}}{4}+E_{\rm{J}} \rm{cos} \left (\dfrac{\varphi_{\rm{ext}}}{4} \right ) \right ){\varphi^{\prime}_{\rm{Z}}}^2 \nonumber \\
& -4E_{\rm{J}}\rm{cos} \left ( \dfrac{\varphi_{\rm{ext}}}{4} \right ), \label{Hjrm} 
\end{align}

\noindent where $\varphi_{\rm{ext}}=\Phi_{\rm{ext}}/\varphi_0$ is the reduced external flux, $E_{\rm{L}}=\varphi^2_{0}/L$ is the inductive energy of the large inner JJs. The first term of $H_{\rm{JRM}}$ is a three-wave mixing term, whereas the second and third terms are quadratic in the mode fluxes and therefore renormalize the mode frequencies. The fourth term is a constant (for a given external flux), that is independent of the mode fluxes. 

To calculate $g_3$ we rewrite the mixing term of $H_{\rm{JRM}}$ in the format of the right-hand side of Eq.\,(\ref{g_3_K}), which gives  

\begin{equation}
g_3=-\dfrac{E_{\rm{J}}}{\hbar}\rm{sin} \left (\dfrac{\varphi_{\rm{ext}}}{4} \right )\dfrac{p_ap_bp_c}{\varphi^{3}_0}\Phi^{0}_{\rm{X}}\Phi^{0}_{\rm{Y}}\Phi^{0}_{\rm{Z}}. \label{g_3_inter} 
\end{equation}

Finally, substituting $\Phi^{0}_{\rm{X,Y,Z}}=\sqrt{\hbar Z_{a,b,c}/2}=\sqrt{\hbar \omega_{a,b,c} L_{a,b,c}/2}$ in Eq.\,(\ref{g_3_inter}) yields

\begin{equation}
g_3=-\rm{sin} \left (\dfrac{\varphi_{\rm{ext}}}{4} \right ) \sqrt{\dfrac{p_{a}p_{b}p_{\rm{c}}\omega_{a} \omega_{b} \omega_{c}}{E^{\rm{eff}}_{J}/\hbar}}, \label{g_3} 
\end{equation}

\noindent where $E^{\rm{eff}}_J$ is the effectively available Josephson energy, which is proportional to $E_{J}$ with a numerical prefactor. This result shows that (1) the sign of the coupling constant $g_3$ is opposite of the external flux $\varphi_{\rm{ext}}$ in the range $-1.4 \cdot 2 \pi \leq \varphi_{\rm{ext}} \leq 1.4\cdot2 \pi$ (corresponding to the JPCs being flux-biased on the primary flux lobe), (2) the magnitude of $g_3$ varies with $\left | \rm{sin} \left (\dfrac{\varphi_{\rm{ext}}}{4} \right ) \right |$, for example it vanishes (no mode coupling) when $\varphi_{\rm{ext}}=0$. In Fig.\,\ref{JRM}(f),(g),(h), we show illustrations of the direction of the dc current flowing in the JRM for $\varphi_{\rm{ext}}<0$, $\varphi_{\rm{ext}}=0$, $\varphi_{\rm{ext}}>0$, respectively, which follows the relation $\rm{sgn}(\varphi_{\rm{ext}})=\rm{sgn}(I_{\rm{circ}})$. The black discs in Fig.\,\ref{JRM}(g) corresponding to $\varphi_{\rm{ext}}=0$, indicate that the circulating current is zero. 

Based on Eq.\,(\ref{g_3}), we can rewrite Eq.\,(\ref{H_int}) in the general form,

\begin{equation}
H_{\rm{3wave}}={\hbar}|g_{ab}|(e^{i\phi^{\prime}_{p}}\textit{a}\textit{b}^{\dagger}+e^{-i\phi^{\prime}_{p}}\textit{a}^{\dagger}\textit{b}), \label{H_int_gen} 
\end{equation}

\noindent where $\phi^{\prime}_{p}=\phi_{p}+n_{g}\pi$ and

\begin{equation}
n_{g}=\begin{cases} 
0 & \varphi_{\rm{ext}}\leq 0 \\
1 & \varphi_{\rm{ext}}>0 
\end{cases}.
\label{ng_sm}
\end{equation}

\section{Dependence of the MPIJIS response on applied flux parity}

In Fig.\,\ref{ParityMag}, we experimentally demonstrate that the direction of the transmission/isolation of the MPIJIS device is determined not only by the sign of the phase difference of the pumps feeding the two JPCs but also by the orientation parity of the magnetic fields biasing the two JPCs or alternatively the parity of the circulating currents flowing in the JRMs, which, in turn, determines the sign of the coupling between the eignmodes of the JPCs. As seen in the measurement results of Fig.\,\ref{ParityMag}, for the same pump, feeding the same pump port $\rm{P}_1$ or $\rm{P}_2$, the direction of the transmission/isolation can be reversed by flipping the magnetic flux in one JRM loop or preserved by flipping the magnetic flux threading two loops. This result opens the door for magnetic-field detection applications, such as the detection of the orientation parity of weak magnetic sources using simple microwave transmission measurements as outlined in Appendix K. It is important to point out here that although the parity of the magnetic fields biasing the JPCs plays a role in setting the directionality of the MPIJIS, it does not generate its nonreciprocal response, which is primarily induced by the pump phase difference. It is also important to outline here that such a systematic investigation of the effect of the parity of the applied fluxes on the MPIJIS directionality is made possible owing to the single-pump feature of the device. Without it, pinning down this effect is considerably more challenging.   

Note that in the experiment, we cannot determine the sign of $I_{\rm{circ}}$ or $\varphi_{\rm{ext}}$ flux biasing the JPCs directly. We only know the sign of the dc current $I_{\rm{coil}}$ generated by the room-temperature current source, which biases the small superconducting coil attached to each JPC. Therefore, to determine the parity of $\varphi_{\rm{ext1,2}}$, we take the following steps: (1) we make an arbitrary assumption regarding $\varphi_{\rm{ext}}$ of one JPC, for example $\rm{sgn}(\varphi_{\rm{ext1}})=\rm{sgn}(I_{\rm{circ1}})=\rm{sgn}(I_{\rm{coil1}})$, where $I_{\rm{coil1}}$ is considered negative only if it is smaller than $I_{\rm{offset1}}$, which is the dc current for which the JPC is biased at the maximum frequency of the primary flux lobe (corresponding to $\varphi_{\rm{ext1}}=0$). In general, $I_{\rm{offset}}$ is slightly off zero when there is a nonzero magnetic field within the cryoperm magnetic shield can enclosing the device. (2) We flux bias $\rm{JPC}_1$ at $I_{\rm{coil1}}<I_{\rm{offset1}}$ for which we assume $\varphi_{\rm{ext1}}<0$. (3) We set $I_{\rm{coil2}}$ such that it is negative, i.e., $I_{\rm{coil2}}<I_{\rm{offset2}}$, and aligns the resonance frequency of $\rm{JPC}_2$ with that of $\rm{JPC}_1$ at $I_{\rm{coil1}}$. (4) We apply a pump tone to $\rm{P}_1$ (for which $\varphi_p=-\pi/2$) and tune its frequency and power to yield a near unity transmission in one direction and isolation in the other direction. (5) Since   $\rm{sgn}(g_3)=-\rm{sgn}(\varphi_{\rm{ext}})$, the sign of $g_3$ of both JPCs is the same if $\rm{sgn}(\varphi_{\rm{ext1}})=\rm{sgn}(\varphi_{\rm{ext2}})$. Thus, if we obtain in step (4) $|S_{21}|\rightarrow 1$ and $|S_{12}|\rightarrow 0$, this implies based on Eqs.\,(\ref{S21}), (\ref{S12}) that the sign of $g_{3}$ of both JPCs is the same (i.e., $p=0$), otherwise the response is reversed. Lastly, (6) once we find in step (5) a pair $\left ( I_{\rm{coil1}},I_{\rm{coil2}} \right )$ which gives rise to the same $g_3$ sign based on the device response, we can determine the other parity configurations in a self-consistent manner including for a pump fed through $\rm{P}_2$.   

\section{Calibration of MPIJIS scattering parameters}
\begin{figure*}
	[tb]
	\begin{center}
		\includegraphics[
		width=1.5\columnwidth 
		]%
		{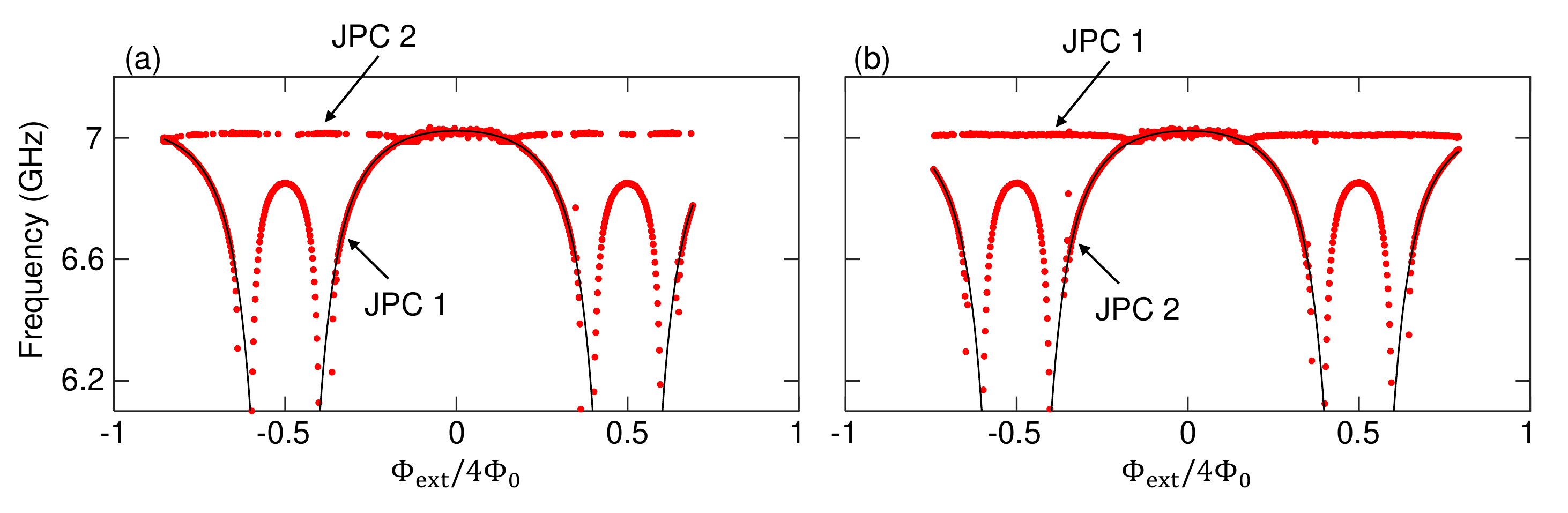}
		\caption{JPC resonance frequency versus external magnetic flux. (a) and (b) exhibit the resonance frequency of mode `a' of the two JPCs that make the MPIJIS as a function of the external magnetic flux threading the first and second JRMs, respectively. In measurements (a) and (b), mode `a' of $\rm{JPC}_2$ and $\rm{JPC}_1$ is parked at its maximum frequency via a constant magnetic flux applied to $\rm{JRM}_2$ and $\rm{JRM}_1$, respectively. The solid black curves represent theory fits. The fits employ the same device parameters in both plots.        
		}
		\label{FluxDependence}
	\end{center}
\end{figure*}

\begin{figure}
	[tb]
	\begin{center}
		\includegraphics[
		width=\columnwidth 
		]%
		{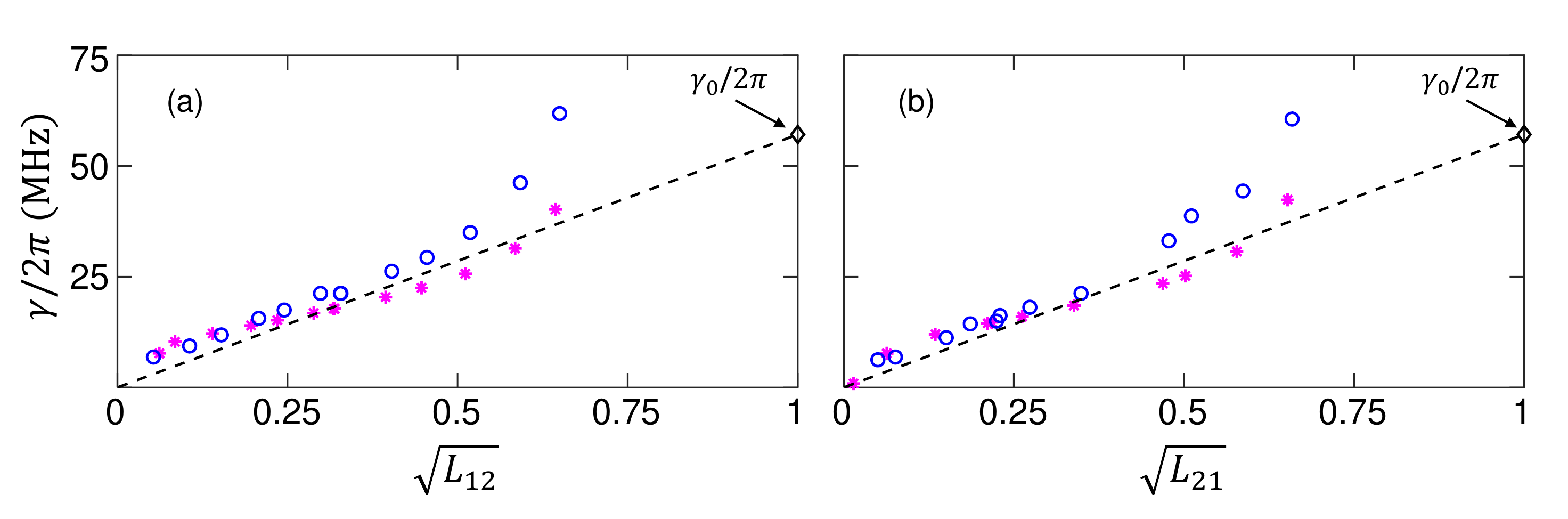}
		\caption{Bandwidth inverse-amplitude-attenuation product measurement. Dynamical bandwidth of MPIJIS ($\gamma/2\pi$) versus its amplitude attenuation $\sqrt{L_{12}}=|S_{12}|$ and $\sqrt{L_{21}}=|S_{21}|$ depicted in plots (a) and (b), respectively. The blue circles are data extracted from the measurements of Fig.\,\ref{VaryPp} for different pump drives fed to $\rm{P}_1$ and $\rm{P}_2$. The magenta stars are theoretical results extracted from the calculated curves. The black diamond corresponds the effective linear bandwidth of MPIJIS when it is off ($L=1$). The dashed black line corresponds to the relation $\gamma=\gamma_0\sqrt{L}$.           
		}
		\label{BWvL}
	\end{center}
\end{figure}

\begin{figure}
	[tb]
	\begin{center}
		\includegraphics[
		width=\columnwidth 
		]%
		{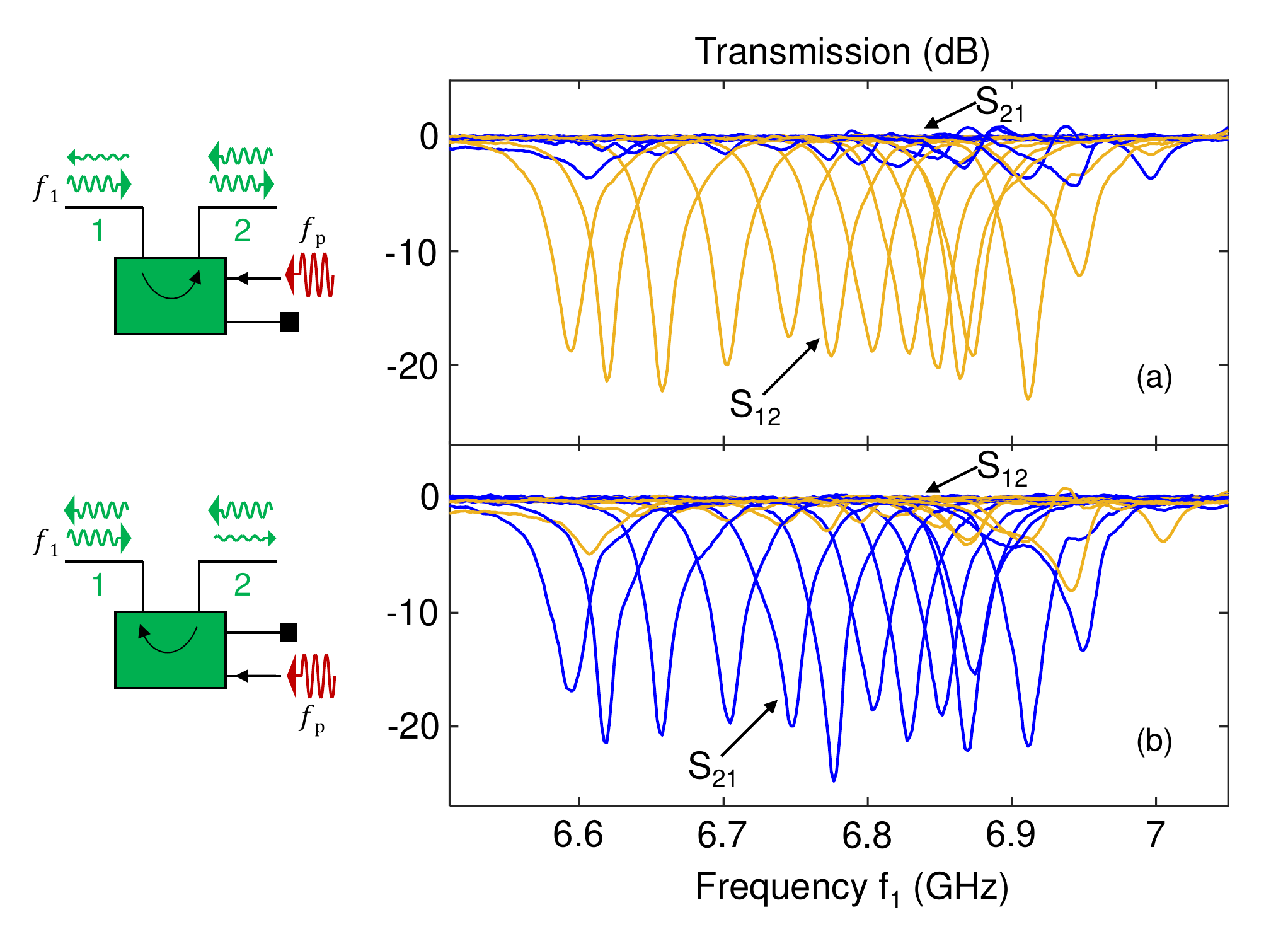}
		\caption{Tunable bandwidth measurement. The resonance frequencies of the JPCs are varied in-tandem by varying the magnetic fluxes threading the two JRMs. For each flux working point, the pump frequency and power are adjusted to yield isolation of more than $18$ dB in the relevant signal direction depending on which pump port is driven, i.e., $\rm{P}_1$ in (a) and $\rm{P}_2$ in (b).     
		}
		\label{Tunablebw}
	\end{center}
\end{figure}

\begin{figure}
	[tb]
	\begin{center}
		\includegraphics[
		width=\columnwidth 
		]%
		{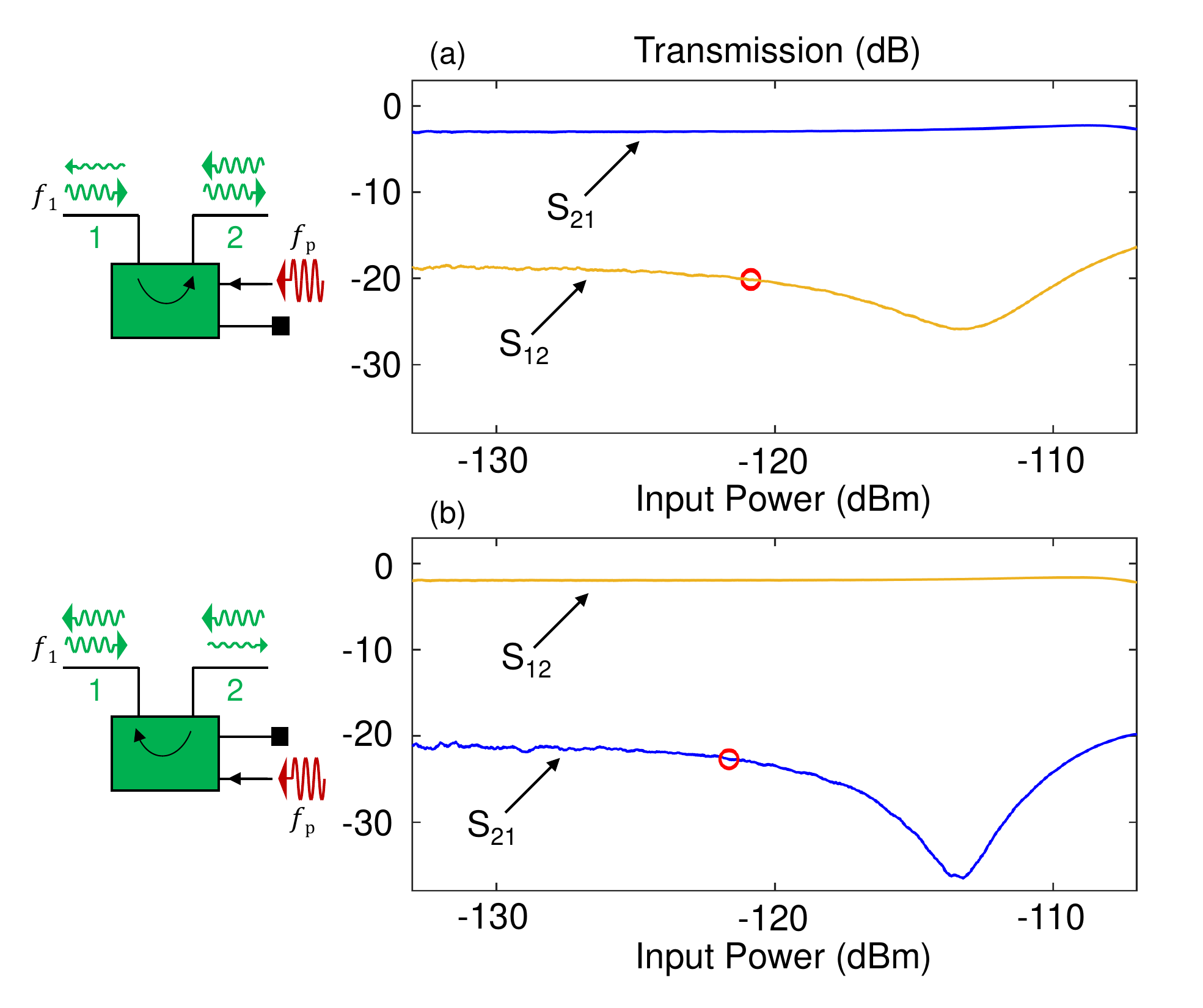}
		\caption{Saturation power measurement of the MPIJIS. (a) and (b), corresponding to pump drives applied to port $\rm{P}_1$ and $\rm{P}_2$, respectively, exhibit transmission parameters of the MPIJIS, i.e., $|S_{21}|^2$ (blue), $|S_{12}|^2$ (orange), measured as a function of the signal input power. The red circles indicate the maximum input power for which the transmission magnitude of the MPIJIS changes by 1 dB relative to the low-power value.        
		}
		\label{DR}
	\end{center}
\end{figure}

The reference level for unity (or close to unity) transmission of the MPIJIS device (in the range $-0.5 - 0$ dB) is determined by its off-resonance transmission response when the pump is off. This is because JPCs function as perfect mirrors when the pump is off, and the on-chip superconducting hybrids are lossless. More specifically, the off-resonance $0$ dB reference value, which we later subtract from the device transmission measurement versus frequency, is calculated as the average of about $100-300$ transmission data points away from the resonance. This is done in order to dilute the effect of frequency-dependent ripples in the transmission response when determining the $0$ dB reference value. These relatively low-frequency and small amplitude ripples in the scattering parameter measurements, as seen in Figs.\,\ref{wpt},\ref{ParityMag}, are generally difficult to avoid and originate from reflections in the coax lines that connect the device under test and the circulator ports. Such a calibration method, based on the off-resonance, pump-off response, or a similar variation in which the pump-off response is subtracted from the pump-on curve is commonly used in determining the amplification gain of JPAs and JPCs.

To calibrate the reflection parameters of the MPIJIS device, we rely on the fact that the two input lines that are coupled to the MPIJIS ports are nominally identical by design. That is, we set them to have the same attenuation and cable lengths both inside and outside the fridge down to the device ports. Furthermore, we verify that the transmission magnitudes versus frequency, measured through these lines at room temperature, agree within less than 1 dB. 

Using this fact, we calibrate the reflection magnitude for the pump-off case by employing the following useful relations. Let $I$ represent the attenuation of the input lines as a function of frequency in dB (takes negative values) and $O_1$ ($O_2$) represent the gain of the output line $1$ ($2$) as a function of frequency in dB (take positive values), respectively. Let also $|S_{11}|$ ($|S_{22}|$) represent the calibrated reflection off port $1$ ($2$) of the MPIJIS in dB (take negative values) and $|S_{21}|$ ($|S_{12}|$) represent the calibrated transmission magnitude from port $1$ to $2$ ($2$ to $1$) in dB (take non-positive values). Then the reflection parameters, i.e., $|R_{11}|$ (dB) and $|R_{22}|$ (dB), as measured by the vector network analyzer are given by

\begin{align}
|R_{11}|=I+O_1+|S_{11}|, \label{R11} \\
|R_{22}|=I+O_2+|S_{22}|. \label{R22} 
\end{align}

Similarly, the transmission parameters, i.e., $|T_{21}|$ (dB) and $|T_{12}|$ (dB), are given by 

\begin{align}
|T_{21}|=I+O_2+|S_{21}|, \label{T21} \\
|T_{12}|=I+O_1+|S_{12}|. \label{T12} 
\end{align}

Note that once $|S_{21}|$ is determined using the off-resonance method (described above), $|S_{22}|$ can be extracted from $|R_{22}|$ measurement. This is because the sum $I+O_2$ is shared by both $|R_{22}|$ and $|T_{21}|$. The same applies to finding $|S_{11}|$ after $|S_{12}|$ is determined. This is because the sum $I+O_1$ is shared by both $|T_{12}|$ and $|R_{11}|$ measurements. 

\section{MPIJIS characterization measurements}

\begin{figure*}
	[tb]
	\begin{center}
		\includegraphics[
		width=1.8\columnwidth 
		]%
		{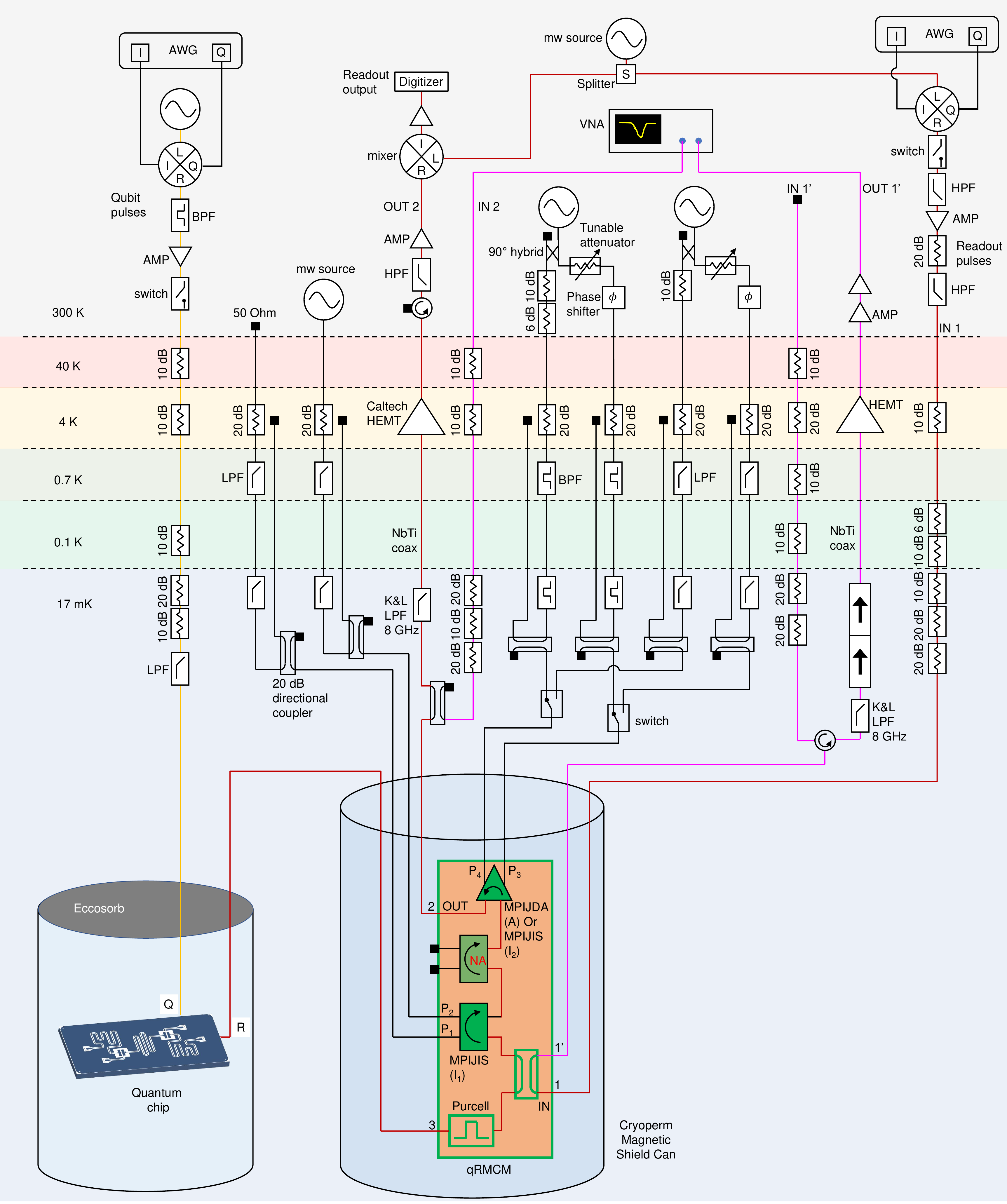}
		\caption{The experimental setup used for taking the measurement results of figures \ref{IsoMagData}, \ref{DoubleIsoData}, \ref{DephasingData}. Input and output lines colored red carry readout signals. Input line colored yellow carry qubit pulses. Input and output lines colored magenta are used to measure the qRMCM in the forward and backward direction. Input and return lines colored black carry pump drives. See text for details.
		}
		\label{Round2FullSetup}
	\end{center}
\end{figure*}

\begin{figure}
	[tb]
	\begin{center}
		\includegraphics[
		width=\columnwidth 
		]%
		{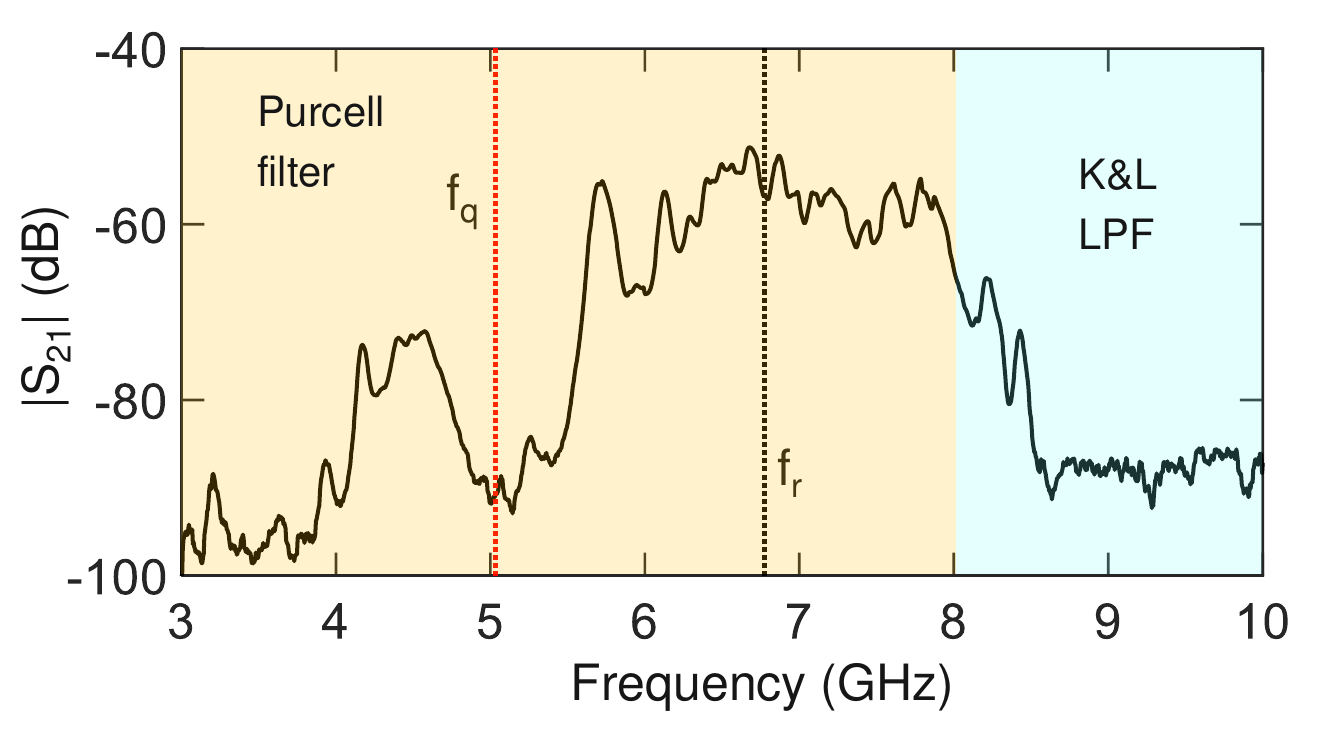}
		\caption{Uncalibrated qRMCM transmission measurement. The transmission measurement is taken using a vector network analyzer in the frequency range $3-10$ GHz. It includes the attenuation of the input line, the transmission of the qRMCM, and the amplification of the output line. The red and black dashed vertical lines indicate the location of the qubit and readout frequencies, respectively. The left- and right-side shaded areas highlight the parts of the transmission landscape of the qRMCM that are dominated by the response of the Purcell filter and the off-the-shelf lowpass filter, respectively.                  
		}
		\label{BroadTransmission}
	\end{center}
\end{figure}

\begin{figure*}
	[tb]
	\begin{center}
		\includegraphics[
		width=1.8\columnwidth 
		]%
		{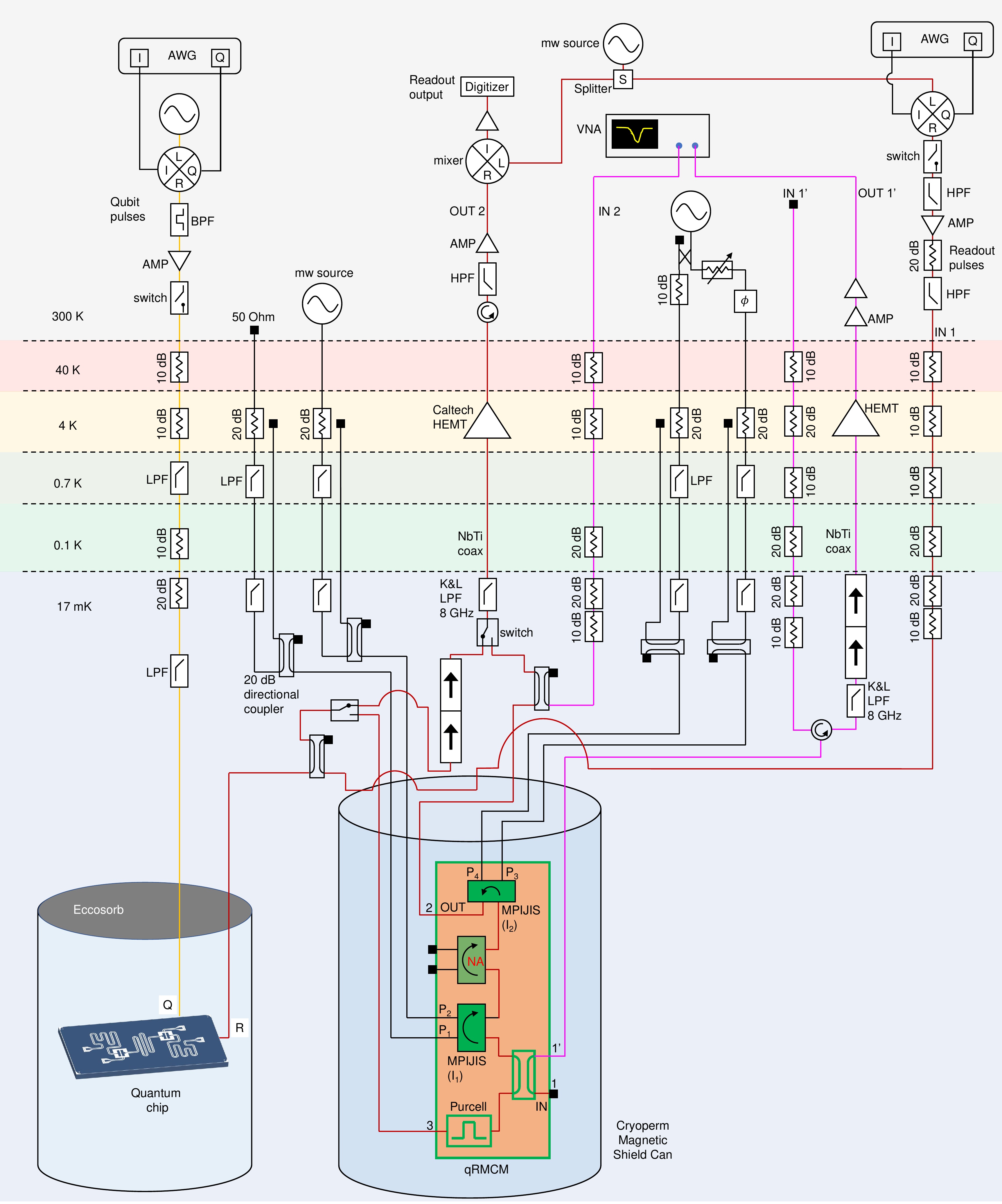}
		\caption{The experimental setup used for taking the measurement results of Fig.\,\ref{magIsolJISsetupSimple}(b)-(c) and  Fig.\,\ref{magIsoJISComp}. Input and output lines colored red carry readout signals. Input line colored yellow carry qubit pulses. Input and output lines colored magenta are used to measure the qRMCM in the forward and backward direction. Input and return lines colored black carry pump drives. See text for details.    
		}
		\label{IsoMagComFullSetup}
	\end{center}
\end{figure*}

In Fig.\,\ref{FluxDependence}(a),(b), we exhibit the dependence of the resonance frequency of mode `a' of $\rm{JPC}_1$ and $\rm{JPC}_2$ on the applied external flux threading the respective JRM. In Fig.\,\ref{FluxDependence}(a) (Fig.\,\ref{FluxDependence}(b)), the resonance frequency of mode `a' of $\rm{JPC}_1$ ($\rm{JPC}_2$) is varied as a function of the external flux applied to $\rm{JRM}_1$ ($\rm{JRM}_2$), while parking the resonance frequency of mode `a' of $\rm{JPC}_2$ ($\rm{JPC}_1$) at its maximum (corresponding to a fixed applied flux threading $\rm{JRM}_2$ ($\rm{JRM}_1$)). The filled red circles represent measured data points, while the black solid curves represent theory fits to the major flux lobes of the JPCs. The theoretical model used to produce the fits is the same as in Refs. \cite{Roch,hybridLessJPC}. It is worthwhile noting that both fits employ the same device parameters. This suggests that fabrication variations on the same chip are small and that the JPCs are indeed nominally identical. From the fits, we extract the following device parameters: $I_0=2.82$ $\mu$A the critical current of the outer Josephson junctions of the JRM, $f_{\rm{max}}=7.0232$ GHz the maximum resonance frequency of mode `a', $Z_{\rm{res}}=51.1$ Ohm the characteristic impedance of the JPC resonators, $L_{\rm{J0}}/L=3.1$ the Josephson inductance ratio between the outer and inner JRM junctions, and $L_{\rm{J0}}/L_{\rm{s}}=5$ the inductance ratio between the JRM outer Josephson junctions and the parasitic series inductance of the superconducting wire in each arm.

The fact that the measured maximum frequencies of the JPCs of the MPIJIS and MPIJDA devices integrated into the qRMCM (all taken from the same wafer) fall within a window of less than $40$ MHz (less than $10$ MHz on the same chip), indicates that the JRM inductance is quite uniform across the wafer and that it is quite feasible, in the future, to operate multiple broadband versions of the MPIJIS/MPIJDA devices using a single microwave pump to further reduce the number of microwave generators required per board or multiple boards.

To investigate the dependence of the dynamical bandwidth of the MPIJIS $\gamma/2\pi$ on the transmission dip in the isolated direction $L_{ij}=\rm{min}{|S_{ij}|^2}$, we plot in Fig.\,\ref{BWvL} using blue circles the extracted bandwidths corresponding to the frequency difference of the $3$ dB points above the minimum of the measured curves (some of which are presented in Fig.\,\ref{VaryPp}). We also plot using magenta stars the calculated bandwidths extracted from the theory curves. As seen in Figs.\,\ref{BWvL}(a),(b), in which $\gamma/2\pi$ is plotted as a function of $\sqrt{L_{12}}$ and $\sqrt{L_{21}}$, respectively, corresponding to the MPIJIS operated in the forward $1\rightarrow2$ and backward $2\rightarrow1$ directions, the calculated bandwidths exhibit a good agreement with the measured data, especially in the limit of vanishing $L_{ij}$, which correspond to large isolations. This behavior mimics the shrinking bandwidths of Josephson amplifiers, such as JPAs and JPCs, in the limit of large gains due to the bandwidth-amplitude gain product, which is characteristic of resonant-structure based parametric devices \cite{JPCreview,microstripJPC}. To highlight this observed behavior in the MPIJIS case, we plot using black dashed line the relation $\gamma=\gamma_0\sqrt{L}$, where $\gamma_0=2\gamma_{a}\gamma_{b}/(\gamma_{a}+\gamma_{b})$, indicated by a black diamond, is the effective linear bandwidth of the JPC without pump $\gamma_0/2\pi=57$ MHz, where $\gamma_{a}/2\pi=40$ MHz and $\gamma_{b}/2\pi=100$ MHz.         

Another related figure of merit is the tunable bandwidth of the device. In Fig.\,\ref{Tunablebw}, we exhibit a tunable bandwidth measurement of the MPIJIS. In this measurement, we vary the magnetic fluxes threading the two JRMs in tandem and adjust the pump power and frequency to yield isolation of more than $18$ dB in the attenuated signal direction $|S_{12}|^2$ (orange curves) or $|S_{21}|^2$ (blue curves) depending on which pump port is driven, i.e., $\rm{P}_1$ in  Fig.\,\ref{Tunablebw}(a) and $\rm{P}_2$ in Fig.\,\ref{Tunablebw}(b). As seen from this measurement, the MPIJIS possesses a tunable bandwidth of about $300$ MHz in both operation modes. This bandwidth is determined and limited by the tunable bandwidths of the two JPCs, the resonance frequency matching of both modes `a' and `b' as a function of applied flux, as well as the bandwidths of the on-chip $90^{\circ}$ hybrids.

Lastly, we measure in Fig.\,\ref{DR} the saturation power of the MPIJIS in the forward (plot a) and backward (plot b) directions for the working point presented in Fig.\,\ref{wpt}. This figure of merit represents the maximum input signal power, which the device can handle for a fixed pump frequency $f_p=2.758$ GHz and pump power before its transmission/isolation response increases or decreases by $1$ dB. As seen in the figure, the saturation power of the MPIJIS, indicated by the red circles, is about $-121$ dBm and it is determined by a change in the isolation magnitude, which, notably, occurs at about $14$ dB lower than the onset of observed change in the transmission magnitude. While this measured saturation power is smaller than the one reported in the proof-of-principle PCB-integrated MPIJIS \cite{MPIJIS} by about $12$ dB, it falls within the measured saturation power range of JPCs operated in amplification or frequency conversion \cite{JPCreview,Conv}. Since the power handling capacity of JPCs are the limiting factor in this case, we expect this figure of merit of the MPIJIS to significantly improve as high-saturation power JPCs become available in the future. 

\section{qRMCM experimental setup}

A detailed diagram of the setup used for measuring the qubit-qRMCM system is shown in Fig.\,\ref{Round2FullSetup}. 
The main results taken with setup are exhibited in Figs.\,\ref{IsoMagData}, \ref{DoubleIsoData}, \ref{DephasingData}.

As seen in Fig.\,\ref{Round2FullSetup}, we mount the qubit chip and the qRMCM in two separate cryoperm magnetic-shield cans. With the qubit shield covered with an eccosorb lid. The setup includes five main set of lines inside the fridge: (1) a qubit input line, colored yellow, that carries qubit pulses and which couples to the qubit via a feedline on the chip. (2) Input and output lines for readout, colored red, which carry input readout signals through the qRMCM towards the readout resonator, and carry the reflected readout signals back through the qRMCM towards the output line that includes, a commercial wideband directional coupler that allows measuring the qRMCM in the opposite direction, an off-the-shelf low-pass filter with a cutoff at $8$ GHz at the mixing chamber and a NbTi superconducting coaxial line that connects the filter to a standard Caltech (SN850D) low-noise HEMT amplifier mounted at the $4$ K stage (spec'd to have a noise temperature of about $3-5$ K in the range $4-11$ GHz when measured at $18.6$ K). Following the HEMT, the readout signals are filtered, amplified, downconverted and digitized at the room-temperature stage using standard electronic equipment. (3) Input and output lines, colored magenta, that are used for setting the working points of the MPIJIS and MPIJDA devices and measuring the qRMCM in the forward and backward direction. (4) Input and return lines, colored black, for feeding the pump drives to the MPIJIS and MPIJDA devices. These input lines include $20$ dB attenuator at the $4$ K and two identical commercial filters, located at the still and the mixing chamber stages, and a commercial wideband directional coupler for attenuating the pump tones by routing unused portions through the return lines, where they get dissipated at $50$ Ohm terminations at the $4$ K stage. To enable the operation of the reconfigurable directional device as an amplifier (MPIJDA) or as an isolator (MPIJIS), we connect it to two sets of pump lines that have different filters incorporated into them. Depending on the desired mode of operation, we switch between pump lines that include lowpass filters with a $5.5$ GHz cutoff that allow the passage of low-frequency pump tones in the MPIJIS case, and bandpass filters that allow the transmission of high-frequency pump tones in the band $11-18$ GHz suitable for the MPIJDA case. The pump lines connected to the single-pump MPIJIS device are hardwired with lowpass filters only (the same kind used for the reconfigurable device). Since the reconfigurable directional device requires feeding two pump signals into its two pump ports, we use a $90$-degree hybrid to evenly split the power of a single microwave generator (in each operation mode). To set the required phase shift between the two pump signals and compensate for any asymmetry in the attenuation of the lines, we incorporate a tunable attenuator and a phase shifter into one of the pump lines at the room-temperature stage. (5) DC lines (not shown) used for flux biasing the small superconducting coils attached to the directional Josephson devices. These lines consist of two parts,  resistive dc lines from room-temperature to the $4$ K stage followed by superconducting dc lines from the $4$ K stage to the mixing chamber.  

Figure\,\ref{BroadTransmission}, shows a broadband transmission measurement result of the qRMCM experiment setup taken using a vector network analyzer. The measurement includes the attenuation of the input line, the transmission of the qRMCM, and the amplification of the output line. The main features seen in this measurement result is set by the Purcell filter incorporated into the qRMCM, namely the transmission peak plateau seen surrounding the readout frequency and the strong attenuation encompassing the qubit frequency. The locations of these two frequencies are indicated using black dashed vertical lines. 

Further details about the design, fabrication, packaging, and performance of the Purcell filter and the superconducting directional coupler can be found in Refs.\cite{MPIJIS,Bronn2015b}.  \\  

\subsection{qRMCM versus magnetic isolators setup}

Figure\,\ref{IsoMagComFullSetup} exhibits a detailed diagram of the setup used for comparing the performance of the qRMCM and two commercial broadband magnetic isolators in the output chain. In particular, this setup is used for taking the data results exhibited in Fig.\,\ref{magIsolJISsetupSimple}(b)-(c), and Fig.\,\ref{magIsoJISComp}. In general, this setup is very similar to the one exhibited in Fig.\,\ref{Round2FullSetup}. It mainly differs from the previous setup in the arrangement of the attenuation and filtering on some of the input lines and in the incorporation of two microwave switches and another wideband directional coupler, which enable us to measure the qubit using the qRMCM and the magnetic isolators in the same cooldown.      

\subsection{Qubit measurement parameters}
The qubit used in the qRMCM experiment is a single Josephson junction transmon with anharmonicity of $330$ MHz capacitively coupled to a superconducting waveguide resonator measured in reflection. The qubit chip has a separate input port through which we control the qubit, which is separate from the readout input line connected to the qRMCM. The qubit and readout frequencies are $f_q=5.033703$ GHz and $f_r=6.77698$ GHz. The readout resonator bandwidth is $\kappa/2\pi=1.1$ MHz, while the qubit-state dependent readout frequency shift is $\chi/2\pi=0.5$ MHz. The readout pulse duration and integration time applied in the MPIJIS-MPIJDA, MPIJIS-MPIJIS, and variable isolation experiments is $T_{\rm{m}}=0.75$ $\mu$s with an average photon number $\bar{n}_{\rm{m}} \backsimeq 30$. The qubit data is averaged over $2000$ iterations. The pump 
frequency applied to the MPIJIS device in the MPIJIS-MPIJDA, MPIJIS-MPIJIS, and variable isolation experiments is $2.712$ GHz, while the pump frequency applied to the reconfigurable directional device in the first two experiments is $16.452$ GHz and $2.648$ GHz, respectively. In the comparison experiment (against the magnetic isolators, see Fig.\,\ref{magIsolJISsetupSimple}) the pump frequencies applied to the $\rm{MPIJIS_1}$ and $\rm{MPIJIS_2}$ are $2.654$ GHz and $2.68$ GHz, respectively, and we set $T_{\rm{m}}=1.5$ $\mu$s. \\

\section{Detection scheme of orientation parity of weak magnetic sources}

\begin{figure*}
	[tb]
	\begin{center}
		\includegraphics[
		width=1.4\columnwidth 
		]%
		{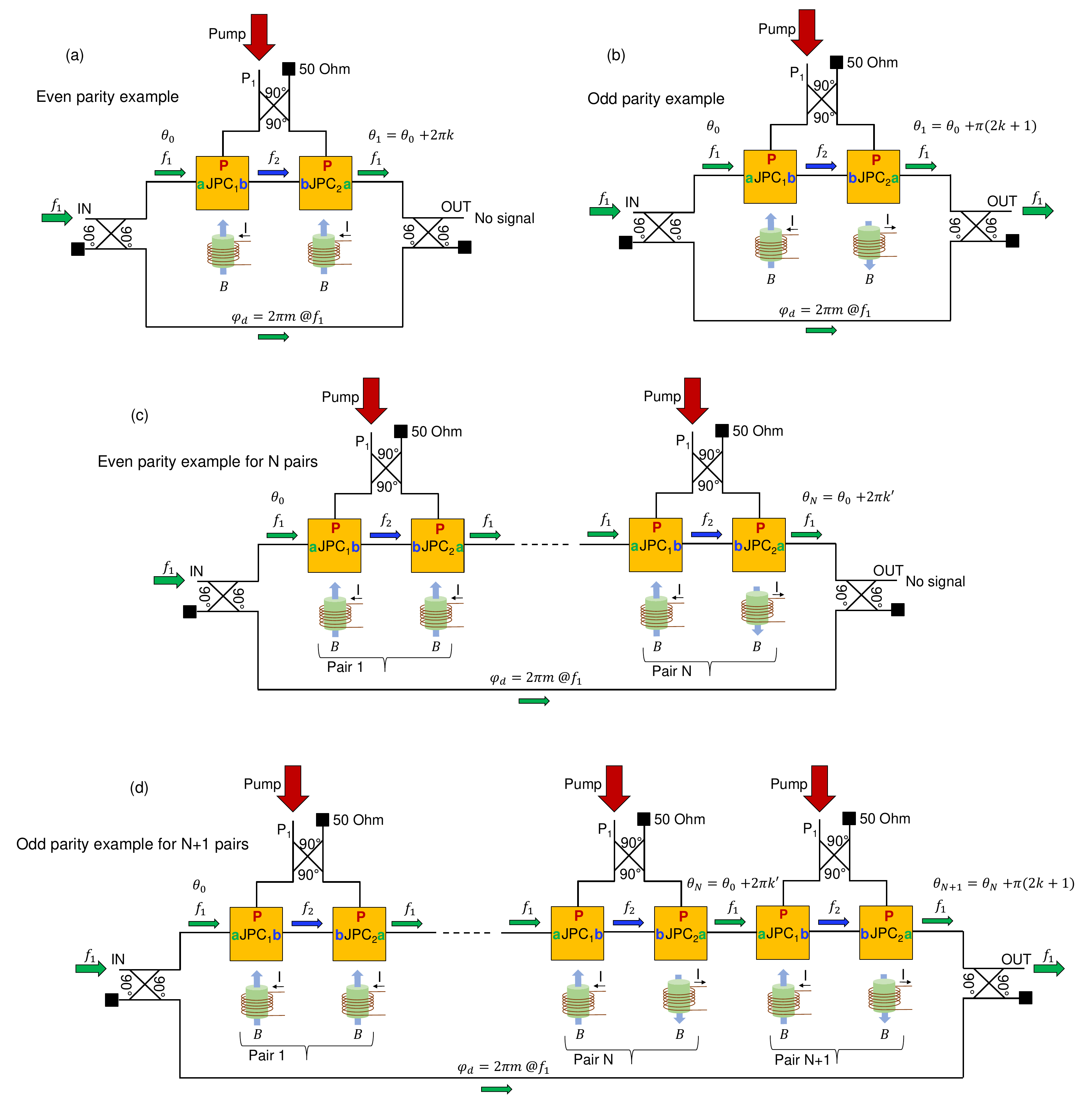}
		\caption{Parity detection scheme of the orientation of weak magnetic sources. (a) and (b) depict interferometers formed by two $90^{\circ}$ hybirds, which interfere waves propagating in two paths, i.e., top and bottom. The top path includes a pair of JPCs coupled back-to-back via their `b' mode. The two JPCs are operated in full frequency conversion mode. They are driven by a single pump tone, which is fed through a third $90^{\circ}$ hybrid that couples to the pump ports of the JPCs. The bottom path includes a lossless delay line, whose length is equal to an integer number of wavelengths at $f_1$. The scheme measures the transmission of a probe signal at $f_1$ input on the IN port of the left hand-side hybrid and output at the OUT port of the right hand-side hybrid. Minimum or maximum transmission though the device, resulting from respective destructive and constructive wave interference, is obtained if the orientation parity of the magnetic fluxes flux-biasing the JPCs is even (a) or odd (b). (c) A generalized detection scheme for N pairs of magnetic sources, which flux-bias N pairs of JPCs that are incorporated into the top path. The scheme response is illustrated for the case of an even parity. (d) An extended version of the scheme in (c) with a total of N+1 pairs of JPCs and magnetic sources in the top path. In this case, the scheme response is illustrated for an odd parity measurement.
		}
		\label{ParityDetection}
	\end{center}
\end{figure*}

As demonstrated in Fig.\,\ref{ParityMag}, the isolation direction of MPIJIS is determined not only by the phase gradient between the pumps feeding the two JPCs but also by the parity of the magnetic fields flux biasing their JRMs. In this section, we show how this interesting property could potentially be used to detect the orientation parity of weak magnetic sources via direct microwave transmission measurements. In Fig.\,\ref{ParityDetection}(a),(b), we show an exemplary detection setup that relies on this property. The setup acts as a two-path interferometer for coherent microwave signals at frequency $f_1$. It consists of an input and output quadrature couplers, i.e., $90^{\circ}$ hybrids, whose inner ports are coupled through two paths. The first (top) path, incorporates a gyrator formed by two JPCs coupled via their `b' mode \cite{JPCgyrator}. The second (bottom) path, includes a lossless transmission line, whose electrical length is an integer multiple of the signal wavelength at $f_1$. The signal frequency $f_1$ used in the scheme lies within the bandwidth of mode `a' of the JPCs. The two JPCs are operated in frequency conversion mode without photon gain, i.e., signals entering port `a' (`b') are transmitted to port `b' (`a') with frequency conversion, and driven through a dedicated $90^{\circ}$ hybrid using a single pump drive, whose frequency is set to $f_p=f_2-f_1$, where $f_2$ lies within the bandwidth of mode `b'. Assuming, without loss of generality, that the pump driving the gyrator is fed through port $\rm{P}_1$ of the hybrid ($\varphi_p=-\pi/2$), the microwave signals transversing the gyrator acquire a differential phase, i.e., $2\pi k$ or $(2k+1)\pi$, $k \in \mathbb{Z}$, depending on the propagation direction of the signals (i.e., from $\rm{JPC}_1$ to $\rm{JPC}_2$ or vice versa) and the direction parity of the magnetic fields flux-biasing the JPCs, as illustrated in Figs.\,\ref{ParityDetection}(a),(b). If we further restrict ourselves, for simplicity, to transmission measurements of signals entering the upper external port of the left $90^{\circ}$ hybrid and exiting via the upper external port of the right $90^{\circ}$ hybrid, we obtain zero transmission (no output signal, `0') in the case of even parity of the biasing magnetic fields (Fig.\,\ref{ParityDetection}(a)) and unity transmission (an output signal, `1') in the case of odd parity of the biasing magnetic fields (Fig.\,\ref{ParityDetection}(b)). In the former (latter) case, the zero (unity) transmission is the result of a destructive (constructive) interference between waves propagating in the upper and bottom paths of the interferometer, which acquire opposite (equal) phases.   

It is important to point out that this proposed magnetic field parity detection scheme is not limited to two magnetic sources. It can be applied, in principle, to N pairs of weak independent magnetic sources as shown in Fig.\,\ref{ParityDetection}(c), which flux bias N gyrators incorporated in series into the upper path of the interferometer. To prove the validity of this generalization, we apply the induction method. The first (base) step of the proof, in which we demonstrate the validity of the measurement setup for a given N, is shown in Figs\,\ref{ParityDetection}(a),(b) for the case N=1. In the second (inductive) step, we demonstrate that the detection scheme is valid for N+1 pairs, given that it is valid for N, as partially illustrated in Fig.\,\ref{ParityDetection}(d). To do that we consider four different cases: 

Case 1: the first N pairs of magnetic sources have an even number of aligned magnetic fields and the two magnetic fields biasing the (N+1)th gyrator are aligned in the same direction (i.e., have an even parity as well). Since the parity detection scheme works for the first N pairs (resulting in a destructive interference at the output), this means that the accumulated phase of the transmitted signal via the upper path is $\theta_N-\theta_0=2 \pi k'$. By adding the (N+1)th pair having an even parity (similar to the case of Fig.\,\ref{ParityDetection}(a)), the total accumulated phase at the output of the (N+1)th pair is given by $\theta_{N+1}-\theta_0=2 \pi k''$, where $k''=k+k'$ and $k, k', k'' \in \mathbb{Z}$. Since the bottom path (including the diagonal lines of the hybrids) always yields an accumulated phase of $(2m+1) \pi$, where $m \in \mathbb{Z}$, the waves propagating in the upper and bottom paths will destructively interefere at the output, thus confirming that the setup is valid for N+1 pairs in this case. 

Case 2: the first N pairs of magnetic sources have an even number of aligned magnetic fields and the two magnetic fields biasing the (N+1)th gyrator are aligned in the opposite directions (i.e., have an odd parity), which means that the N+1 pairs have odd parity in total. By concatenating the (N+1)th pair having an odd parity (similar to the case of Fig.\,\ref{ParityDetection}(b)), the total accumulated phase at the output of the (N+1)th pair is given by $\theta_{N+1}-\theta_0=(2k''+1)\pi$. Consequently, the waves propagating along the upper and bottom paths will constructively interfere at the output (indicating an odd parity of the whole chain), thus confirming that the setup is valid for N+1 pairs in this case as well.

Case 3: the first N pairs of magnetic sources have an odd number of aligned magnetic fields and the two magnetic fields biasing the (N+1)th gyrator are aligned in the opposite directions (i.e., have an odd parity). This means that the N+1 pairs have even parity in total. It is straightforward to show that the total accumulated phase at the output of the (N+1)th pair is $\theta_{N+1}-\theta_0=2\pi k''$, which yields a destructive interference (no signal) at the device output, which is indicative of even parity. 

Finally, the fourth and last case is similar to case 3 except that the (N+1)th gyrator is flux biased with two aligned magnetic fields (i.e., having an even parity as in Fig.\,\ref{ParityDetection}(a)). In this case, the whole chain (N+1 pairs) will preserve the odd parity of the N pairs. Since the total accumulated phase at the output of the (N+1)th pair is given by $\theta_{N+1}-\theta_0=(2k''+1) \pi$, the resultant wave interference at the output of the device is constructive, which, in turn, yields an output signal indicative of odd parity.         

Such interferometric microwave detection schemes could potentially be used in a variety of applications. For example, they can be used to perform a \textit{Z} operator parity checks on flux qubits or capacitively shunted flux qubits (CSFQs) using direct microwave transmission measurements (as opposed to (1) using ancilla qubits coupled to data qubits, (2) entangling them via single- and two-qubit gates, and (3) measuring the final state of the ancilla qubits as done in the surface code architecture). In such an application, each CSFQ is coupled to a JRM via a suitable mutual inductance. Since the qubit state is encoded in the direction of the persistent current in the CSFQ loop, opposite circulating currents would introduce opposite magnetic flux biases through the JRMs, which would determine the transmitted signal through the interferometric device based on the parity of the qubit states. As another example, they can be used to determine the strength and orientation of weak magnetic sources. This can be done by (1) coupling the source of the unknown magnetic field to one JRM of the interferometric gyrator device, (2) flux biasing the other JRM using a controlled and well-characterized magnetic field, (3) injecting a pump tone to either $\rm{P}_1$ or $\rm{P}_2$, (4) applying a microwave signal to the device input, and (5) maximizing the transmitted microwave through the device by varying the power and frequency of the applied pump as well as the magnetic field of the controlled magnetic source. For a pump tone applied to $\rm{P}_1$, for example, a maximum output signal is expected when the two magnetic fluxes threading the two JRMs have equal magnitudes but opposite signs (as seen in Fig.\,\ref{ParityDetection}(b)).

We can estimate the range of magnetic fields $B_{\rm{ext}}$ that are typically applied to JRM loops of area $A$ at the JRM plane, by considering the bounds $0.1\Phi_0/A<|B_{\rm{ext}}|<\Phi_0/A$. For a JRM loop area of $A=$ 100 $\mu$m x 100 $\mu$m, the range is $2\cdot 10^{-8} \rm{T} < |B_{\rm{ext}}| < 2\cdot 10^{-7} \rm{T}$. Much smaller field bounds can be detected using a larger loop. However, bigger loops could add series linear inductance to the JRM, which is undesired (especially when it is comparable to the Josephson inductance of the JJs).

\end{document}